\documentclass[letterpaper,11pt]{article}
\pdfoutput=1
\usepackage{jheppub}
\usepackage{hyperref}

\usepackage{setspace}
\usepackage{inputenc}
\usepackage{cancel}
\usepackage{amsmath, amssymb}
\usepackage{xfrac} 
\usepackage{amsfonts} 
\usepackage{graphicx}
\usepackage{multirow} 
\usepackage{hyperref}
\usepackage{xcolor}
\usepackage{pifont}
\usepackage{floatrow}
\usepackage{slashed} 
\usepackage{cancel} 
\usepackage{comment} 
\usepackage{footmisc} 
\usepackage{physics} 
\usepackage{makecell} 
\usepackage{subcaption}
\usepackage[toc,page]{appendix}
\usepackage{placeins} 

\definecolor{nicered}{rgb}{0.6,0,0}
\definecolor{nicegreen}{rgb}{0.1,0.5,0.1}
\definecolor{niceblue}{rgb}{0,0.4,0.8}
\definecolor{newgreen}{rgb}{0,0.667,0}
\hypersetup{
	colorlinks=true,      
	linkcolor=black,      
	citecolor=blue,       
	filecolor=magenta,     
	urlcolor=blue
}

\newcommand{\cmark}{\textcolor{green}{\ding{51}}}
\newcommand{\xmark}{\textcolor{red}{\ding{55}}}
\newcommand{\gdm}{g_\text{DM}}
\newcommand{\mdm}{m_\text{DM}}

\title{\boldmath Impact of Sommerfeld Effect and Bound State Formation in Simplified $t$-Channel Dark Matter Models}

\author[a]{Mathias Becker,}
\author[a]{Emanuele Copello,}
\author[a]{Julia Harz,}
\author[b]{Kirtimaan A. Mohan,}
\author[c,d]{and Dipan Sengupta}

\affiliation[a]{Physik Department T70, Technische Universit\"at M\"unchen, 85748 Garching, Germany}
\affiliation[b]{Department of Physics and Astronomy, \\ 567 Wilson Road, East Lansing, Michigan-48824, USA}
\affiliation[c]{Department of Physics and Astronomy, 9500 Gilman Drive, \\ University of California, San Diego, USA}
\affiliation[d]{ARC Centre of Excellence for Dark Matter Particle Physics, Department of Physics, \\ University of Adelaide, South Australia 5005, Australia}

\emailAdd{mathias.becker@tum.de}
\emailAdd{emanuele.copello@tum.de}
\emailAdd{julia.harz@tum.de}
\emailAdd{kamohan@msu.edu}
\emailAdd{disengupta@physics.ucsd.edu}

\abstract{The existence of a dark matter model with a rich dark sector could be the reason why WIMP dark matter has evaded its detection so far. For instance, colored co-annihilation naturally leads to the prediction of heavier dark matter masses. Importantly, in such a scenario the Sommerfeld effect and bound state formation must be considered in order to accurately predict the relic abundance. Based on the example of the currently widely studied $t$-channel simplified model with a colored mediator, we demonstrate the importance of considering these non-perturbative effects for correctly inferring the viable model parameters. We emphasize that a flat correction factor on the relic abundance is not sufficient in this context. Moreover, we find that parameter space thought to be excluded by direct detection experiments and LHC searches remains still viable. Additionally, we illustrate that long-lived particle searches and bound-state searches at the LHC can play a crucial role in probing such a model. We demonstrate how future direct detection experiments will be able to close almost all of the remaining window for freeze-out production, making it a highly testable scenario.} 
\begin{document}

\maketitle
\flushbottom






\section{Introduction}
\label{sec:Introduction}

One of the prominent outstanding puzzles of particle physics and cosmology is the origin and nature of the most abundant matter component of the universe, namely Dark Matter (DM).
Apart from its manifestation through gravitational effects affecting several astrophysical observations of ``ordinary'' matter (assuming the DM has a particle physics origin), we do not have a clear picture about its nature.
For decades, one of the most promising candidates for DM have been `weakly interacting massive particles' (WIMPs), cold non-relativistic particles that are able to interact with the primordial thermal bath of Standard Model (SM) particles (and potentially with other species beyond the SM)  in the early universe.
In this period of time, they can annihilate into SM particles and, as the universe expands and cools, they freeze-out leaving behind a relic that we can observe today as a clear imprint on, for example, the spectrum of the cosmic microwave background \cite{Planck2018}.
A number of beyond the standard model (BSM) physics scenarios predict particles that can serve the purpose of WIMP dark matter, allowing us to formulate strategies to try to detect them \cite{Arcadi:2017kky}.

Over the last few decades, a large variety of experiments have been searching for Dark Matter (DM), ranging from the production of dark sector particles at colliders \cite{Abulaiti:2799299} to direct detection of DM-nucleon scattering on earth \cite{XENON:2017vdw,PICO:2017tgi} as well as the observation of astrophysical signals of annihilation and/or decay products of dark sector particles \cite{HESS:2013rld}.
A conclusive positive (direct) signal for DM is still missing, while the data from these experiments can be used to constrain models of DM. 

One possibility for why WIMP dark matter has evaded detection so far, could be the existence of a rich dark sector. 
Such scenarios open up for the possibility of colored co-annihilation~\cite{Harz:2012fz,Ellis:2014ipa,Harz:2014gaa,Ibarra:2015nca,Baker:2015qna,Harz:2016dql,ElHedri:2018atj,Schmiemann:2019czm,Branahl:2019yot}, which naturally lead to a heavier expected dark matter mass, evading current experimental searches.
These types of models are currently intensively studied by the theoretical and experimental community in form of so-called $t$-channel simplified models~\cite{Arina:2020tuw,Arina:2020udz,Abdallah:2015ter}. 
In particular, the interplay of the aforementioned experimental searches and the requirement not to overproduce dark matter allows for complementary constraints on the model parameters.

If the DM-SM coupling is sizable, the evolution of DM proceeds via the thermal freeze-out of DM, which typically takes place at temperatures of $T  \sim \mdm/30$ when DM itself is non-relativistic. 
In scenarios where dark sector particles interact via a light mediator, long-range effects such as the Sommerfeld Effect (SE) \cite{Sommerfeld:1931qaf,Sakharov:1948plh} will enhance (diminish) the annihilation cross-section \cite{Hisano:2002fk} due to the resulting attractive (repulsive) potential, altering the theoretically predicted dark matter abundance~\cite{Drees:2009gt,Beneke:2014hja,Beneke:2019qaa,Hisano:2006nn,Harz:2014gaa}.

Additionally, dark sector bound states can radiatively form via the emission of the light mediator particle.
Their decays effectively deplete the dark-sector particle densities, acting as an additional effective annihilation channel for DM and therefore affecting the theoretical prediction of the dark matter relic density~\cite{vonHarling:2014kha,Ellis_2015,An:2016gad,Asadi:2016ybp,Petraki:2016cnz,Liew:2016hqo,Binder:2020efn}. 
Based on the methods developed in \cite{Petraki_2015}, it has been shown that Sommerfeld enhancement and bound state formation (BSF) arising from non-Abelian interactions can lead to $\mathcal{O}(50\%-300\%)$ corrections to the theoretically predicted dark matter abundance \cite{Harz:2018csl,toolbox,ElHedri:2017nny}.
Given this sizeable correction, we demonstrate that taking into account the aforementioned non-perturbative effects is an imperative task in order to derive correct exclusion limits in colored simplified $t$-channel models. 
We also stress that the same argument could be extended to any model featuring colored co-annihilation or similar conditions. 

In particular, we show, compared to previous works on this subject, that several important effects which had not been previously taken into account significantly expands the scope of both the model phenomenology and the complementary ways of experimental detection. Specifically, we point out the following salient features of this work:

\begin{itemize}
    \item So far, previous work considering non-perturbative effects in co-annihilation scenarios assumed that the DM annihilation cross-section is solely determined by the annihilations of the mediators charged under a non-Abelian gauge group \cite{Harz:2018csl,toolbox}. 
    We also take into account the effects of a non-zero coupling of the mediator to the DM candidate, which is mandatory for annihilations of the mediator to impact the DM relic abundance. 
    Furthermore, the presence of this coupling is crucial to test the model at various experiments. 
    We provide a detailed analysis of the interplay of current and future experimental searches and point out the areas of parameter space where bound-state effects are large.
    \item When computing the BSF rate, we include the effects of the three-gauge-boson vertex, as described in \cite{Harz:2018csl}, which was neglected previously \cite{toolbox,Ellis_2015}. 
    Additionally, we estimate the effects of a non-negligible mediator decay width on the efficiency of DM depletion induced by bound-state effects.
    \item In \cite{Mohan_2019}, two of the authors of this work analyzed the simplified $t$-channel models described in this paper by calculating direct detection and collider limits at next-to-leading order (NLO) and showed that spin-independent direct detection limits at NLO can be significantly more important than the corresponding tree-level spin-dependent part. 
    However, in the aforementioned work, the co-annihilation scenario and the impact of the Sommerfeld enhancement as well as of bound state formation were not taken into account in the calculation of the DM relic density. 
    The present paper aims at showing the full impact of both these improvements. 
    \item While recently the effects of BSF on the relic density have been considered in the context of the conversion-driven freeze-out \cite{garny2021bound} and the superWIMP mechanism \cite{Bollig:2021psb}, we investigate in this article, the scenario where DM is in thermal equilibrium with the dark sector and the SM.
    \item In order to derive exclusion limits, we fix the DM-SM coupling by the requirement not to exceed the observed dark matter abundance. This differs from previous works on the subject \cite{Arina:2020tuw,Arina:2020udz}, where the DM-SM coupling was fixed requiring the mediator decay width to equal $5 \%$ of its mass. The latter approach, however, would overestimate the experimental conclusion limits in the mass-compressed region, which is the main focus of this work.
\end{itemize}

For this purpose, we focus on a class of simplified $t$-channel DM models called $u_R$, $d_R$ and $q_L$ ~\cite{Mohan_2019} (or \texttt{S3M\_uR}, \texttt{S3M\_dR} and \texttt{S3M\_QL}, in the spirit of~\cite{Arina:2020tuw,Arina:2020udz}), where DM is a Majorana fermion $\chi$ and has a Yukawa interaction via a color triplet scalar $X$ with a SM quark.

Interestingly, the model parameters, the coupling of DM to Standard Model (SM) particles $\gdm$, the DM mass $\mdm$ and the mass splitting in the dark sector $\Delta m$, are constrained by a combination of cosmological, collider and direct detection limits. 
In this work, we demonstrate how the corresponding experimental exclusion limits and expected experimental signatures are altered when considering the effect of Sommerfeld enhancement and bound state formation on the DM relic abundance.

We find that:
\begin{itemize}
    \item 
    When including SE, we find that the maximally allowed DM mass and mass splitting in the co-annihilating region of the parameter space are shifted from $(\mdm, \Delta m) \lesssim (1 \, \mathrm{TeV}, 30\, \mathrm{GeV})$ to $(\mdm, \Delta m) \lesssim (1.4 \, \mathrm{TeV}, 40\, \mathrm{GeV})$, leading to a correction around $\left(40 \% , 33 \% \right)$.
    \item BSF provides a significant correction to the annihilation cross-section when $\gdm < g_s$. 
    In this regime, it adds up to the SE, leading to a stronger impact and must be included in order to obtain a legitimate result.  
    Since BSF effectively acts as an additional annihilation channel, it always increases the effective annihilation cross-section.  
    We find that including the SE and BSF leads to a shift of the maximally allowed DM mass and mass splitting from $(\mdm, \Delta m) \lesssim (1 \, \mathrm{TeV}, 30\, \mathrm{GeV})$ to $(\mdm, \Delta m) \lesssim (2.4 \, \mathrm{TeV}, 50\, \mathrm{GeV})$, hence a correction around $\left(140 \% , 66 \% \right)$.
    \item Depending on the parameters considered, corrections from the SE can be positive or negative, while BSF always increases the annihilation cross-section. 
    As a result, a simple flat correction factor (``K-factor") is not sufficient for an approximate consideration of these effects.
    \item In case of an observation, considering SE and BSF is important in order to infer the correct model parameters ($\mdm$, $\Delta m$, $\gdm$).
    \item SE and BSF extend the parameter region that would lead to underabundant thermal dark matter, where the observed DM relic density can only be produced by non-thermal DM production. Consequently, this region might not be fully probed by long-lived particle searches in contrast to the expectation without the inclusion of non-perturbative effects.
    \item We find that bound state searches at the LHC offer a unique opportunity to constrain couplings in the region of $10^{-6} \lesssim \gdm \lesssim 10^{-2}$, which otherwise typically evades the constraints from prompt collider searches, direct detection or long-lived-particle searches. 
\end{itemize}

The analysis leading to these conclusions is organized as follows: In section \ref{sec:model}, we give a detailed description of the simplified $t$-channel models analyzed and the relevant processes for the relic abundance calculation. 
In section~\ref{sec:long_range_effects}, we illustrate some theoretical aspects of SE and BSF in the simplified $t$-channel model and we discuss their impact on the relic density and the model parameters in~\ref{sec:relicdensity}.
In section~\ref{sec:DD}, we summarize the constraints utilized from spin-independent and spin-dependent searches, while in section~\ref{sec:colliders}, we explain how we exploit prompt collider searches, including the search for BSF at the LHC, and long-lived particle signatures.
Finally, in section~\ref{sec:Results}, we present our combined results and we elaborate on the interplay of the various constraints and their potential to exclude parts of the parameter space. 
Most importantly, we discuss the impact of SE and BSF on the estimation of the correct exclusion limits. 
Moreover, we show the corresponding projected exclusion constraints from future experiments and highlight the potential reach of long-lived particle searches and of searches for dark sector bound states at the colliders. 
We conclude in section \ref{sec:Conclusions}.
\section{Simplified $t$-channel models and Dark Matter Cosmology}
\label{sec:model}
In this section, we briefly describe the $t$-channel simplified model \cite{DiFranzo_2013,Mohan_2019,Arina:2020tuw,Arina:2020udz,Abdallah:2015ter} as well as the various thermally averaged cross-sections that are relevant for evaluating the DM relic abundance. The $t$-channel model we consider consists, in addition to the SM,  a SM-singlet Majorana fermion $\chi$ which is  the lightest dark sector particle, and three color-triplet complex scalar fields $X_i$ ($i$ indicates the generation) which interact with $\chi$ and the SM quarks via a Yukawa coupling $g_{\text{DM}}$. The scalars are charged under the SM gauge group $(SU(3)\times SU(2))_Y$ and its simplest form there are three possible quantum number assignments possible:
\begin{equation}
    (3,1)_{2/3},\quad(3,1)_{-1/3},\quad(3,2)_{-1/6}.
\end{equation}
The three possible choices of the mediator's quantum numbers correspond to three different models, which we label as the $u_R$, $d_R$ and $q_L$ models, respectively.

The dark sector features a $\mathbb{Z}_2$ symmetry such that $\chi$ is the lightest stable particle and our DM candidate.
The interaction Lagrangian of the dark sector particles is thus given by:
\begin{equation}
    \mathcal{L}\supset\sum_{i}(D_\mu X_i)^\dagger(D^\mu X_i)+\sum_{i,j} \left(g_{\text{DM},ij}X_i^\dagger \Bar{\chi}P_R q_j + g_{\text{DM},ij}^* X_i\Bar{q}_j P_L \chi\right)\ ,
    \label{eq:Lagr}
\end{equation}
where $D_\mu$ is the covariant derivative and the index $i$ runs over the quark and mediator flavours of the model considered (up-type right-handed quarks, down-type right-handed quarks and left-handed quarks). $P_L$ and $P_R$ are the left and right handed projectors respectively. The Yukawa couplings,$g_{DM}$, are chosen to be real valued, flavour-diagonal and flavour-universal  for simplicity, implying that $g_{\text{DM},ij}=g_\text{DM}\delta_{ij}$ \footnote{ In general it is possible to go beyond this approximation by allowing for off-diagonal Yukawa couplings, which can be constrained by flavor observables as for instance top quark flavor changing neutral currents \cite{Liu:2021crr}.}. 

\begin{figure}[!t]
\centering
\begin{subfigure}{.3\textwidth}
\centering
\includegraphics[width=0.49\textwidth]{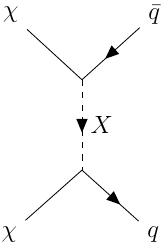}
\caption{DM Annihilation} \label{fig:processCC}
\end{subfigure}
\begin{subfigure}{.3\textwidth}
\centering
\includegraphics[width=0.7\textwidth]{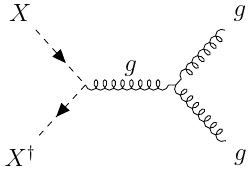}
\caption{Colored Annihilation}\label{fig:processXXd1} 
\end{subfigure}
\begin{subfigure}{.3\textwidth}
\centering
\includegraphics[width=0.55\textwidth]{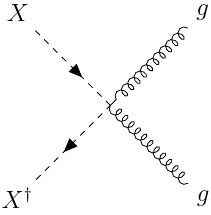}
\caption{Colored Annihilation}\label{fig:processXX4} 
\end{subfigure}
\\[2pt]
\begin{subfigure}{.3\textwidth}
\centering
\includegraphics[width=0.49\textwidth]{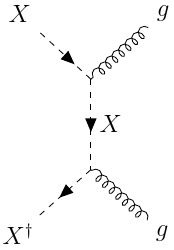}
\caption{Colored Annihilation}\label{fig:processXXd3}
\end{subfigure}
\begin{subfigure}{.3\textwidth}
\centering
\includegraphics[width=0.7\textwidth]{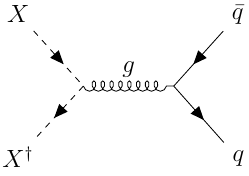}
\caption{Colored Annihilation}\label{fig:processXXd2}
\end{subfigure}
\begin{subfigure}{.3\textwidth}
\centering
\includegraphics[width=0.49\textwidth]{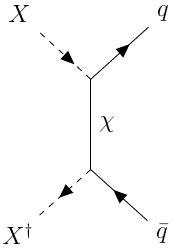}
\caption{Colored Annihilation}\label{fig:processXXdqq}
\end{subfigure}
\\[2pt]
\begin{subfigure}{.3\textwidth}
\centering
\includegraphics[width=0.49\textwidth]{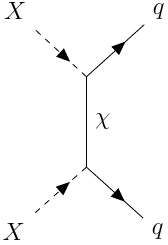}
\caption{Colored Annihilation}\label{fig:processXXqq}
\end{subfigure}
\begin{subfigure}{.3\textwidth}
\centering
\includegraphics[width=0.7\textwidth]{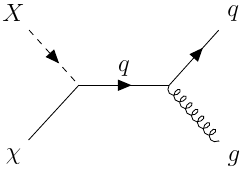}
\caption{Co-annihilation}\label{fig:processXcs}
\end{subfigure}
\begin{subfigure}{.3\textwidth}
\centering
\includegraphics[width=0.49\textwidth]{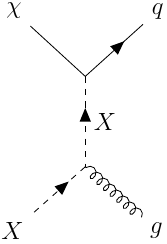}
\caption{Co-annihilation} \label{fig:processXct}
\end{subfigure}
\caption{\small Representative Feynman diagrams corresponding to the main subset of processes contributing to (co-)annihilations described in Tab.~\ref{tab:processes}. For simplicity, we don't show here possible interfering diagrams from crossing symmetries (for example, we would have an $u$-channel for $\chi\chi\rightarrow q\bar{q}$, $XX^\dagger\rightarrow q \bar{q}$ and $XX\rightarrow q q$).
We also only illustrated the gluon gauge vertices, since the strong coupling dominates.
In diagram \ref{fig:processXcs}, we could also have the interaction of any other SM gauge boson with the quarks, while in diagrams \ref{fig:processXXd1}, \ref{fig:processXX4}, \ref{fig:processXXd3}, \ref{fig:processXXd2} and \ref{fig:processXct}, the gluon can be replaced by a photon or a Z boson in the $u_R$ and $d_R$ models and additionally by $W^{\pm}$ bosons in the $q_L$ model.}
\label{fig:feynm_diagr}
\end{figure}

As a result of its Yukawa interaction \footnote{For this work, we do not consider possible renormalizable interactions between the scalars and the Higgs field, for example via a trilinear coupling. As shown in \cite{Harz_2018_HiggsEnh,Harz_2019,Biondini:2018xor}, such interactions can lead to sizeable effects from Sommerfeld enhancement and bound state formation and is subject to a follow-up work~\cite{Becker_?}\label{fn:Higgs_tril}.}, the DM number density depends both on direct pair annihilation of DM, particles, $\chi\chi\rightarrow \text{SM SM}$, as well as co-annihilation and colored annihilations processes into SM particles involving $\chi-X$, $\chi-X^\dagger$, $X-X^\dagger$ and $X-X$ as initial scattering states. The latter determining the density of the scalar mediators. 
A representative class of Feynman diagrams are shown in Fig.~\ref{fig:feynm_diagr}.

Under the assumption that all $\mathbb{Z}_2$-odd particles will finally decay into dark matter and will be in equilibrium with each other until freeze-out \cite{Edsj__1997}, we track the evolution of the total dark-sector comoving number density (or \textit{yield}) $\tilde{Y}=\tilde{n}/s$, where $\tilde{n}$ is the total dark-sector number density and $s$ is the entropy density of the universe, resulting in the sum of the comoving yields of the co-annihilating $\mathbb{Z}_2$-odd species:
\begin{equation}
    \tilde{Y}=Y_\chi+\sum_{i=u,c,t} \bigg(  Y_{X_i}+Y_{X_i}^{\dagger} \bigg) =Y_\chi+2\sum_{i=u,c,t} Y_{X_i}.
    \label{eq:totYield}
\end{equation}
We can then write an effective Boltzmann equation for $\tilde{Y}$ as a function of the variable $x=m_\text{DM}/T$ in the following form:
\begin{equation}
    \dfrac{\dd \tilde{Y}}{\dd x}=-c\,g_{*,\text{eff}}^{1/2}\dfrac{\langle\sigma_{\text{eff}} v_{\text{rel}}\rangle}{x^2}\left(\tilde{Y}^2-\tilde{Y}^2_{\text{eq}}\right),
    \label{eq:BoltzmannEq}
\end{equation}
where 
\begin{align}
&c = \sqrt{\pi/45}\, m_{\text{Pl}}\,m_\chi,\\
&g_{*,\text{eff}}^{1/2}=\dfrac{g_{*S}}{\sqrt{g_*}}\left(1+\dfrac{T}{3g_{*S}}\dfrac{\dd g_{*S}}{\dd T}\right),\\
&Y_{\chi}^{\text{eq}}\simeq\dfrac{90}{(2\pi)^{7/2}}\dfrac{g_\chi}{g_{*S}}x^{3/2}\,e^{-x},\\
&Y_{X}^{\text{eq}}=Y_{X^{\dagger}}^{\text{eq}}\simeq\dfrac{90}{(2\pi)^{7/2}}\dfrac{g_X}{g_{*S}}[(1+\delta)x]^{3/2}\,e^{-(1+\delta) x},
\end{align}
with $g_\chi=2$ and $g_X=3$ being the internal degrees of freedom of the Majorana particle $\chi$ and the colored scalars $X$, $m_\text{Pl}$ the Planck mass, $g_*$ ($g_{*S}$) the number of effective relativistic degrees of freedom for the energy (entropy) density of the Universe and
\begin{equation}
    \delta\equiv\dfrac{m_X-m_\chi}{m_\chi}\equiv\dfrac{\Delta m}{m_\chi},\quad \Delta m\equiv m_X-m_\chi.
\end{equation}
Hereby, the effective annihilation cross-section in Eq.~\eqref{eq:BoltzmannEq} is given by
\begin{equation}
    \langle\sigma_{\text{eff}} v_{\text{rel}}\rangle=\sum_{ij}\langle\sigma_{ij}v_{ij}\rangle \dfrac{Y_i^{\text{eq}}}{\tilde{Y}^{\text{eq}}}\dfrac{Y_j^{\text{eq}}}{\tilde{Y}^{\text{eq}}},
    \label{eq:effective_sigmav_coannih}
\end{equation}
where $\langle\sigma_{ij}v_{ij}\rangle$ comprises of all the annihilation cross-sections of two co-annihilating species $i$ and $j$.
If the scalars are much heavier than DM, their abundance gets quickly Boltzmann-suppressed and the only relevant process for determining the DM density is the direct $\chi-\chi$ annihilation. On the other hand, when $\delta\ll1$, the scalar mediators continue interacting for a longer time and their thermally-averaged (co-)annihilation cross-section contributions to Eq.~\eqref{eq:effective_sigmav_coannih} are significant even around the freeze-out of the DM particle candidates.
We stress that in order to derive Eq.~\eqref{eq:effective_sigmav_coannih}, two crucial assumptions are made. 
First, the rate of elastic scatterings of dark sector particles with the SM bath is much larger than their annihilation rates, due to the fact that the number density of the SM particles are not Boltzmann-suppressed.
This allows us to approximate $f_i\sim f_i^{\text{eq}}$, where $f_i$ is the phase space density of the particle species $i$.
In addition to this, DM and its co-annihilating partners are considered to be in chemical equilibrium, meaning that the processes that convert a dark sector particle into another one are faster than the expansion of the universe.
In other words, the inter-conversion rate $\chi\leftrightarrow X$ is larger than the Hubble rate, which allows us to approximate $Y_i/\tilde{Y}\simeq Y_i^{\text{eq}}/\tilde{Y}^{\text{eq}}~$\footnote{A situation where chemical equilibrium is not guaranteed may occur when the Yukawa coupling that regulates DM interactions is too small. In this case, one  is not allowed to perform the approximation to arrive at Eq.~\eqref{eq:effective_sigmav_coannih} and one needs to take into account the full set of coupled Boltzmann equations \cite{Ellis_2015}. In this regime, the DM-production mechanism is also sometimes called \textit{conversion-driven freeze-out} \cite{Garny_2017, Garny_2018} or \textit{co-scattering} \cite{Dagnolo_2017}, see also the corresponding discussion in Sec.~\ref{sec:relicdensity} and Sec.~\ref{sec:Results}.}.
For concreteness, Eq.~\eqref{eq:effective_sigmav_coannih} can be explicitly written by means of the allowed co-annihilation processes, enlisted in Tab.~\ref{tab:processes} and depicted in Fig.~\ref{fig:feynm_diagr}, as 
\begin{equation}
\begin{aligned}
    &\langle\sigma_\text{eff}v_\text{rel}\rangle= \dfrac{2 Y_X^\text{eq} Y_{X^\dagger}^\text{eq}\langle\sigma_{XX^\dagger}v_\text{rel}\rangle}{\tilde{Y}^2_\text{eq}} 
    + \dfrac{2 Y_X^\text{eq} Y_X^\text{eq}\langle\sigma_{X^{(\dagger)}X^{(\dagger)}}v_\text{rel}\rangle}{\tilde{Y}^2_\text{eq}} 
    + \dfrac{4 Y_\chi^\text{eq} Y_X^\text{eq}\langle\sigma_{\chi X^{(\dagger)}}v_\text{rel}\rangle}{\tilde{Y}^2_\text{eq}} 
    + \dfrac{Y_\chi^\text{eq} Y_\chi^\text{eq}\langle\sigma_{\chi\chi}v_\text{rel}\rangle}{\tilde{Y}^2_\text{eq}} \\
    &= \dfrac{1}{\left[g_\chi+2\sum_i g_{X_i}(1+\delta)^{3/2}e^{-\delta x}\right]^2}\,\times\\
    &\bigg[2\Big(\langle\sigma_{XX^\dagger}v_\text{rel}\rangle+\langle\sigma_{X^{(\dagger)}X^{(\dagger)}}v_\text{rel}\rangle\Big) g_X^2 (1+\delta)^3\,e^{-2x\delta}+4\langle\sigma_{\chi X^{(\dagger)}}v_\text{rel}\rangle g_X g_\chi(1+\delta)^{3/2}e^{-\delta x} + \langle\sigma_{\chi\chi}v_\text{rel}\rangle g_\chi^2\bigg],
\end{aligned}
    \label{eq:effective_sigmav_XXbar}
\end{equation}
where the multiplicities in front of the various terms come from the sum in Eq.~\eqref{eq:effective_sigmav_coannih} and by considering that scalars and antiscalars give the same contribution (we will omit the $\dagger$ in processes other than for $XX^\dagger$ pair annihilations). 
From this expression, we can appreciate how, for $\delta \simeq 0$, colored annihilation and co-annihilation processes can provide a sizable contribution to the total annihilation cross-section. On the contrary, for large $\delta$, these processes are negligible so that the $t$-channel pair annihilations of DM particles into SM quarks is the dominant channel.
In the degenerate limit, $\delta=0$, the relative weights of Eq.~\eqref{eq:effective_sigmav_XXbar} respectively amount to
\begin{equation}
    \dfrac{2 g_X^2}{(g_\chi+2\sum_i g_{X_i})^2}, \quad  \dfrac{2 g_X^2}{(g_\chi+2\sum_i g_{X_i})^2}, \quad \dfrac{4 g_X g_\chi}{(g_\chi+2\sum_i g_{X_i})^2}, \quad \dfrac{ g_\chi^2}{(g_\chi+2\sum_i g_{X_i})^2}.
    \label{eq:effective_sigmav_XXbar_delta0}
\end{equation}
In the $u_R$ and $d_R$ models, we consider three copies of SU(2)-singlet scalars, each of which carrying only colored degrees of freedom; this implies that $g_X=3$ for each scalar, while for the Majorana singlet we only have two spin configurations and $g_\chi=2$.
The $q_L$ model, instead, features SU(2)-doublet mediators;
this implies the existence of six copies of color-triplet scalars which make the denominator in Eq.~\eqref{eq:effective_sigmav_XXbar_delta0} even larger, resulting in a smaller effective annihilation cross-section.
In the co-annihilating region, we will see that this will be the main difference between the $q_L$ model and the $u_R$ and $d_R$ models in computing the relic abundance (cf. Sec.~ \ref{sec:relicdensity}).
%
While the relative contributions to the total effective cross-section depend on the internal degrees of freedom of the particles involved, the individual cross-sections can have very different magnitudes. 
\begin{center}
\begin{table}[!t]
	\centering
	\begin{tabular*}{\textwidth}{l @{\extracolsep{\fill}} ccccc}
		\hline
		& & & \\
		Process& Contribution to $\langle\sigma v\rangle$ & $v_\text{rel}$ & Color Structure & BSF \\
		& & & \\
		\hline
		& & & \\
		$\chi \chi \rightarrow q_i \bar{q}_i$ & $g_\text{DM}^4$ & \makecell{$v_\text{rel}^2$ ($m_q=0$)\\[1pt]$v_\text{rel}^0$ ($m_q\neq0$)} & none & \xmark  \\
		& & & \\
		\hline
		& & & \\
		$X_i X^\dagger_j \rightarrow gg$ & $g_s^4 e^{- 2 x \delta}$ & $v_\text{rel}^0$ & $|\mathcal{M}|^2 \sim \frac{2}{7}[\mathbf{1}] + \frac{5}{7} [\mathbf{8}]$ &  \\
		& & & &  \\
		& & & & \cmark \\
		$X_i X^\dagger_j \rightarrow q_i \bar{q}_j$ & $(\alpha g_\text{DM}^2 + \beta g_s^2)^2 e^{-2x\delta}$ & \makecell{$v_\text{rel}^2$ ($m_q=0$)\\[1pt]$v_\text{rel}^0$ ($m_q\neq0$)} &\makecell{ $|\mathcal{M}|^2 \sim f_1 \left(g_\text{DM},g_s \right) [\mathbf{1}]$ \\$+ f_8 \left(g_\text{DM},g_s \right) [\mathbf{8}]$} &   \\
		& & & \\
		\hline
		& & & \\
		$X_i X_j \rightarrow q_i q_j$ & $g_\text{DM}^4 e^{- 2 x \delta}$ & $v_\text{rel}^0$& $|\mathcal{M}|^2 \sim \frac{1}{3} [\mathbf{\bar{3}}] + \frac{2}{3} [\mathbf{6}]$ &   \\
		& & & &  \\
		& & & & \textcolor{gray}{\ding{51}} \\
		$X_i X_i \rightarrow q_i q_i$ & $g_\text{DM}^4 e^{- 2 x \delta}$ & $v_\text{rel}^0$ & $|\mathcal{M}|^2 \sim [\mathbf{6}]$ &  \\
		& & & \\
		\hline
		& & & \\
		$X_i \chi \rightarrow q_i A$ & $g_\text{DM}^2 g_\text{gauge}^2 e^{- x \delta} $ & $v_\text{rel}^0$  & none & \xmark  \\
		& & & \\
		\hline
	\end{tabular*}
	\caption{Summary of the most relevant processes contributing to the (co-)annihilation of DM (the corresponding Feynman diagrams are depicted in Fig.~\ref{fig:feynm_diagr}). In the second column, we present the coupling structure combined with the suppression factor caused by the mass splitting $\delta$ in the dark sector. 
	Here, $g_\text{gauge}$ generically represent the gauge coupling in the vertex $q\bar{q}A$, with $A$ being the involved gauge boson (these are $g$, $\gamma$, $Z$, and $W^\pm$, although the latter applies only to the $q_L$ model).
	In the $XX^\dagger\rightarrow q\bar{q}$ process, $\alpha$ and $\beta$ generically indicate the contribution from the Yukawa- and the gluon-mediated process. The third column shows the dominant velocity dependence of $\sigma v_\text{rel}$ in the limit of low relative velocities.
	A distinction between massive and massless quark is also highlighted, with the latter being the most relevant case for our analysis.
	The fourth column gives the decomposition of the squared matrix element into the individual contributions of different color configurations $[\hat{\mathbf{R}}]$, with $f_1$ and $f_8$ being some functions of the coupling constants. The fifth column indicates the ability of the initial-state particles to form (\cmark) or not to form (\xmark) a bound state and a \textcolor{gray}{\ding{51}} corresponds to the suppressed BSF when these are not in conjugate representations.}
	\label{tab:processes}
\end{table}
\end{center}
In Tab.~\ref{tab:processes}, we highlight the contributions to $\langle\sigma_\text{eff} v_\text{rel}\rangle$ as a function of the couplings involved and of the relative mass-splitting $\delta$ (second column). Moreover, we show the dominant velocity-dependent contribution in the limit of small relative velocities (third column)
\footnote{In this work we avoid referring to the term in the velocity expansion of $\sigma v_\text{rel}$ that has a velocity dependence of $\left( v^0, v^2, v^4, ...\right)$ as $(s, p, d, ...)$-wave and so on.
This terminology is borrowed from the expansion of the probability amplitude into partial-waves with different values of the orbital angular momentum $\ell$, $\mathcal{M} = \sum_\ell \mathcal{M}_\ell$, where the terms with $\ell=(0, 1, 2, ...)$ are called $(s, p, d, ...)$-wave and so on.
However, it is inconsistent to identify a term in the velocity expansion of $\sigma v_\text{rel}$ with a certain partial-wave, as it has been discussed in detail in \cite{toolbox}.
In fact, the momentum expansion of the matrix element for a given partial-wave results in $\mathcal{M}_\ell \sim \sum_n p^{\ell+2n}$. 
Consequently, one finds $\sigma v_\text{rel} \sim |\mathcal{M}|^2 = \sum_\ell |\mathcal{M}_\ell|^2 \sim \sum_{\ell,n,n'} p^{2 \left( \ell + n + n' \right)}$ \cite{toolbox,Just2008}. 
This implies that terms proportional to $p^2$ are not purely ``\textit{p}-wave'' and actually receive contributions from states with $\ell=0$ (e.g., when $n=1$ and $n'=0$).
For the successive terms in the momentum expansion scaling with $p^{2m}$, the contribution emerges from $m+1$ different partial-waves so that utilizing the partial-wave terminology fails for $m \neq 0$.
This clarification is of particular importance when dealing with Sommerfeld corrections to processes that have no velocity-\textit{independent} terms \cite{toolbox}.\label{fn:nwave}}.
In the co-annihilating parameter region, we see that the $t$-channel direct annihilation of DM into quarks as well as the pair annihilation of the scalar-antiscalar triplets into quark-antiquark pairs are velocity-suppressed in the limit of massless quarks and therefore subdominant for almost all the parameter space we are interested in.
Clearly, for lighter DM and small mass-splittings, the top quark mass would not be negligible.
Nevertheless, as we will see (cf. Section~\ref{sec:relicdensity}), typical Yukawa couplings in this region of the parameter space are $g_\text{DM}\lesssim 10^{-1}$ and the process is therefore suppressed with powers of $g_\text{DM}^4$.
In fact, as a general feature, in those regions of the parameter space where the Yukawa coupling $g_\text{DM}$ is smaller than the strong coupling $g_s$, processes that do not involve vertices with gluons are negligible even if they are not velocity-suppressed (e.g., $XX\rightarrow qq$ annihilations).

Finally, as we will thoroughly describe in the next section, the individual thermally-averaged cross-sections can receive large corrections from non-perturbative effects generated by long-range, colored interactions between color-charged, initial states, so that the actual contribution in percentage of one channel can be quite sizable.
For this reason, in the fourth and fifth columns we also show the color structure of the amplitude and the possible contributions from BSF to the effective annihilation process, as we will explain in more detail in the following subsection.

\section{Long-range effects in the simplified $t$-channel model}
\label{sec:long_range_effects}

Processes that involve two massive, colored particles in the initial states are subject to the long-range behaviour of the gluonic Coulomb potential generated by the strong force exerted between them.
This influence distorts their scattering wavefunctions in such a way that they can experience a further attraction or repulsion, depending on the color charge they carry, that is to say on their color representation, which directly influences the gauge coupling $\alpha_g$ related to the gluon potential (see the definition later in Eq.\ref{eq:alpha_g}).
The impact of this effect becomes important when the incoming particles travel at non-relativistic speeds.
The typical interaction distance is given by the inverse of the average relative momentum $k=\mu v_\text{rel}$ exchanged between two scattering particles ($\mu$ being their reduced mass). When the interaction distance  approaches the Bohr radius $(\mu\alpha_g)^{-1}$ of the interacting system, implying $\alpha_g\sim v_\text{rel}$, non-perturbative effects from the continuous exchange of gluons between the two interacting states become relevant.
This is the so-called Sommerfeld effect \cite{Sommerfeld:1931qaf}. 
Moreover, the presence of an attractive potential can naturally lead to the formation of bound states of the color-charged particles that also affect interaction rates.
In the following sections we summarize the pivotal results related to these effects in a non-Abelian gauge theory and apply them to our simplified model.

\subsection{Color decomposition and Sommerfeld effect}
\label{sec:colorDec_Sommerfeld}
Let us indicate with $\textbf{R}_\mathbf{1}$ and $\textbf{R}_\mathbf{2}$ the color representations under which two incoming particles transform.
Their interaction can then be decomposed into a direct sum of irreducible representations,
\begin{equation}
    \textbf{R}_\mathbf{1}\otimes\textbf{R}_\mathbf{2}=\bigoplus_{\hat{\textbf{R}}}\hat{\textbf{R}}\ .
\end{equation}
The gluonic interaction in the non-relativistic regime is described at leading order by a static Coulomb-like potential that depends on the irreducible representation as follows
\begin{equation}
    V_{[\hat{\textbf{R}}]}(r)=-\frac{\alpha_g^{[\hat{\textbf{R}}]}\left( Q \right)}{r}.
    \label{eq:Vr}
\end{equation}
Here, $Q$ represents the averaged momentum transfer in the interaction.
The effective gluon coupling constant in a given irreducible representation of the unbound scattering state $\alpha_g^{[\hat{\textbf{R}}]} (Q)$ is related to the strong coupling constant $\alpha_s(Q)$ by
\begin{equation}
    \alpha^{[\hat{\textbf{R}}]}_g \left( Q \right)=\alpha_s \left( Q \right)\times\frac{1}{2}[C_2 (\textbf{R}_\mathbf{1})+C_2 (\textbf{R}_\mathbf{1})-C_2 (\hat{\textbf{R}})]\equiv\alpha_s \left( Q \right) \times k_{[\hat{\textbf{R}}]},
    \label{eq:alpha_g}
\end{equation}
$C_2(\textbf{R})$ being the quadratic Casimir invariant of the given representation \textbf{R}.
Notice that the $k_{[\hat{\textbf{R}}]}$ factor can be either positive or negative, meaning that the potential can be either attractive or repulsive.
We evaluate the coupling constant $\alpha_g^{[\hat{\textbf{R}}]}(Q)=k_{[\hat{\textbf{R}}]}\alpha_s(Q)$ at the scale of the average transferred momentum of the scattering states, $Q=\mu v_\text{rel}$, by considering corrections to the RGE of the $\beta$-function of $\alpha_s(Q)$ up to NNLO, as given in \cite{Zyla:2020zbs} and as employed by numerical codes used for our scans (e.g., in \texttt{micrOMEGAs} \cite{Belanger:2006is}). For simplicity, we suppress in the following, the dependence on the average transferred momentum $Q$, while consistently taking it into account in our numerical calculations. 
Since the scalar fields of our models belong to the fundamental representation \textbf{3} of SU(3) (the antiscalar fields to the conjugate one, $\bar{\mathbf{3}}$), we can easily list the different possible potentials we obtain from Eq.~\ref{eq:Vr}. Since $\mathbf{3}\otimes\Bar{\mathbf{3}}=\mathbf{1}\oplus\mathbf{8}$ and $\mathbf{3}\otimes\mathbf{3}=\mathbf{\bar{3}}\oplus\mathbf{6}$
(the case with $\mathbf{\bar{3}}\otimes\mathbf{\bar{3}}$ gives identical results to the last one since $C_2 (\textbf{R})=C_2 (\bar{\textbf{R}})$), we have the following potentials:
\begin{equation}
  V(r)_{\mathbf{3}\otimes\Bar{\mathbf{3}}}=
    \begin{cases}
    -\dfrac{4}{3}\dfrac{\alpha_s}{r}\quad[\mathbf{1}]\\[8pt]
    +\dfrac{1}{6}\dfrac{\alpha_s}{r}\quad[\mathbf{8}]
    \end{cases}; \quad   V(r)_{\mathbf{3}\otimes\mathbf{3}}=
    \begin{cases}
    -\dfrac{2}{3}\dfrac{\alpha_s}{r}\quad[\mathbf{\bar{3}}]\\[8pt]
    +\dfrac{1}{3}\dfrac{\alpha_s}{r}\quad[\mathbf{6}]
    \end{cases}. \label{eq:ColorPotential}
\end{equation}

We can see how the singlet state accounts for the most attractive potential. In this work, we neglect the effects of the Yukawa potentials generated by electroweak bosons.
For light DM, the Yukawa exponential suppression makes the influence at large distances negligible. In fact, in order for long-range effects to have a sizable impact, the Bohr radius needs to be smaller than the inverse mass of the force mediator, $(\alpha\,m_X/2)^{-1}\lesssim m_A^{-1}$.
For typical electroweak couplings of order $\alpha\sim 0.02$, effects of the electroweak potential start to be important for $m_X\gtrsim 8 $~TeV. However, thermal freeze-out typically takes place at $T \sim \mdm/30$, which results in $T \sim m_X/30$ in the coannihilating regime. Since electroweak symmetry breaking (EWSB) occurs at $T_\text{EWSB} \sim 150\,\mathrm{GeV}$, electroweak gauge bosons are massive during freeze-out only if $m_X \lesssim 4.5\, \mathrm{TeV}$. Thus, below this threshold, effects of the electroweak potential can be safely omitted\footnote{For works addressing electroweak forces in the long-range regime, we refer e.g. to Refs.~\cite{Hisano:2006nn,Beneke:2009rj,Beneke:2014hja,Beneke:2019qaa,Oncala:2021tkz}.}. 

If the dominating term in the bare annihilation cross-section $\sigma_0$ is velocity-independent, the Sommerfeld effect results in a simple multiplicative factor ( multiplying the ``\textit{s}-wave'' term, which is the only one in the partial-wave expansion) \cite{Cassel_2010,toolbox}:
\begin{equation}
    \sigma_{\text{SE},[\mathbf{R}]}v_\text{rel}=c_{[\textbf{R}]} S_{0,[\textbf{R}]}\,\sigma_0,
\end{equation}
where
\begin{equation}
    S_{0,[\textbf{R}]}=S_0\left(k_{[\textbf{R}]}\dfrac{\alpha_s}{v_{\text{rel}}}\right)
    \label{eq:somfact_color}
\end{equation}
is the $s$-wave Sommerfeld factor and where $c_{[\textbf{R}]}$ is a factor coming from the color decomposition of the amplitude and marks the relative contribution of the $\hat{\textbf{R}}$ initial state to the total cross-section\footnote{This factor actually depends on the orbital $\ell$ and spin $s$ angular momenta of the initial scattering state, as shown in \cite{toolbox}.
Since we are interested only in color-charged scalars and by only considering the dominant $s$-wave contribution(indicated by the subscript $0$), we can directly employ the results for the $\ell=0$, $s=0$ case.}.
For the potentials illustrated in Eq.~\eqref{eq:ColorPotential}, one finds \cite{toolbox}
\begin{align}
    \sigma_{\mathbf{3}\otimes\bar{\mathbf{3}}\rightarrow gg} v_\text{rel} &=\sigma_{\mathbf{3}\otimes\bar{\mathbf{3}}\rightarrow gg,0}\left(\dfrac{2}{7}S_{0,[\textbf{1}]}+\dfrac{5}{7}S_{0,[\textbf{8}]}\right), \\ 
    \sigma_{\mathbf{3}\otimes\bar{\mathbf{3}}\rightarrow q \bar{q}} v_\text{rel} &=\sigma_{\mathbf{3}\otimes\bar{\mathbf{3}},0}\left(f_{[\textbf{1}]}(g_s,\gdm)S_{0,[\textbf{1}]} +f_{[\textbf{8}]}(g_s,\gdm)S_{0,[\textbf{8}]}\right), \\
    \sigma_{\mathbf{3}\otimes\mathbf{3}\rightarrow qq} v_\text{rel} &=\sigma_{\mathbf{3}\otimes\mathbf{3}\rightarrow qq,0} S_{0,[\textbf{6}]}, \\
    \sigma_{\mathbf{3}_i\otimes\mathbf{3}_j\rightarrow q_i q_j} &=\sigma_{\mathbf{3}_i\otimes\mathbf{3}_j\rightarrow q_i q_j,0}\left(\dfrac{1}{3}S_{0,[\bar{\mathbf{3}}]}+\dfrac{2}{3}S_{0,[\mathbf{6}]}\right).
    \label{eq:SigmaVSomEf_ColFact}
\end{align}
The case of annihilation into a quark-antiquark pair is more involved, as it is not straightforward to decompose the color structure of the amplitude, due to the fact that both QCD- and Yukawa-dominated diagrams contribute and interfere.
In general, one needs to compute the single color contribution of each term in the squared matrix element and then determine the color weight factors, that we indicate here as general functions $f_{[\mathbf{1}]}$ and $f_{[\mathbf{8}]}$ of the couplings involved (e.g., $g_s$ and $g_\text{DM}$).
Since in our model we encounter this situation only for processes that are velocity suppressed across the vast majority of the parameter space, we choose to neglect the Sommerfeld effects for them.

In the last process in Eq.~\eqref{eq:SigmaVSomEf_ColFact} we consider the case of distinguishable particles both in the initial and in the final states.
This is the situation, for example, for a $X_i+X_j\rightarrow q_i+q_j$, where $i$ and $j$ refer to two distinct flavours.
By employing Eq.~\eqref{eq:somfact_color} in our model, the four types of Sommerfeld factors needed are:
\begin{equation}
    S_{0,[\mathbf{1}]}=S_0\left(\frac{4\alpha_s^S}{3v_\text{rel}}\right),\; S_{0,[\mathbf{8}]}=S_0\left(\frac{-\alpha_s^S}{6v_\text{rel}}\right),\; S_{0,[\bar{\mathbf{3}}]}=S_0\left(\frac{2\alpha_s^S}{3v_\text{rel}}\right),\; S_{0,[\mathbf{6}]}=S_0\left(\frac{-\alpha_s^S}{3v_\text{rel}}\right).
\end{equation}
For the Coulomb potential, the function $S_0(\zeta_s)$ is given by
\begin{equation}
    S_0(\zeta_s)=\dfrac{2\pi\zeta_s}{1-e^{-2\pi\zeta_s}}
    \label{eq:S0_coulomb}
\end{equation}
with $\zeta_s=\alpha_{g,[\textbf{R}]}/v_{\text{rel}}=k_{[\textbf{R}]}\,\alpha_s/v_{\text{rel}}$ regulated by the color factor as in Eq.~\eqref{eq:alpha_g}.
At small velocities, $S_0\sim\zeta_s\sim \alpha_{g,[\textbf{R}]} v_{\text{rel}}^{-1}$  and, depending on its sign, results in either an enhancement ($S_{0,[\textbf{1}]}$ and $S_{0,[\bar{\mathbf{3}}]}$) or a decrease ($S_{0,[\textbf{8}]}$ and $S_{0,[\textbf{6}]}$) of the perturbative, tree-level cross-section.
Therefore, as soon as $v_\text{rel}\lesssim \alpha_{g,[\textbf{R}]}$, the Sommerfeld factor can have a sizable impact on the cross-sections involved, so that the annihilation processes we discussed in Sec.~\ref{sec:model} are affected significantly.

For annihilation processes that are dominated by velocity-dependent terms (see Table~\ref{tab:processes}) we should also take into account the Sommerfeld corrections to the different partial-wave contributions.
In fact, one can show (see, e.g.,  \cite{Cassel_2010,Iengo2009,toolbox}) that each term of the squared amplitude at a given $\ell$ must be multiplied by
\begin{equation}
S_\ell(\zeta)=S_0(\zeta)\prod_{k=1}^\ell \left(1+\frac{\zeta^2}{k^2}\right).
\label{eq:SommerfeldPartialWave}
\end{equation}
Therefore, as described in footnote \ref{fn:nwave}, at a given order in the momentum expansion, one should in principle consider for each partial-wave term the corresponding $S_\ell$.
For instance, in the $v_\text{rel}^2$ term of the expansion of $\sigma v_\text{rel}$ one needs to consider both an $s$-wave and a $p$-wave term, which are corrected by $S_0$ and by $S_1=S_0(1+\zeta_s^2)$, respectively.
In what follows, we will only take into account the dominant $s$-wave Sommerfeld corrections and will omit these higher partial-wave contributions, since the latter come with factors that are equal to 1 plus a correction $\zeta_s^2\propto\alpha_s^2$, which is of order $\mathcal{O}(10^{-2})$.

\subsection{Bound-state formation, ionization and decay}
\label{sec:BSF_ion_dec}
\begin{figure}[!t]
    \centering
    \includegraphics[width=0.6\textwidth]{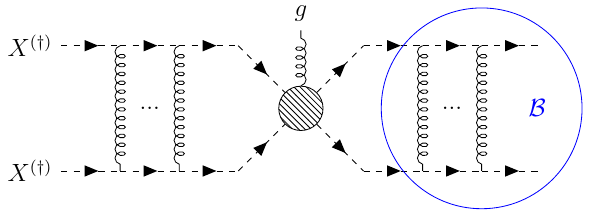}
    \caption{\small Feynman diagram for the radiative capture into a bound state of a scalar triplet pair (either $XX^\dagger$, $XX$, or $X^\dagger X^\dagger$. It consists of the (non-perturbative) initial and final state wavefunctions of the incoming particles, and the perturbative 5-point function (the grey blob) that includes the radiative vertex.
    The final state is then the bound state $\mathcal{B}$ and the radiated (on-shell) gluon $g$.}
    \label{fig:bsf}
\end{figure}

In a similar regime where the Sommerfeld effect is relevant, color charged particles can also form unstable bound states (BS) via the emission of a gluon (cf. Fig.~\ref{fig:bsf}):
\begin{equation}
    X_1 + X_2 \rightarrow \mathcal{B}(X_1 X_2) + g.
\end{equation}

\begin{figure}[!t]
    \centering
    \includegraphics[width=0.8\textwidth]{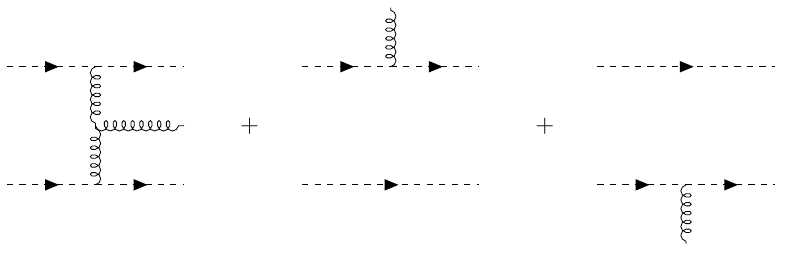}
    \caption{\small Tree-level contributions to the radiative vertex in Fig.~\ref{fig:bsf}. The leftmost diagram shows the radiation from the gluon propagator, allowed by the non-Abelian nature of QCD. The other two contributions come directly from the emission from an external leg.}
    \label{fig:BSF_vertices}
\end{figure}

Due to the non-Abelian nature of the strong force, gluon radiation can be emitted both from the external legs (i.e., the wavefunctions of the colored particles forming the bound state), or from a gluon propagator exchanged between the two incoming states, what we call a radiative vertex, see Fig.~\ref{fig:BSF_vertices}.   
The wavefunction of the bound state is determined by solving its corresponding Schr\"odinger equation. This results in a  discrete  energy spectrum, with energy eigenvalues given by the binding energies ${\mathcal{E}_{n\ell m}=-\kappa^2/(2\mu\,n^2)}$ with $n$ being the principal quantum number and $\mu$ the reduced mass of the two-particle system\footnote{In the definition of the binding energy, we include for completeness $\ell$ and $m$, respectively the orbital and the magnetic quantum numbers. These are not relevant for us, as for non-relativistic hydrogen-like systems the energy of the bound states depend only on $n$.}.

The strong coupling is evaluated at the scale of the average momentum exchange of the particles in the process.
In particular, in the ladder diagram (cf. Fig.~\ref{fig:bsf}), the average momentum transfer between the bound-state wavefunctions is given by the Bohr momentum $\kappa_{[\hat{\textbf{R}}]}\equiv\mu\,\alpha^B_{g,[\hat{\textbf{R}}]}=\mu\,k_{[\hat{\textbf{R}}]}\,\alpha^B_{s,[\hat{\textbf{R}}]}$ (cf. also Eq.~\eqref{eq:alpha_g}). 
For the strong coupling in the radiative vertex in the BSF diagrams in Fig.~\ref{fig:BSF_vertices}, $\alpha_{s,[\hat{\textbf{R}}]}^{\text{BSF}}$, the expectation value of the momentum exchanged is the same as the one of the emitted gluon $|\textbf{P}_g|=\omega$, with $\omega$ being the gluon energy.
From energy-momentum conservation, in the non-relativistic regime, $\omega$ must be approximately equal to the difference between the relative kinetic energy of the initial scattering states and the binding energy of the final bound state, neglecting their total kinetic energies\footnote{From energy-momentum conservation, one would actually have $\omega=\mathcal{E}_\textbf{k}-\mathcal{E}_{n\ell m}+(\textbf{K}^2-\textbf{P}^2)/2M$, with $M$ the total mass of the system and \textbf{K} and \textbf{P} the total scattering and bound state three-momenta, respectively.
However, in the non-relativistic regime, the total three-momenta are much smaller than the total energy and the masses, so that the approximation shown holds.}.
\begin{equation}
   Q^\text{BSF} \equiv \omega \simeq\mathcal{E}_\textbf{k}-\mathcal{E}_{n\ell m}=\frac{\mu}{2}\left[v_\text{rel}^2+(\alpha^B_{g,[\hat{\textbf{R}}]})^2\right].
\end{equation}
Finally, for the emission of radiation directly from a mediator exchanged by the scattering states, as shown in Fig.~\ref{fig:BSF_vertices}, the momentum transfer is $Q^\text{NA}\simeq|\textbf{p}-\textbf{q}|\simeq\sqrt{\rule{0pt}{2ex}k^2+\kappa^2}=\mu\sqrt{\rule{0pt}{2ex}v_\text{rel}^2+\alpha_{g,[\hat{\textbf{R}}]}^B{^2}}$.
Since $k<<\kappa$ when BSF is relevant, one can approximate this last expression as $Q^\text{NA}\simeq \kappa_{[\hat{\textbf{R}}]}$ and $\alpha_{s,[1]}^\text{NA}\simeq\alpha_{s,[1]}^B$.
For more details on the computation of the transition amplitudes, we refer the reader to \cite{Petraki_2015,Harz:2018csl}, on which the formalism employed in this work is based on. Note that the effects of BSF on the relic density have also been considered in the context of non-relativistic effective field theories including non-zero temperature corrections \cite{Biondini:2018pwp,Biondini:2018ovz,Biondini:2019int,Biondini:2018xor,Binder:2018znk,Binder:2019erp,Binder:2020efn,Binder:2021otw}.

As described earlier in Sec.~\ref{sec:model}, the dark-sector (anti)scalars are color (anti)triplets.
From group algebra, the relevant decompositions for a scalar anti-scalar interaction into irreducible representations of SU(3) for the radiative transitions are $\mathbf{3}\times\overline{\mathbf{3}}=\mathbf{1}+\mathbf{8}$, $\mathbf{3}\times\mathbf{3}=\overline{\mathbf{3}}+\mathbf{6}$ and $\mathbf{8}\times\mathbf{8}=\mathbf{1}_S+\mathbf{8}_S+\mathbf{8}_A+\mathbf{10}_A+\overline{\mathbf{10}}_S+\mathbf{27}_S$, such that the only allowed capture processes for a scalar-antiscalar initial state are
\begin{subequations}
\begin{align}
    (X+X^\dagger)_{[\textbf{8}]}&\rightarrow\big\{\mathcal{B}(XX^\dagger)_{[\textbf{1}]}+g\big\}_{[\mathbf{8}]},\label{eq:RelevantCapture}\\
    (X+X^\dagger)_{[\textbf{1}]}&\rightarrow\big\{\mathcal{B}(XX^\dagger)_{[\textbf{8}]}+g\big\}_{[\mathbf{1}_S]},\\
    (X+X^\dagger)_{[\textbf{8}]}&\rightarrow\big\{\mathcal{B}(XX^\dagger)_{[\textbf{8}]}+g\big\}_{[\mathbf{8}_S] \text{ or } [\mathbf{8}]_A}.
\end{align}
\label{eq:parantiparBSF}
\end{subequations}
Here assume that the two-particle initial state is either in the singlet or the octet representation. 
The final-state particles are the bound state $\mathcal{B}$ which belong to an irreducible color representation and the emitted gluon $g$, is in the adjoint representation \textbf{8}, by definition.
The combination of color algebras in the final states must match the one in the initial state.
For example, we see that when the initial state is a color singlet, we can only form an octet bound state. The additional gluon that is emitted, combines with the bound state to form a symmetric singlet representation $\mathbf{1}_S$.
However, as noted earlier, only the singlet potential is attractive, and hence only the first of the processes above in Eq.~\ref{eq:parantiparBSF} is relevant\footnote{The octet bound state could still have a significant role when a further attractive long-range interaction is present.
One example are Higgs-mediated processes in the limit of heavy initial state particles.
In fact, being the Higgs a real scalar, it always leads to an attractive force, which counteracts (enhances) the repulsiveness (attractiveness) of the gluonic potential \cite{Harz_2019,Harz_2018_HiggsEnh}. \label{fn:Higgs}}.

For particle-particle interactions the following bound state formation processes are possible (analogous antiparticle-antiparticle interactions exist that we do not reproduce here):
\begin{subequations}
\begin{align}
    (X+X)_{[\bar{\mathbf{3}}]}\rightarrow\big\{\mathcal{B}(XX)_{[\mathbf{6}]}+g\big\}_{[\bar{\mathbf{3}}]},\\
    (X+X)_{[\bar{\mathbf{3}}]}\rightarrow\big\{\mathcal{B}(XX)_{[\bar{\mathbf{3}}]}+g\big\}_{[\bar{\mathbf{3}}]},\\
    (X+X)_{[\mathbf{6}]}\rightarrow\big\{\mathcal{B}(XX)_{[\bar{\mathbf{3}}]}+g\big\}_{[\mathbf{6}]},\\
    (X+X)_{[\mathbf{6}]}\rightarrow\big\{\mathcal{B}(XX)_{[\mathbf{6}]}+g\big\}_{[\mathbf{6}]}.
\end{align}
\label{eq:parparBSF}
\end{subequations}
In this work, we only take into account the effects of singlet bound states, since they constitue the dominant effects (cf. Eq.~\ref{eq:ColorPotential}). In fact, the BSF cross-section of a  ($[\mathbf{\bar{3}}]$) bound state is subject to cancellations arising on the level of the squared matrix element of the formation process caused by the color structure of the process \footnote{ We illustrate the effect of these cancellations in Appendix \ref{sec:AppendixA}.}.
Hence, in the following, we will focus on the dominant effect of the capture into the color-singlet ground state $(n\ell m)=(100)$ (see Eq.~\eqref{eq:RelevantCapture}).
Generally, excited states, can also open additional annihilation channels through their direct decays into radiation or via bound-to-bound transitions and subsequent decays.
We neglect these effects in the present work, and leave them for future improvement\footnote{During the completion of this manuscript, two preprints including and discussing excited-state contributions on DM abundance appeared \cite{garny2021bound,Binder:2021vfo}. 
As shown there, the qualitative prediction is not altered significantly, apart for the zero mass-splitting case.
For example, in the single-generation $d_R$ model analyzed in Ref.~\cite{garny2021bound}, the shift in the largest possible DM mass is found to be at the percent level for mass splittings $\Delta m\gtrsim \text{few}$ GeV, while only for $\Delta m = 0$ the same amounts to  $\mathcal{O}(25\%)$.}.

For a scalar-antiscalar pair transforming in the fundamental representation and with degenerate masses $m_X$, the bound-state formation cross-section reads as \cite{Harz:2018csl}:
\begin{equation}
        \sigma_{\{100\}}^{[\mathbf{8}]\rightarrow[\mathbf{1}]} v_\text{rel}=\dfrac{2^7\,17^2}{3^5}\dfrac{\pi\alpha_{s,[\mathbf{1}]}^{\text{BSF}}\alpha_{s,[\mathbf{1}]}^B}{m_X^2}\;S_{\text{BSF}}(\zeta_S,\zeta_B).
    \label{eq:BSF_sigmav_singlet}
\end{equation}
Here, $\alpha_{s,[\mathbf{1}]}^{\text{BSF}}$ represents the strong coupling constant from the gluon emission vertex, while $\alpha_{s,[\mathbf{1}]}^B$ arises from the ladder diagrams involving the bound state wavefunction. 
We will omit the subscript $[\mathbf{1}]$ in the following.
The $S_\text{BSF}$ function, arises from the overlap integrals involving the scattering-state and the bound-state wavefunctions and reads as
\begin{equation}
    S_\text{BSF}(\zeta_S,\zeta_B)=\left(\dfrac{2\pi\zeta_S}{1-e^{-2\pi\zeta_S}}\right)(1+\zeta_S^2)\dfrac{\zeta_B^4 e^{-4\zeta_S\arccot(\zeta_B)}}{(1+\zeta_B^2)^3}.
    \label{eq:BSF_factor}
\end{equation}
Here, $\zeta_S\equiv\alpha_g^S/v_\text{rel}$ and $\zeta_B\equiv\alpha_g^B/v_\text{rel}$ parameterize the ratios between the strong coupling and the relative velocity for the scattering ($S$) and bound state ($B$), respectively\footnote{See also Tab.~1 in Ref.~\cite{Harz:2018csl} for the various definitions of the strong gauge coupling constants.}.
The first parenthesis corresponds to the $s$-wave Coulomb Sommerfeld factor (cf. Eq.~\eqref{eq:S0_coulomb}), coming from the normalization of the scattering wavefunction, while the second term accounts for its $p$-wave correction (cf. Eq.~\eqref{eq:SommerfeldPartialWave}). The presence of this second factor is due to the fact that the radiative capture is, at leading order, a dipole transition between the initial scattering state and the final bound state.
As it is well-known, these transitions impose a selection rule $\Delta\ell=\pm 1$ on the orbital angular momentum, such that in order for the final bound state to be in the ground state $(100)$, the initial scattering state must be in a $\ell=1$ state, hence the $p$-wave correction to the Sommerfeld effect.

Bound-state effects give their peculiar contribution in the last factor, which accounts for the convolution of the bound-state wavefunction with the radiative vertex. 
Similarly to the simple Sommerfeld factor, at large velocities we have $S_\text{BSF}\sim\zeta_B^4\sim (\alpha_g^B/v_\text{rel})^4\ll 1$ and the cross-section in Eq.~\eqref{eq:BSF_sigmav_singlet} gets suppressed; at low velocities we obtain again the typical $S_\text{BSF}\sim \alpha_g^S/v_\text{rel}$ scaling when the interaction between the scattering state is attractive, such that $\zeta_S\gtrsim 1$ and $\zeta_B\gtrsim 1$.
The BSF cross-section in this case is enhanced and can compete with the (co-)annihilation processes.
However, when the interaction is repulsive and $\zeta_S\lesssim1$, the BSF probability becomes exponentially suppressed, $S_\text{BSF}\propto\exp(-2\pi|\zeta_S|)$, as it becomes harder to form bound states when two incoming particles experience a strong repulsive force.
Overall, the BSF cross-section is maximized when $\alpha_g^S/\alpha_g^B=0.5$ and when $\alpha\sim v_\text{rel}$.

We may define a thermally-averaged BSF cross-section as
\begin{equation}
    \langle\sigma_\text{BSF} v_\text{rel}\rangle=\left(\dfrac{\mu}{2\pi T}\right)^{3/2}\int \dd^3 v_\text{rel} \exp(-\mu v_\text{rel}^2/2T) \left[1+f_g(\omega)\right]\sigma_\text{BSF}v_\text{rel},
\end{equation}
where $\omega=\mu/2 \left[(\alpha_g^B)^2+v_\text{rel}^2\right]$ is the energy emitted by the radiated gluon, which equals its kinetic energy minus the (negative) binding energy $\mathcal{E}_{100}=-\mu (\alpha_g^B)^2/2$ of the newly formed bound state, and~$f_g(\omega)=(\exp(\omega/T)-1)^{-1}$ is the gluon distribution function, accounting for the Bose-enhancement factor $1+f_g(\omega)$ from the final state gluon.
This term is necessary to ensure the detailed balance between bound-state formation and ionization reactions, i.e. \textit{ionization equilibrium}.

In fact, if bound-states are successfully formed, they can still be ionized by energetic gluons in the thermal plasma and dissociate into their constituents, or can directly decay into radiation.
The former process usually dominates at temperatures larger than the binding energy, when particles in the thermal plasma are sufficiently energetic to disrupt the bound state.
The ionization rate of a bound state can be written as
\begin{subequations}
\begin{align}
    \expval{\Gamma_{\text{ion},[\mathbf{1}]}}&= g_g \int_{\omega_\text{min}}^\infty \dd\omega \frac{\omega^2}{2\pi^2}\frac{1}{e^{\omega/T}-1}\sigma_\text{ion},\label{eq:gamma_ion}\\
    \sigma_\text{ion}&=\dfrac{g_X^2}{g_g g_\mathcal{B}}\left(\frac{\mu^2 v_\text{rel}^2}{\omega^2}\right)\sigma_\text{BSF},
\label{eq:sigma_ion}
\end{align}
\end{subequations}
where the relation between the ionization cross-section $\sigma_\text{ion}$ and the BSF cross-section follows from the Milne relation (see for example Appendix D of \cite{Harz:2018csl}).
Here, $g_X=3$, $g_g=8$ and $g_\mathcal{B}$ are the internal degrees of freedom of the scalar triplets, the gluons and the bound-states, respectively, and where $\mu=m_X/2$ in our model.
For the capture into the singlet-state, we simply have $g_{\mathcal{B},[\mathbf{1}]}=1$.
Importantly, we see that the ionization rate becomes exponentially suppressed for $T\lesssim \omega$, as already anticipated.

The decay rate $\Gamma_\text{dec,[\textbf{R}]}$ of a $\ell=0$ bound state in a given representation \textbf{R} into gauge bosons is computed by taking the $s$-wave perturbative annihilation cross-section times relative velocity of the corresponding scattering states and multiplying it with the bound state wavefunction evaluated at the origin:
\begin{equation}
    \Gamma_\text{dec,[\textbf{R}]}=(\sigma^{s\text{-wave}}_{\text{ann},[\textbf{R}]}\,v_\text{rel})|\psi_{n00}^{[\textbf{R}]}(0)|^2.
    \label{eq:decayrate_general}
\end{equation}
Considering the color-singlet ground state decaying into a pair of gluons, one obtains from Eq.~\eqref{eq:BSF_sigmav_singlet} and from\footnote{For the ground-state $(n\ell m)=(100)$, we have $\psi_{100}(r)=\sqrt{\kappa^3/\pi}\exp(-\kappa r)$, where $\kappa=\mu\alpha_g^B$.} $|\psi_{100}^{[\mathbf{1}]}(0)|^2=8m_X^3(\alpha_{s,[\mathbf{1}]}^B)^3/27\pi$ that
\begin{equation}
    \Gamma_\text{dec,[\textbf{1}]}=\frac{32}{81}m_X(\alpha_s^{\text{ann}})^2(\alpha_{s,[\mathbf{1}]}^B)^3.
\end{equation}
The formation and subsequent decay of bound states involving dark sector particles can therefore affect the DM relic abundance by effectively opening up a new annihilation channel.
In fact, these effects must be incorporated as additional terms into the system of coupled Boltzmann equations that controls the evolution of the number densities of bound and unbound particles.
By assuming that bound-states are meta-stable and close-to-equilibrium (implying that their number density $Y_\mathcal{B}$ is almost constant, $\dd Y_\mathcal{B}/\dd x\approx 0$) one can describe the convoluted system of Boltzmann equations in an effective manner \cite{Ellis_2015}.
This assumption is reasonable as long as chemical equilibrium is assured between free particles (a premise we already made) and that the processes involving the bound-states (formation, ionization, rapid decays, level-transitions, etc.) are fast enough to exceed the Hubble expansion, which is indeed the case for temperatures larger than their binding energies (see also Ref.~\cite{Binder:2021vfo} for a recent more detailed argumentation).

In a very similar fashion as for the bare co-annihilation scenario, we can now consider a single Boltzmann equation for the total number density of the dark sector particles (cf. Eq.~\eqref{eq:totYield}).
The effects introduced by BSF can in fact be reabsorbed into an effective thermally-averaged BSF cross-section affecting the evolution of the $X$ number density and, thus, the total one, as shown in \cite{Ellis_2015}.
This effective contribution is given by
\begin{equation}
    \langle\sigma_\text{BSF} v_\text{rel}\rangle_\text{eff}\equiv\langle\sigma_\text{BSF}^{[\mathbf{8}]\rightarrow[\mathbf{1}]} v_\text{rel}\rangle\;\dfrac{\expval{\Gamma_{\text{dec}[\mathbf{1}]}}}{\expval{\Gamma_{\text{dec}[\mathbf{1}]}}+\expval{\Gamma_{\text{ion},[\mathbf{1}]}}},
    \label{eq:effectiveBSFsigma}
\end{equation}
where angular parentheses indicate thermal averaging.

We notice that, at large temperatures the ionization processes are much more efficient than bound state decays, hence the effective contribution of bound states in the dark sector evolution is negligible at these temperatures so that the relic density calculation is actually independent on the BSF cross-section.
As the universe cools down, the ionization rate becomes exponentially suppressed and, eventually, decay processes will dominate, efficiently depleting the dark sector scalars, so that the effect of BSF on the Boltzmann equation will be potentially relevant even at temperatures close to the bound state binding energy ($T\gtrsim \mathcal{E}_\mathcal{B}$).

Since in our model we consider bound states formed by three (six) types of colored scalar pairs in the $u_R$ and $d_R$ ($q_L$) models, the BSF contribution will be given by the sum of three (six) terms like Eq.~\eqref{eq:effectiveBSFsigma}.
In this spirit, we can write the \textit{total} annihilation cross-section for the $X_i$ particle of flavour $i$ by adding these terms to each $\langle\sigma_{X_iX_i^\dagger}v_\text{rel}\rangle$.
Summing over all the flavours, we obtain the following \textit{total} effective annihilation cross-section for the $X$ color triplet scalars:
\begin{align}
\langle \sigma_{X X^\dagger} v_{\text{rel}} \rangle_\text{eff} = \sum_i \left( \langle \sigma_{X_i X_i^\dagger} v_\text{rel} \rangle +  \langle\sigma_\text{BSF}^{[\mathbf{8}]\rightarrow[\mathbf{1}]} v_\text{rel}\rangle\;\dfrac{\Gamma_{\text{dec}[\mathbf{1}]}}{\Gamma_{\text{dec}[\mathbf{1}]}+\Gamma_{\text{ion},[\mathbf{1}]}}\right)\, .
\end{align}
This quantity supersedes the naive $XX^\dagger$ annihilation cross-section in the first term in Eq.~\eqref{eq:effective_sigmav_XXbar}.
Again, the importance of the new term becomes relevant at later times, where BSF can efficiently deplete the relic density way beyond the typical scales of thermal freeze-out via late decays of the dark sector bound states (see, e.g., Fig. 6 in \cite{Harz:2018csl}).

At this point, we would like to stress a non-trivial consideration.
Given the fact that the color triplet scalars forming the bound states are unstable against Yukawa-mediated decays into DM and a quark, one could potentially encounter the situation in which the lifetime of a constituent particle is comparable or even shorter than the bound state lifetime itself, for example for large values of $\gdm$.
In this situation, the bound state could decay to $\mathcal{B}(X X^\dagger)\rightarrow X^\dagger+(\chi+q)$ and this eventuality needs to be included when writing the Boltzmann equation.
Although this additional decay channel might appear to play only the role of a bound state destroyer, thereby reducing the effect of BSF in depleting the DM abundance, by exploiting the principle of detailed balance in equilibrium, also the opposite reaction $X^\dagger +\chi+ q\rightarrow \mathcal{B}(XX^\dagger)$ must be considered. This sort of bound-state inverse decay provides an additional competing formation channel: 
its contribution can be recast into an effective BSF cross-section, with a form equivalent to Eq.\eqref{eq:effectiveBSFsigma}.
We refer the reader to Appendix \ref{app:B}, for a treatment of the problem.
Effectively, by assuming that the constituent-decay of the bound state is regulated by $\gdm$, we show in Eq.\eqref{eq:appB_dsigmadgDM} that the effective cross-section must be a monotonically-increasing function of the coupling $\gdm$.
In this sense, by neglecting the constituents decay, one is effectively considering the most conservative estimate of the effect of bound states on the DM abundance evolution.
We leave a more detailed analysis of this effect for future work.

Notice that, in principle, bound states between colored scalars of same and different flavor $i$ and $j$ could exist, since their formation does not depend on the flavour structure.
This is the case, for example, for a $\mathcal{B}(X_i,X_j)$ particle-particle bound state, which however falls in the class of same-representation bound-states that have zero BSF amplitude in the degenerate limit (cf. Eqs.~\eqref{eq:appA_3bar3bar}-\eqref{eq:appA_6anti3_Mnoteq}).
However, due to the flavor-diagonal structure of the Lagrangian considered in Eq.~\eqref{eq:Lagr}, different-flavour bound states, including particle-antiparticle bound state, cannot decay into gluons, but only into a quark-antiquark pair.
Nevertheless, as Tab.~\ref{tab:processes} illustrates, this decay is less efficient than the decay into gluon-pair.
First, for most of the parameter space the annihilation cross-section into quarks is velocity-suppressed.
Second, even when considering the top-quark mass, the velocity-independent part of the perturbative cross-section $X X^\dagger\rightarrow q\bar{q}$ that would determine the decay rate $\Gamma_{\mathcal{B}\rightarrow q\bar{q}}$ becomes larger than the velocity-independent part of the $X X^\dagger\rightarrow gg$ only when $\gdm$ is large\footnote{For the parameter space here considered, one can check that, within the co-annihilating region, $\Delta\lesssim 0.2\,\mdm$, in order to ensure $\Gamma_{\mathcal{B}\rightarrow q\bar{q}}\gtrsim \Gamma_{\mathcal{B}\rightarrow gg}$ one would need $\gdm>2$. For these values, we would always find under-abundant DM (e.g., cf. Fig.~\ref{fig:gDMRelic_uR}).}\label{fn:quarkdecay}.
Therefore, the decay rate into $q\bar{q}$ can be safely neglected.
The remaining decay channels that do not depend on the DM coupling (e.g., into $\gamma\gamma$, $Z\gamma$ or $ZZ$), one can see that they account for less than $1\%$ of the total decay width of the bound state.

Thus, since $\Gamma(\mathcal{B}(X^\dagger_i,X_j) \rightarrow SM) \ll \Gamma(\mathcal{B}(X^\dagger_i,X_i) \rightarrow SM)$ for the parameter space of interest, we do not consider the contribution of $\mathcal{B}(X^\dagger_i,X_j)$ particle-antiparticle to the effective annihilation cross-section, given in Eq.~\eqref{eq:effectiveBSFsigma}.

\section{Impact of the long-range effects on the $t$-channel model parameters}
\label{sec:relicdensity}
In the following, we investigate the impact of the Sommerfeld effect and bound state formation on the parameter space of the simplified $t$-channel models. In order to solve the Boltzmann equation, we employ the publicly available program \texttt{micrOMEGAs 5.2.7} \cite{Belanger:2006is} and modify it to account for non-perturbative long-range effects on the thermally-averaged cross-section.
Our code extends \cite{toolbox} by accounting for the emission of a gluon from the mediator (cf. Fig.~\ref{fig:BSF_vertices}) that was previously neglected in the literature and using the strong coupling constant according to the scale involved. Moreover, our implementation encompasses the depletion of the DM relic density by bound states that form and decay way beyond the typical time scales of thermal freeze-out of $x \simeq 30$, a feature not included in \cite{toolbox}. One can indeed check (see, e.g., Fig.6 in \cite{Harz:2018csl}) that the maximum of the bound state contribution to the effective annihilation cross-section can lie at temperatures of $x\simeq1000$, allowing for DM relic density depletion at $x \gg 30$.
Finally, by making use of the assumptions illustrated in Sec.~\ref{sec:long_range_effects}, our implementation fully exploits the ability of \texttt{micrOMEGAs} to automate the calculation of cross-sections by using the \texttt{CalcHEP} package \cite{Belyaev_2013}, without the need to externally calculate them\footnote{The code will be made available in an upcoming work \cite{Becker_?}.}.
\begin{figure}[!t]
\centering\lineskip=0pt
\begin{subfigure}[b]{0.6\textwidth}
   \includegraphics[width=1\linewidth]{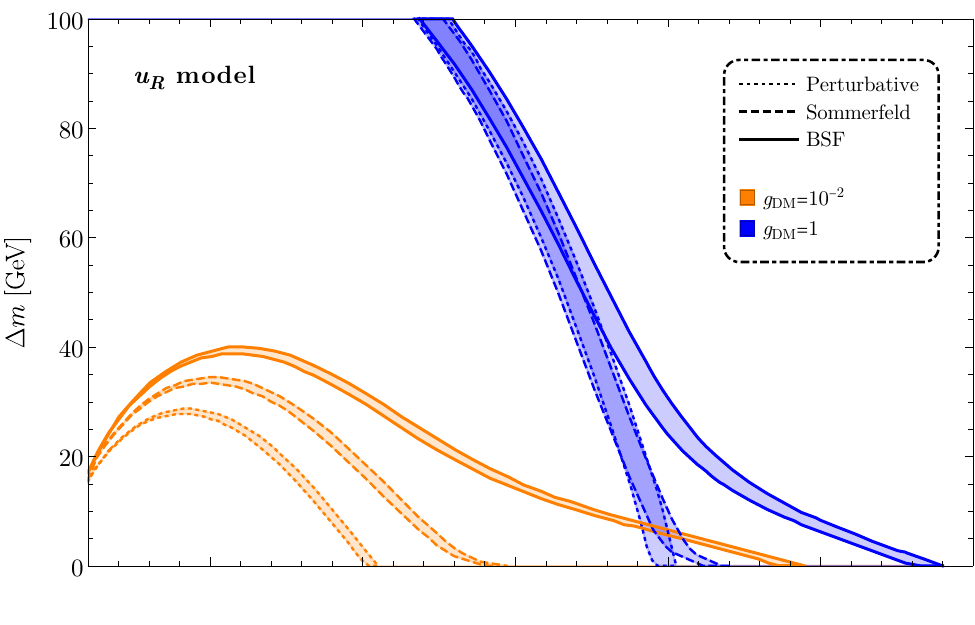}
\end{subfigure}
\\[-3.0ex]
\begin{subfigure}[b]{0.6\textwidth}
   \includegraphics[width=1\linewidth]{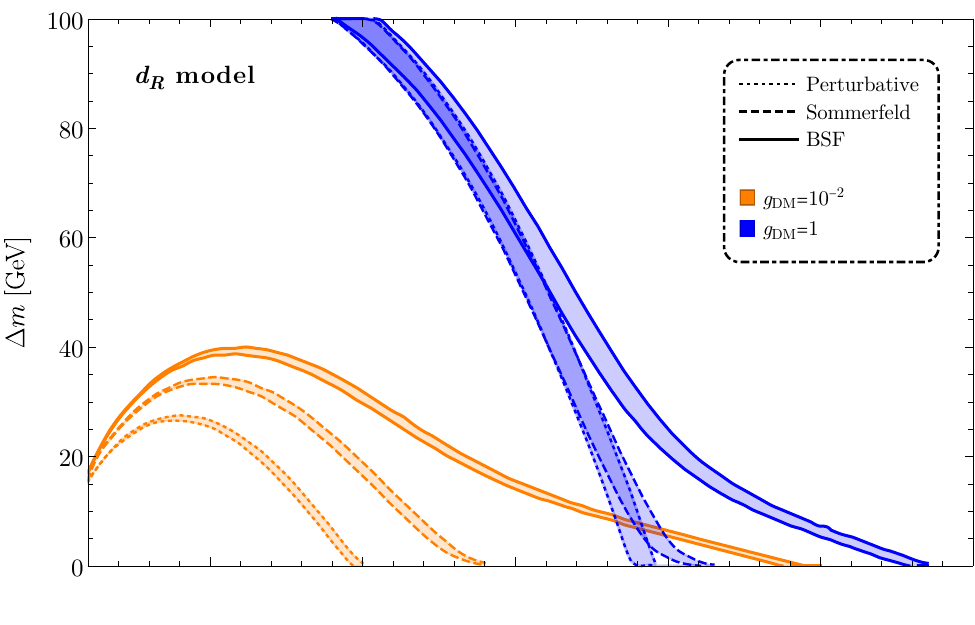}
\end{subfigure}
\\[-3.0ex]
\hspace{3.15pt}
\begin{subfigure}[b]{0.616\textwidth}
   \includegraphics[width=1\linewidth]{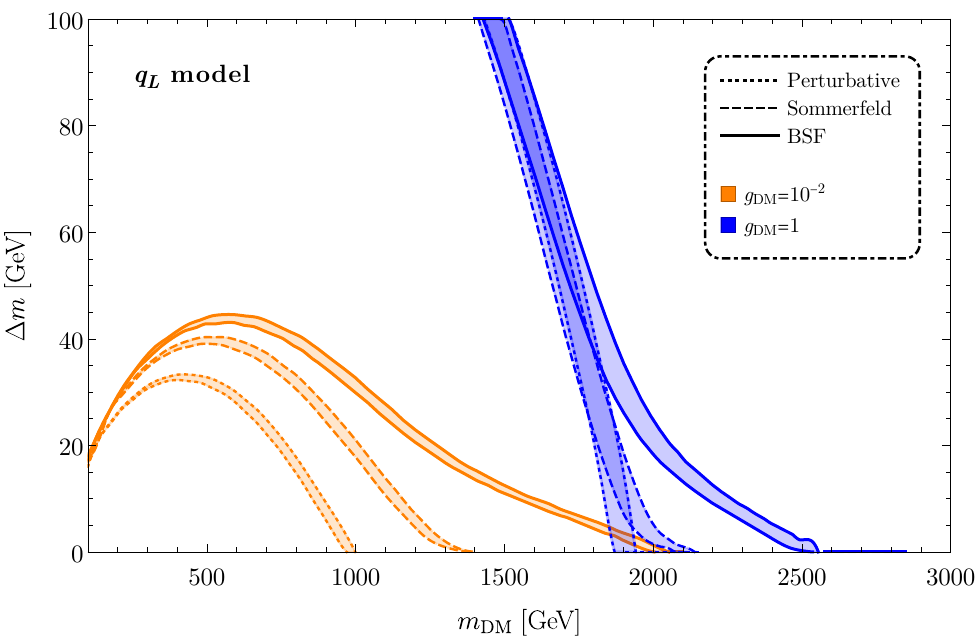}
\end{subfigure}
\caption{\small Mass splitting $\Delta m = m_X-m_\text{DM}$ vs. DM mass ($m_\text{DM}$) in the three models here considered: $u_R$ (top), $d_R$ (middle), and $q_L$ (bottom).
The bands correspond to the observed DM relic density within a $5\sigma$ uncertainty according to \cite{Planck2018}. 
We show the results by progressively including the perturbative contribution only (dotted lines), Sommerfeld corrections (dashed lines) and further inclusion of formation and decay of bound states of the colored triplet scalars (solid lines). 
The two-color scheme refers to two benchmark values of the Yukawa coupling: $g_\text{DM}=1$ (blue) and $g_\text{DM}=10^{-2}$ (orange).}
\label{fig:3models_RD}
\end{figure}

In Fig.~\ref{fig:3models_RD}, we show the impact of non-perturbative effects on the parameter space of the $u_R$, $d_R$ and $q_L$ models described in \ref{sec:model}.
For a fixed value of the Yukawa coupling $g_\text{DM}$, each band represents the parameter range that is in agreement with the observed relic density of DM within $5\sigma$ acceptance $\Omega_\text{DM}h^2=~0.120\pm~0.005$ \cite{Planck2018}.
The two different colors refer to two different benchmark values for $g_\text{DM}$, $10^{-2}$ (orange) and 1 (blue). The different style of the lines identify the type of interactions considered: perturbative cross-section only (dotted), with Sommerfeld effect (dashed) and with Sommerfeld+BSF (solid). 
We notice, as expected, that non-perturbative effects are more pronounced for smaller values of $\gdm$ and smaller mass splittings $\Delta m$. Table~\ref{tab:processes} shows that for a smaller $\Delta m/\mdm$ the suppression of processes involving two colored scalars (cf. Figs. \ref{fig:processXXd1}-\ref{fig:processXXqq} + BSF) is lifted such that BSF and processes that are affected by the SE become more relevant. Additionally, BSF is purely governed by the strong gauge coupling and thus less relevant if the competing processes are enhanced by a large $\gdm$.\footnote{In fact, the strong coupling constant $g_s=\sqrt{4\pi \alpha_s}$ lies in between 1.0 and 1.6 in the region of interest, depending on the average momentum transfer considered. Thus, for $\gdm=1$ interactions regulated by $g_s$ are comparable in strength to interactions regulated by $\gdm$.} Therefore, BSF and the SE have an larger impact on the viable parameter space for a smaller $\gdm$ and for smaller $\Delta m/\mdm$.  

When fixing $\gdm$, the total effective annihilation cross-section of DM is maximized for $\Delta m \rightarrow 0$, since the contributions from co-annihilations and colored annihilations are maximized. Thus, we find the maximal DM mass in agreement with the observed relic density for $\Delta m \rightarrow 0$.
In the $u_R$ model, for small $g_\text{DM}$, the maximal dark matter mass is shifted from roughly 1.0 TeV to 2.4 TeV.
For larger couplings, $g_\text{DM} = 1$, the maximal dark matter mass is expected at 2.9 TeV instead of 2.0 TeV.

For the $d_R$ model, the overall behavior is very similar and entails a shift of the upper bound on the dark matter mass from 1.2 TeV to 2.4 TeV for the smaller $g_\text{DM}$ and from  1.9 TeV to roughly 2.8 TeV for $g_\text{DM}=1$.
Finally, in the $q_L$ case, the effect of BSF on the relic density is milder, as already anticipated at the beginning of this section, with a shift from 1.0 TeV to 2.0 TeV ($g_\text{DM}=10^{-2}$) and from 1.8 TeV to 2.5 TeV ($g_\text{DM}=1.0$), respectively.
This effect is caused by the effective co-annihilation cross-section (see, e.g., Eq.~\eqref{eq:effective_sigmav_XXbar_delta0} for the $\Delta m=0$ case). With more co-annihilating species involved, the effective contribution of each channel is smaller as illustrated in section \ref{sec:model} (cf., e.g., Eq.~\eqref{eq:effective_sigmav_XXbar_delta0}). In the following, for conciseness, we will restrict ourselves to discuss the $u_R$ model, since the basic features of the $d_R$ and $q_L$ cases are very similar.\\

\begin{figure}[!t]
    \centering
    \includegraphics[width=0.8\textwidth]{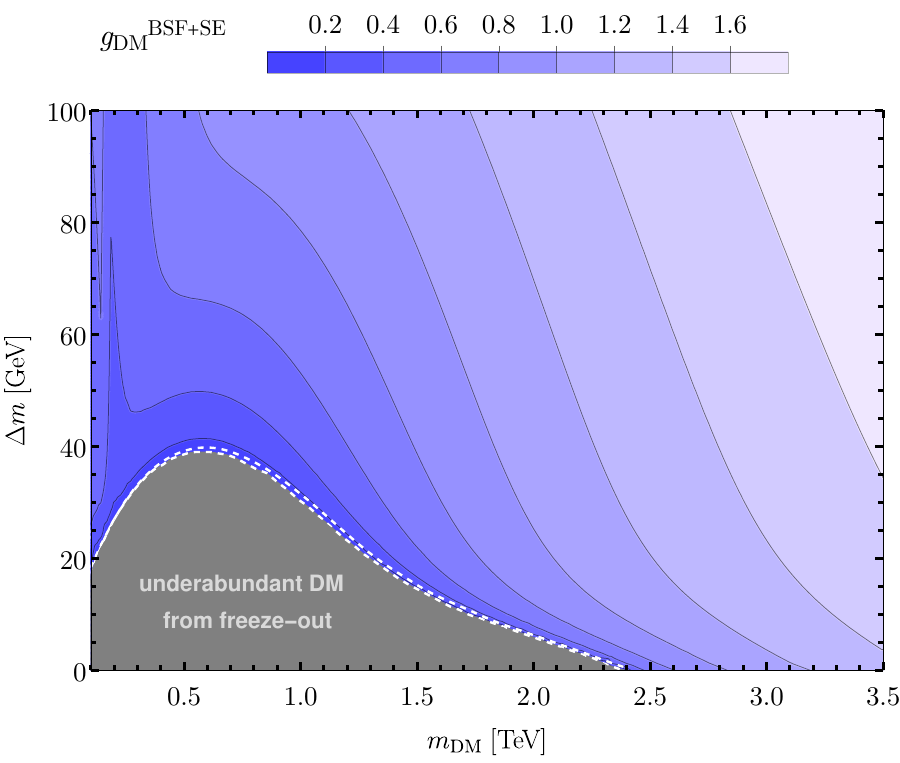}
    \caption{\small Values of the Yukawa coupling that ensures the correct $\Omega_\text{DM}$ for different mass-splitting $\Delta m = m_X-M_\text{DM}$ and DM mass combinations. 
    Here, we take into account BSF+Sommerfeld processes, which become more relevant the closer we are to the co-annihilating region (small mass-splitting).
    The dashed white lines are in correspondence of the region where $10^{-7}\lesssim g_\text{DM}\lesssim 10^{-2}$. Below this interval, it is not possible to ensure chemical equilibrium between the unbound particles in the dark sector and the co-annihilation freeze-out assumption breaks down.}
    \label{fig:gDMRelic_uR}
\end{figure}

In Fig.~\ref{fig:gDMRelic_uR}, we display the values of $g_\text{DM}$ in the $u_R$ model that are able to account for the maximally viable observed DM density parameter $\Omega_\text{DM}h^2=0.125$ (at $+5 \sigma$) in the $(m_\text{DM},\Delta m)$ plane, when BSF and Sommerfeld corrections are taken into account. 
In this sense, the displayed couplings serve as a lower bound on the Yukawa coupling $\gdm$, since a smaller coupling would lead to an overproduction of DM.

We obtain a more stringent lower bound on $\gdm$ for an increasing DM mass, since a larger DM mass implies a smaller relic abundance, and therefore larger effective annihilation cross-section, in order to keep the density parameter at a constant value.

This behavior does not apply for points in the parameter space where the relative mass-splitting is larger than roughly 20~\%.
In fact, for larger relative mass splittings, annihilation processes involving one or more colored scalars (cf. Figs. \ref{fig:processXXd1}-\ref{fig:processXct}) are not efficient due to their Boltzmann suppression. This change of mechanism can be recognized by the sudden change in the bands in the small mass regime around the region corresponding to $\Delta m \simeq 0.2\, m_\text{DM}$.

Furthermore, decreasing the mass splitting implies a smaller $\gdm$, as not only DM annihilations are less suppressed by the (smaller) mediator mass but, additionally, the exponential suppression of the colored annihilations is lifted; these latter are, in turn, enhanced by the Sommerfeld correction and BSF.
Eventually, we encounter a region of parameter space where annihilations mediated by the strong gauge coupling $g_s$ are (almost) efficient enough to deplete the DM density sufficiently on their own. 
This region is enclosed by the white dashed lines and the lower bound on $\gdm$ lies in the interval $10^{-7}\lesssim g_\text{DM} \lesssim10^{-2}$.

For even smaller mass splittings (in the gray region below the lower white dashed line), we are not able to find a lower bound on $\gdm$ anymore, because the corresponding coupling does not suffice to establish chemical equilibrium within the dark sector, so that the freeze-out production of DM does not apply.
In this region, DM freeze-out leads to a relic density (much) smaller than observed in the measurement of the cosmic microwave background.\footnote{The correct relic density could be generated via conversion-driven freeze-out or freeze-in production of DM \cite{Garny_2017,Dagnolo_2017,Garny:2018ali}.}

We estimate the smallest coupling $\tilde{g}_\text{DM}$ ensuring chemical equilibrium in the dark sector by demanding that the interaction rate $\Gamma_X$ of the (inverse) decays $X \leftrightarrow \chi q$ fulfills
\begin{equation}
    \Gamma_X \frac{Y_X^\text{eq}}{Y_\chi^\text{eq}} > H \, ,
    \label{eq:chemeqcond}
\end{equation}
for $T \simeq m_\chi/30$. This condition ensures that chemical equilibrium between $X$ and $\chi$ is established at the time of thermal freeze-out\footnote{Note that out-of-chemical equilibrium effects might already occur at couplings larger than $\tilde{g}_\text{DM}$, when chemical equilibrium is established at $x=30$ but $X$ and $\chi$ leave chemical equilibrium for times slightly before or after thermal freeze-out.}. When neglecting the quark mass, assuming a small mass splitting $\Delta m$ and only considering two-body (inverse) decays of $X$, we obtain from Eq.~\eqref{eq:chemeqcond} 
\begin{align}
 \tilde{g}_\text{DM} \gtrsim \sqrt{\frac{m_\text{DM}}{\mathrm{GeV}}} \left(1 \cdot 10^{-9} + 6.8 \cdot 10^{-11} \frac{m_\text{DM}}{\Delta m}  \right) \, ,  \label{eq:CDFOcoupling}   
\end{align}
\begin{figure}[!t]
    \centering
    \includegraphics[width=0.8\textwidth]{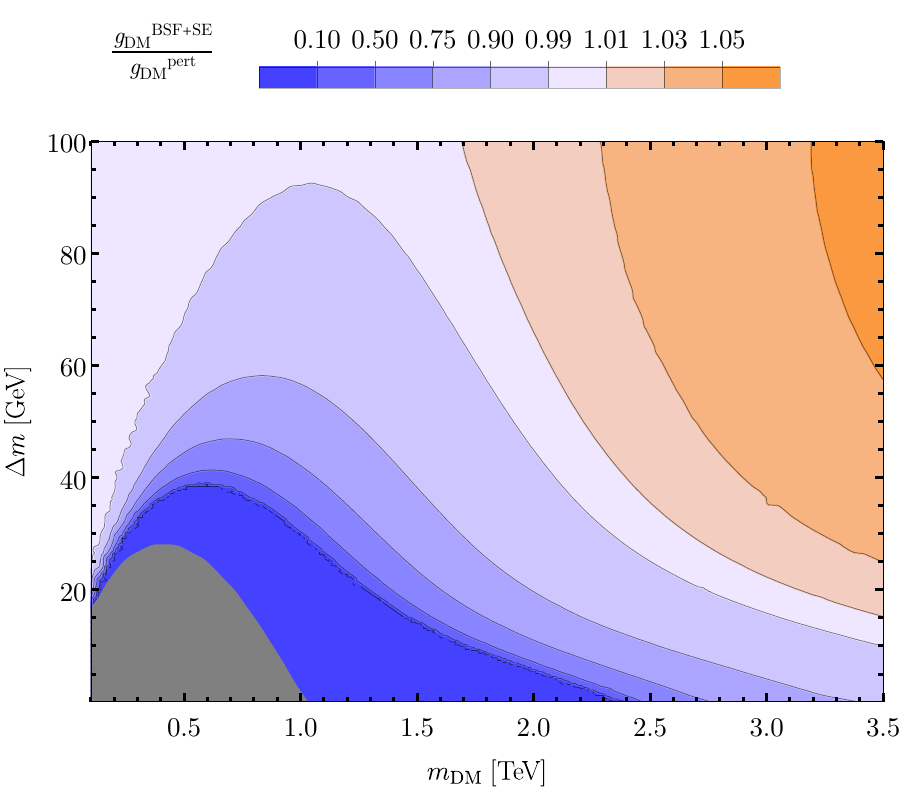}
    \caption{\small The values of the Yukawa coupling ensuring the correct $\Omega_\text{DM}$ including non-perturbative effects (BSF+SE) are compared to the ones including perturbative annihilations only in the ($\mdm,\Delta m$)-plane. 
    Blue-tone colors imply that including non-perturbative effects requires a smaller coupling than in the purely perturbative scenario, while orange-tones indicate the opposite.}
    \label{fig:BSFvspert_uR}
\end{figure}
Note, that Eq.~\eqref{eq:CDFOcoupling} only applies when $\mdm \gg \Delta m \gg m_q$ and thus does not describe the chemical equilibrium condition of the mediator $X$ associated with the top and DM where conversions proceed at next-to-leading order. We conclude that DM production does not proceed via thermal freeze-out for couplings $\gdm \lesssim \tilde{g}_\text{DM}$.

In order to demonstrate the error on the lower bound on $\gdm$ (cf. Fig.~\ref{fig:gDMRelic_uR}) made when not including the SE and BSF, we compare in Fig.~\ref{fig:BSFvspert_uR}, the corresponding lower bound on the coupling $\gdm^\text{SE+BSF}$ obtained when including BSF and SE with the lower bound on $\gdm^\text{Pert}$ obtained considering perturbative annihilations only. 
It is important to note that the correction can lead to higher and lower values such that a dedicated analysis is necessary in order to estimate its implication on the parameter space.

Blue regions in Fig.\@ \ref{fig:BSFvspert_uR} imply that a smaller $\gdm$ is expected in these scenarios when including SE and BSF, while orange regions indicate the opposite. 
The SE can, depending on the sign of the potential, enhance or decrease the effective annihilation cross-section, while BSF, which effectively provides a new annihilation channel, can only increase the effective annihilation cross-section of DM. 
Furthermore, the SE provides a correction to any process featuring two colored particles in the initial state (cf. Figs. \ref{fig:processXXd1}-\ref{fig:processXXqq}), even if they are exclusively mediated by $\gdm$. Thus, the SE has an effect even in the case $\gdm \gg g_s$. BSF, on the other hand, as a new annihilation channel purely mediated by $g_s$, becomes less important as soon as $\gdm \gg g_s$. 
With that in mind, and by comparing Figs.\@ \ref{fig:BSFvspert_uR} and \ref{fig:gDMRelic_uR}, we find that orange regions, where the coupling required to not overproduce DM gets enhanced, coincide with regions of the parameter space where $\gdm \gtrsim g_s \approx 1$. As we can extract from Table \ref{tab:processes}, this region of parameter space is dominated by direct DM annihilations (cf. Fig.\@ \ref{fig:processCC}) and colored annihilations involving a $XX$ initial state (e.g., cf.\@ \ref{fig:processXXqq}), which results in a repulsive potential. 
For parameter points with $\Delta m \gtrsim 0.2\,\mdm$, colored annihilations are strongly Boltzmann suppressed and non-perturbative effects are irrelevant.
Blue regions, where the coupling required is reduced when considering the SE and BSF, are a result of strong bound state effects due to a small relative mass splitting $\Delta m/\mdm$ and a smaller $\gdm \lesssim g_s$. Furthermore, for $\gdm \ll g_s$, colored annihilations involving a $X X^\dagger-$pair mediated by $g_s$ (cf. Fig.\@ \ref{fig:processXXd1}-\ref{fig:processXXdqq}) dominate the effective DM annihilation cross-section, resulting in a attractive potential. While the effects induced by SE for a repulsive potential are on the percent level, the corrections to the coupling for regions with both SE for an attractive potential and efficient BSF are significantly more sizable.
Due to this non-trivial behaviour throughout the parameter space, for a definite exclusion or discovery of the model, BSF and SE have to be considered. We want to emphasize that a constant correction factor (``K-factor") is not sufficient.

\section{Limits on $g_\text{DM}$ from Direct Detection}
\label{sec:DD}
To calculate direct detection constraints on the parameter space, we follow Ref.~\cite{Mohan_2019}\footnote{See also \cite{Belanger:2021smw}.}. Direct detection (DD) constraints arise from the non-observation of DM-nuclei scattering on earth. 
The constraints on the DM-nucleon cross-section come from spin-independent (SI) and spin-dependent (SD) interactions. We use current spin-independent limits from Xenon-1T~\cite{XENON:2017vdw} and spin-dependent limits from the PICO-60 experiment~\cite{PICO:2017tgi}. Future projections are considered for the planned DARWIN experiment~\cite{DARWIN:2016hyl}. 
\begin{figure}
    \centering
    \begin{subfigure}{0.49 \textwidth}
    \centering
    \includegraphics[width=0.5\textwidth]{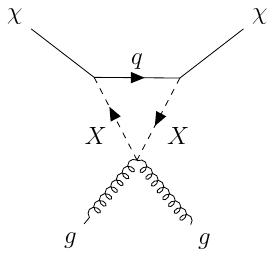}
    \end{subfigure}
    \hspace{-2cm}
    \begin{subfigure}{0.49 \textwidth}
    \centering
    \includegraphics[width=0.5\textwidth]{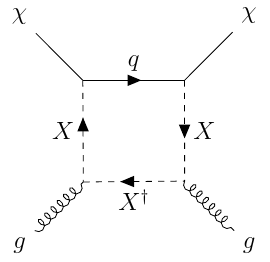}
    \end{subfigure} \\ \vspace{5mm}
    \begin{subfigure}{0.49 \textwidth}
    \centering
    \includegraphics[width=0.5\textwidth]{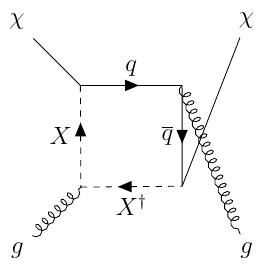}
    \end{subfigure}
    \hspace{-2cm}
    \begin{subfigure}{0.49 \textwidth}
    \centering
    \includegraphics[width=0.5\textwidth]{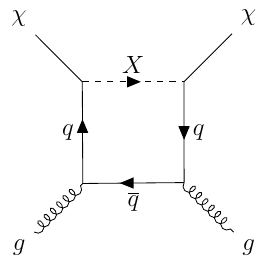}
    \end{subfigure}
    \caption{1-loop diagrams for the DM-gluon elastic scattering, which contribute to the spin-independent DM-nucleon cross-section.}
    \label{fig:SIDD_Diagrams}
\end{figure}
In our model, SD DM-nucleon scattering is mediated at tree-level by the s-channel exchange of a colored mediator $X$ and the SD DM-nucleon cross-section increases with $\gdm^4$. 
For SI scattering however, due to the Majorana nature of the DM candidate, the velocity unsuppressed tree-level contribution is absent.
Thus, SI DM-nucleon scattering is induced at the one-loop level, where it receives its dominant contribution from the diagrams shown in Fig.~\ref{fig:SIDD_Diagrams}.  
Just as in SD scattering, the parametric dependence to the Yukawa coupling for the SI DM-nucleon cross-section also scales as $\gdm^4$. 
To compute the spin-independent DM-nucleon scattering cross-section in this simplified model, we perform a complete one-loop matching of the relevant Wilson coefficients. 
Hereby, we consider all possible diagrams and interference effects. 
We also perform a renormalization group evolution (RGE) from the scales of the mediator mass to the low-energy scale ($\simeq1$~GeV), relevant for DM scattering with the heavy nucleon.
A detailed account of this is provided in \cite{Mohan_2019}.
Including the RGE evolution leads to an enhancement of roughly a factor of two at the amplitude level, hence roughly a factor of four at the cross-section level.
Therefore, since the experimental limits on SI cross-sections are about six orders of magnitude stronger than for the SD ones, and since the SI cross-section, although one-loop-suppressed, is enhanced by the RGE-evolution, both SI and SD constraints are relevant in our study.

In Section \ref{sec:Results}, we will illustrate the impact of the DD constraints on the parameter space of the $u_R$ model here examined.
Given the dependence of both SI and SD cross-sections on $\gdm^4$, from DD experiments we generically obtain upper limits on $\gdm$ for each data point $(\mdm,\Delta m)$.
This upper limit combined with the lower limit on $\gdm$ from preventing an overabundance of DM, will set the allowed and excluded areas of the parameter space, as we will discuss in detail in Ref.~\ref{sec:Results_expconstraints}.
\section{Collider Constraints} \label{sec:colliders}
Since the mediator and dark matter have masses around the TeV-scale, the class of models considered are also subject to constraints from collider experiments.
There are three possible searches that can be used to place constraints on the parameter space:
\begin{enumerate}
    \item Prompt searches: here the mediators are produced and decay promptly to dark matter and jets.
    \item Long-lived particle (LLP) searches: here the mediators that are produced in at the LHC do not decay promptly and may decay inside or outside the detector and require a dedicated search different from the prompt decay searches.
    \item Bound state formation at LHC: here, mediators that are pair produced can form bound states and subsequently decay to SM gauge bosons.
\end{enumerate}
We will show that prompt searches, LLP searches and bound state searches are relevant and complement each other in constraining the parameter space of the model. In the following subsections, we describe each of these LHC searches and our procedure for recasting them for the t-channel model. 




\subsection{Prompt Searches}
\label{sec:promptsearches}
Representative Feynman diagrams for the most relevant processes are shown in Fig.~\ref{fig:fdcol} and include:
\begin{figure}
    \centering
    \begin{subfigure}{0.3\textwidth}
    \centering
    \includegraphics[width=0.6\textwidth]{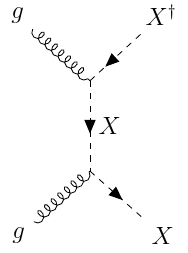}
    \end{subfigure}
    \begin{subfigure}{0.3\textwidth}
    \centering
    \includegraphics[width=0.8\textwidth]{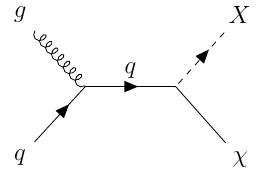}
    \end{subfigure}
    \begin{subfigure}{0.3\textwidth}
    \centering
    \includegraphics[width=0.7\textwidth]{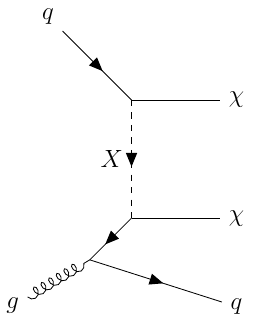}
    \end{subfigure}
    \caption{Feynman diagrams for processes at the LHC. The lines follow the same notation as in Fig.~\ref{fig:SIDD_Diagrams}. In particular, from left to right, we draw the pair-production of mediators, the associated production of DM with a mediator, and the DM pair production with ISR jet.}
    \label{fig:fdcol}
\end{figure}
\begin{enumerate}
\label{colproc}
    \item pair-production of mediators followed by decay to a quark and dark matter.
    \item Associated production of dark matter with a mediator.
    \item Pair-production of dark matter with an Initial State Radiated (ISR) jet.
\end{enumerate}
The production cross-section of the first of these processes contribution purely mediated by the QCD strong coupling, while the latter processes are controlled by the product of dark matter coupling $g_\text{DM}$ and the strong coupling constant.  
Finally, processes scaling with $\gdm^4$ such as pair production of DM and  $q q \to X X$ are also included in the analysis. However, for the region of interest in this paper\footnote{For our analysis, we set the DM coupling $\gdm$ by the requirement not to overproduce DM, resulting in large values of $\gdm$ only if the masses of the dark sector particles are large.}, the phase space suppression ensures that these cross sections are also subdominant.

The cross-sections for these processes are calculated at the Next to Leading Order (NLO) using $\mathrm{MG5\_aMC@NLO}$~\cite{Alwall:2014hca}.
We use the CT14NLO \cite{Dulat:2015mca} PDF set and set the renormalization and the factorization scale at $\mu_{r}=\mu_{f}= \frac{H_{T}}{2}$, where $H_{T}$ is the scalar sum of transverse momenta of all final state particles\footnote{Note that all cross section calculations including the scale setting is performed to match the conventions adopted by the experimental search, for example in \cite{ATLAS:2020syg}}. 
The details of the cross-section calculation is provided in \cite{Mohan_2019}. As detailed in \cite{Mohan_2019}, the K-factors for these processes are fairly flat and range between 1.5-1.6.  

Having described the Monte-Carlo procedure for the generation of events for the collider analysis, we now look at the relevant searches. Given the processes relevant for LHC referred to in Fig. \ref{fig:fdcol} and enumerated above,
prompt searches at the LHC constrain a bulk of the parameter space, as will be shown later.
Both mono and multi-jet + $\cancel{E}_{T}$ searches at the LHC are relevant, especially for the region of parameter space where the  mass difference ($\geq 5 $ GeV) and couplings are sizeable. The mono-jet + $\cancel{E}_{T}$ searches primarily constrain the compressed spectra, and relies on an emitted hard ISR jet, while the multi-jet + $\cancel{E}_{T}$ search constrain moderate to high $\Delta m$ region, due to increased jet activity in the final state.

We will first start by describing the constraints arising from mono and multi-jet + $\cancel{E}_{T}$ searches at the LHC.  They are obtained by recasting the most updated searches at the LHC for these processes, and reinterpreting them in our model. We use the MadAnalysis5 Public Analysis Database (Ma5-PAD) framework \cite{Dumont:2014tja} to recast and reinterpret ATLAS searches. While we only use the latest ATLAS searches to obtain our constraints, we point out that CMS searches have similar coverage, and the constraints only differ slightly depending on the process. 
We briefly describe the validation process for each of the searches. The impact of these searches on the relevant parameter space in question for this work is discussed in Sec.~\ref{sec:Results}.
\subsubsection{Mono-Jet + $ \cancel{E}_{T}$ Searches }
We briefly summarize the recasting of ATLAS search for mono-jet + $ \cancel{E}_{T}$ final state, described in \cite{ATLAS:2021kxv}. In addition to trigger and isolation criteria of objects like leptons and jets, described in \cite{ATLAS:2021kxv}, the search requires the following pre-selection criteria:
\begin{itemize}
     \item A leading jet with $\rm p_{T} > 150 ~GeV$ within a pseudorapidity gap of $|\eta| < 2.4$, and up to additional three jets with $\rm p_{T}> 30$ GeV and $|\eta| < 2.8$. 
     \item A minimum missing transverse energy of 200 GeV. 
     \item A requirement of the azimuthal angle between jets and missing energy of $\rm \Delta\phi(jet,\cancel{E}_{T})>0.4$~for $\cancel{E}_{T} > 200$ GeV and $\rm \Delta\phi(jet,\cancel{E}_{T}) > 0.6$ for ($\rm  \cancel{E}_{T} > 250 ~GeV$), respectively.
     \item Isolated leptons are vetoed with a pre-selection criteria of $\rm p_{T} > 20 ~GeV$ for electrons and $\rm p_{T} > 20 ~GeV$.
\end{itemize}
In addition to the pre-selection criteria, the search is split into 12 exclusive and 12 inclusive search regions depending on the missing energy criteria.
We recast the analysis within the Ma5-PAD framework~\cite{Dumont:2014tja}. The recasted code, as well as the details of the recast process will be made public at~\cite{ma5url}.
A brief description of the recast process following the experimental search is described here.  Jets are reconstructed
using FASTJET \cite{Cacciari:2011ma}, using the $\rm anti-k_{T}$\cite{Dokshitzer:1997in} algorithm with a jet radius of 0.4, and with detector effects simulated by Delphes3 \cite{deFavereau:2013fsa}. All parameters are tuned to match specifications provided by ATLAS. The analysis was validated by reproducing the cutflows for the benchmarks provided in the ATLAS documentation \cite{ATLAS:2021kxv}. 
The cutflows were reproduced to within 10 $\%$ of the official documented results, and therefore we consider this analysis to be validated. 

For the re-interpretation purpose, signal events corresponding to the processes listed above were generated at tree level in MadGraph5~\cite{Alwall:2014hca} with up to two additional jets.  
These signal events were subsequently showered and hadronized using PYTHIA8~\cite{Sjostrand:2014zea}, with matrix element and parton shower (ME-PS) \cite{Hoeche:2005vzu} merging, with a merging parameter set to $m_{X}/4$. 
The event rates after all cuts are normalized to the NLO cross-sections and an integrated luminosity of 139 $fb^{-1}$.
To extract the limits on the models, based on the normalized number of events obtained for each signal region, we apply the log-likelihood method to obtain an exclusion for each point in the $m_{\chi}-\Delta m$ plane. 
 An in-built confidence level calculator within  Ma5-PAD was used to compute the associated upper limit at the 95 $\%$ confidence level (CL) on the signal cross-section according to the CLs prescription \cite{Read_2002}. 
Even though the analysis contains a large number of signal regions, the bulk of the exclusion limit is determined predominantly from signal regions that have large signal rates, low background rates and small uncertainties.

\subsubsection{Multi-Jet + $\cancel{E}_{T}$ Searches}
\label{sec:collider_multijet}
The class of multi-jet + $\cancel{E}_{T}$ searches targets BSM scenarios with at least two jets and a significant amount of missing transverse energy.
We use the latest ATLAS search for this channel at an integrated luminosity of $139$~fb$^{-1}$ for re-interpretation purposes\cite{ATLAS:2020syg}. 
The search targets SUSY particles, in particular gluinos and electroweak gauginos within a simplified model scenario that undergo cascade decays to yield a large multiplicity of jets and $\cancel{E}_{T}$ in the final state. We re-interpret this search to set constraints on the models described in this paper. 

The baseline criteria for this search is,
\begin{itemize}
    \item At least two isolated jets with $p_{T} > 20~\mathrm{GeV}$ and $|\eta| < 2.8$.
    \item A leading jet with $\rm p_{T} > 200~\mathrm{GeV}$ and a subleading jet with $\rm p_{T} > 50~\mathrm{GeV}$. 
    \item A criteria of the azimuthal angle between the leading two jets and missing energy~$\Delta\phi (\mathrm{jet},\cancel{E}_{T}) > 0.2$
    \item An effective mass (sum on the scalar $p_{T}$ of all jets and $\cancel{E}_{T}$ ) $\rm m_{eff}> 800$ GeV.
    \item Electrons (muons) with $p_{T}> 7(6)$ GeV vetoed with the above pre-selection criteria. 
    \end{itemize}
    
    In addition the search is divided into a large number of signal regions depending on the final-state targets and mass gaps.  
    For search regions without final states involving b-jets, which is what we require for our re-interpretation purposes, the signal regions are designed to optimize the signal-to-background ratio in bins of $p_{T}$, the number of jets and $\cancel{E}_{T}$.
    The details of the full search strategy can be found in \cite{ATLAS:2020syg}.
    To recast this analysis, we follow the experimental paper, with the implementation done within the Ma5-PAD framework.
    Pre-selection criteria, detector specifications and signal regions were implemented within the fast simulation platform. 
    For benchmark points documented in \cite{ATLAS:2020syg} for the recast were generated using MadGraph5~\cite{Alwall:2014hca} with at least two additional jets. Parton showering and ME-PS matching was performed with PYTHIA8.
    As was the case with the mono-jet + $\cancel{E}_{T}$ search, jets were constructed with FASTJET with a radius of 0.4 using the  $\mathrm{anti}-\mathrm{K_{T}}$ algorithm. The signal yields, and the subsequent cutflows were compared with the official ATLAS documentation.  
    We find that the recasted search agrees with the official cutflows within 10 $\%$, and hence we consider the analysis to be validated. 
    
To obtain the projections for high-luminosity LHC (HL-LHC) \cite{ZurbanoFernandez:2020cco}, we provide a naive luminosity-rescaled  95 $\%$ confidence level exclusion limit. 
In order to do this, we simply rescale the signal and background yields to HL-LHC in order to estimate the exclusion limit. 
We note that these are rough estimates, while more accurate limits depend on a proper background estimate from data-driven methods, dealing with high pile-ups as a consequence of a larger bunch crossing at high luminosity. 
    
\subsubsection{Bound State Production at LHC} \label{sec:BSLHC}

Colored mediators can be pair produced at the LHC and subsequently form bound states $\mathcal{B}(XX^\dagger)$. 
When the coupling of the mediator to dark matter $(g_\text{DM})$ is small, these bound states predominantly decay to pairs of gauge bosons. 
For the specific scenario we consider here, the decay to pairs of gluons is dominant, with a branching ratio $\sim 99\%$, followed by decay to pairs of photons, $Z\gamma$ and then pairs of Z-bosons.\footnote{Decays to W-bosons are relevant for the $q_L$ model, but not for the $u_R$ model. Furthermore, in the parameter space where these searches are relevant, the coupling $g_{DM}$ is very small and the decay of the mediator to pairs of quarks or dark matter are negligible. 
Finally, decays to $Z\gamma$ are suppressed in the $u_R$ model, but could play a more important role in the $q_L$ model.} 
Here, we consider the production of the bound state $\mathcal{B}(XX^\dagger)$ and its subsequent decay to pairs of photons at the LHC, since the diphoton decay channel provides the strongest constraints for the $u_R$ model. 
Importantly, these constraints do not depend on the mass of dark matter and, for small values of $g_{DM}$, are not sensitive to the mediator and dark matter coupling.

We therefore use results from high-mass ($\gtrsim 100$ GeV) diphoton searches performed by the ATLAS experiment~\cite{ATLAS:2017ayi} to place constraints on the mass of the mediator.
The analysis carried out in Ref.~\cite{ATLAS:2017ayi} provides limits on the fiducial cross-section (times branching ratio) at $95\%$ C.L., which we correct by an almost flat efficiency factor, given by the detector acceptance times the reconstruction efficiency, as described in \cite{ATLAS:2016gzy}, in order to obtain the corresponding total cross-section (times branching ratio).
In order to calculate the theoretical cross-section, we follow the procedure described in Refs.~\cite{Martin:2008sv,Batell:2015zla}, which calculates the bound-state production cross-section at NLO (QCD) and places constraints on the $s$-wave $(J^{PC}=0^{++})$ stoponium production at LHC. 
In the following, we briefly outline the main key points of the analysis. 
The LO production cross-section for a stoponium-like bound state in the Narrow Width Approximation (NWA) is given by
\begin{equation}
    \sigma\big(pp\rightarrow \mathcal{B}(X X^\dagger)\big)= \frac{\pi^2}{8 m_{\mathcal{B}}^3}\Gamma\big(\mathcal{B}(X X^\dagger) \to gg\big)\,\mathcal{P}_{gg}\left(\frac{m_{\mathcal{B}}}{13~\text{TeV}}\right)\ . \label{eq:BSFatLHCcs}
\end{equation}
Here, $\Gamma\big(\mathcal{B}(X X^\dagger\big) \to gg) $ is the decay width of the bound state to pairs of gluons, $\mathcal{P}_{gg}$ is the gluon luminosity for proton-proton collisions at 13~TeV of center-of-mass energy, and $m_{\mathcal{B}} \simeq 2 m_{X}$, is the mass of the bound state\footnote{The relation between the masses is not an exact equality because of the presence of the binding energy. However, its magnitude is much smaller than the mass of the constituents and therefore negligible.}. 
We exploit the NLO calculation derived in \cite{Martin:2008sv} and evaluate the strong coupling $\alpha_s$ at the scale $\mu=m_{\mathcal{B}}$; we then multiply this result by the branching ratio of bound state decays into diphoton resonances (around $0.3\%-0.5\%$).
We refer the reader to ~\cite{Martin:2008sv,Batell:2015zla}, where relevant expressions for the decay width of $\mathcal{B}(X X^\dagger)$ to $gg$, $\gamma \gamma$, $Z\gamma$ and $ZZ$ can be found.

\begin{figure}[!t]
    \centering
    \includegraphics[width=0.8\textwidth]{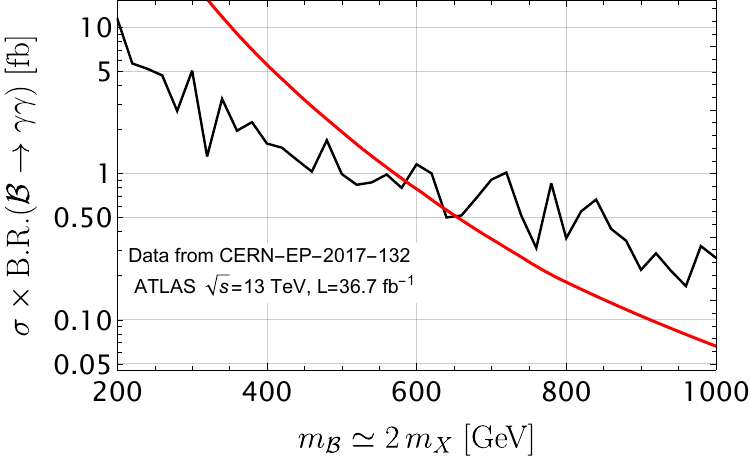}
    \caption{Exclusion limit on the mass of the $\mathcal{B}(XX^\dagger)$ bound state from collider data. 
    The total cross-section for the $pp\rightarrow\mathcal{B}(XX^\dagger)\rightarrow \gamma \gamma$ channel as measured by the ATLAS experiment ($\sqrt{s}=13$ TeV, $\mathcal{L}=36.7$ fb$^{-1}$) is shown as a solid black line \cite{ATLAS:2017ayi}. The red line is our theoretical prediction for the $u_R$ model considered. We exclude masses whenever this line is above the experimental one.}
    \label{fig:stoponium@LHC}
\end{figure}

Accounting for the three generations of stoponium-like bound states of the $u_R$ model, we show in Fig.~\ref{fig:stoponium@LHC} the present exclusion limits on the bound state mass by comparing the experimental bound on the production cross-section times the diphoton-resonance branching ratio (black line) to the theoretical one (red line) for the $u_R$ model.
The figure indicates that the exclusion limits on the mediator mass $m_X\simeq m_\mathcal{B}/2$:
\begin{equation}
    100~\text{GeV}\lesssim m_X \lesssim 290~\text{GeV} \ .
    \label{eq:BS@LHC}
\end{equation}
The lower limit of $100$~GeV is an artifact of the experimental analyses and lower masses can be probed by looking at data from other lower energy experiments. 

In order to provide an estimate of the projected exclusion bounds from the HL-LHC, we perform a rescaling of the integrated luminosity with respect to the data used from ATLAS previously\footnote{We use the asymptotic expressions to determine the significance ($Z =\text{signal}/\sqrt{\text{background}}$). The number of background events are determined from Fig.~2 of \cite{ATLAS:2017ayi} and are rescaled by the luminosity for HL-LHC. We cross-checked that this method is consistent with the experimental bounds obtained from the full experimental likelihood analysis, as has also been noted in \cite{ATLAS:2016gzy}. Additionally, we checked that HL-LHC will have a large number of background events so that the asymptotic form of the significance is a good approximation.  }.
The upper bound on the mass of the mediator increases significantly for HL-LHC and we find that
\begin{equation}
    m_X \lesssim 650~\text{GeV} \ .
        \label{eq:BS@HL-LHC}
\end{equation}

We would like to stress once again that these limits are independent of the precise value of $\gdm$, the Yukawa coupling connecting DM with the mediator and SM particles.

 \subsection{Long-Lived Particle Searches}
 \label{sec:colliders_LLP}
A region of the parameter space of our interest with small mass gaps results in decay widths that can only be probed by LLP searches. 
For this work, we consider searches for heavy stable charged particles (HSCP). 

Pair-production of mediators result in charged particles, that, for certain regions of parameter space, can have small enough decay widths such that they decay outside the detector. 
Since the mediators we study in this model are color triplets, they will form R-Hadrons (neutral or charged).
The exact dynamics of the formation of charged hadrons is governed by QCD. 
For the purposes of this work, we will use the cloud model of hadronization referred in \cite{CMS:2013czn}.
The heavy charged hadron with a velocity $\beta = v/c < 1$ travels through the detector and decays after crossing the tracker or the muon chamber, leaving an ionizing track with an energy loss larger than SM particles. 
For R-hadrons that decay outside the detector, the time-of-flight (TOF) measurement can distinguish BSM particles from the corresponding SM particles. 

ATLAS and CMS have performed HSCP searches at 8 and 13 TeV center-of-mass energies at the LHC, with results presented within supersymmetric models with long-lived gluinos and squarks. 
We follow the method employed in \cite{Belanger:2018sti} to recast and re-interpret CMS searches for HSCP at 8 and 13 TeV LHC~\cite{CMS:2013czn,CMS-PAS-EXO-16-036}. 
The CMS search constrained long lived stops, staus and heavy vector-like leptons utilizing the tracker only and the tracker+TOF analysis. 
Note that the tracker+TOF analysis is relevant only for large values of $c\tau$ ($\geq 10\,\mathrm{m}$). 
We directly translate the cross-section limits on stops and staus obtained by CMS to constraints on our model.  
The regime of maximal sensitivity depends on the lifetime of the parent particles. 
For $c\tau\simeq 1$ cm, a significant fraction of LLPs decay within the tracker, resulting in a suppression of the HSCP signal. 
Therefore, we multiply the pair production cross-section by the fraction of particles decaying either outside the tracker or outside the CMS detector. 
This fraction is dependent on trigger and selection criteria. 
Following previous works~\cite{Belanger:2018sti}, we obtain the effective cross-section by multiplying the efficiency fraction $f_{\mathrm{LLP}}(\tau)$, detailed in \cite{CMS:2015lsu} as, 
\begin{equation}
    \sigma_{\mathrm{eff}}= \sigma \times f_{\mathrm{LLP}}(L,\tau)
\end{equation}
where $L$ is the detector size, which corresponds to $L= 3(11)$ m \footnote{We assume zero efficiency for HSCP tracks if they stop within the tracker (tracker only analysis) or within the detector (TOF analysis).} for the tracker only (tracker+TOF) analysis. 
The effective cross-section is then directly compared to the CMS cross-section upper limits for direct production of stops. 
The cross-sections at leading order is calculated using $\mathrm{MG5\_aMC@NLO}$~\cite{Alwall:2014hca}. 

We finally provide a tentative projection for HL-LHC following previous work in \cite{Belanger:2018sti}. 
As noted there, high-luminosity LHC projections for LLP searches get complicated by the fact that backgrounds are primarily instrumental and therefore cannot be simulated using Monte-Carlo generators. 
Moreover, HL-LHC requires new triggers as well as a better understanding of pile-up rates. 
Nevertheless, following \cite{Belanger:2018sti}, we perform a simple re-scaling of expected signal and observed background events. 
To this end, we  normalize the  signal events obtained after all cuts to the production cross-section and an integrated luminosity of 3000 $\rm fb^{-1}$. For the background, we take the official number of background events from the CMS analysis and normalize it for HL-LHC.
\section{Results and discussion} \label{sec:Results}
\begin{figure}[p]
    \begin{subfigure}{0.48\textwidth}
    \centering
    \includegraphics[width=\textwidth]{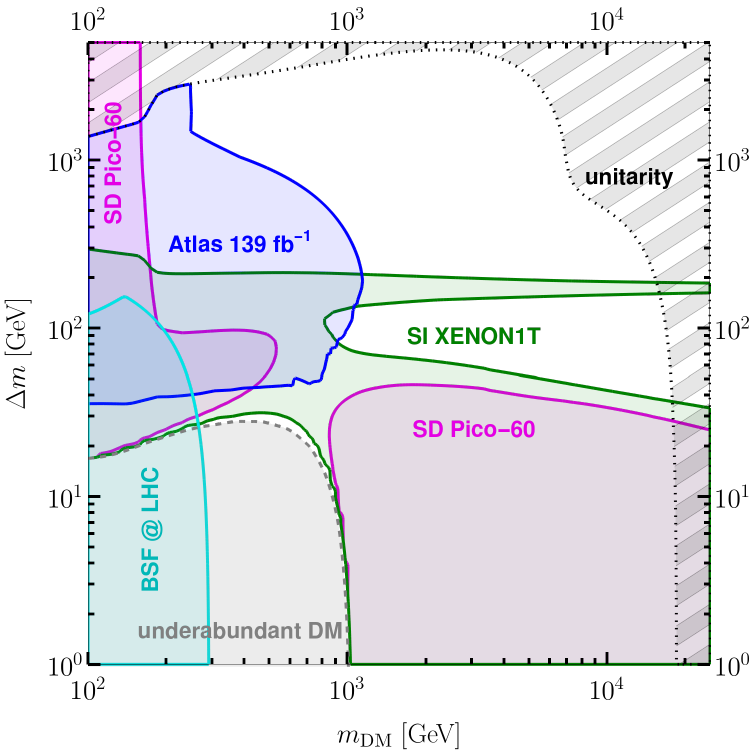} \label{subfig:7A}
    \caption{Perturbative Annihilations}
    \end{subfigure}
    \hspace{1mm}
    \begin{subfigure}{0.48\textwidth}
    \centering
    \includegraphics[width=\textwidth]{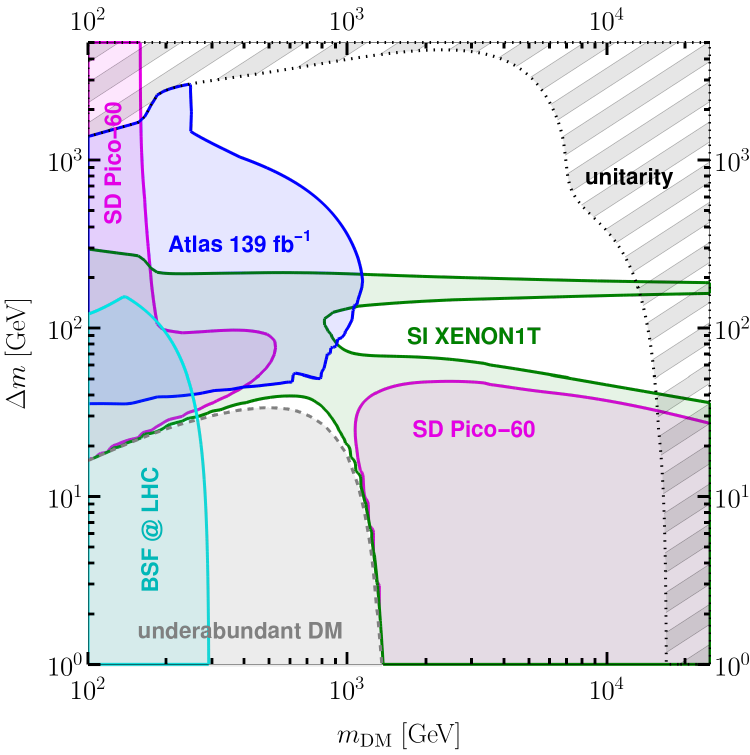} \label{subfig:7B}
    \caption{Sommerfeld}
    \end{subfigure}    \\ \vspace{4mm}
    \begin{subfigure}{0.48\textwidth}
    \includegraphics[width=\textwidth]{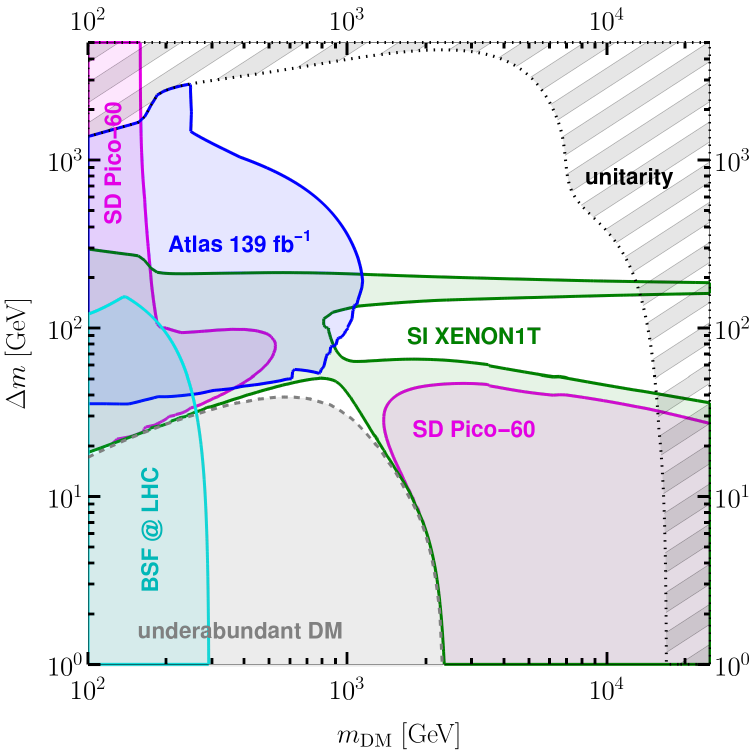} \label{subfig:7C}
    \caption{Sommerfeld+BSF}
    \end{subfigure}
    \vspace{3mm}
    \caption{\small Exclusion limits from various experiments on the $u_R$-version of the model in the $(\mdm,\Delta m)$-plane. For each point in the plane, we determine the smallest value of $\gdm$ such that DM is not overproduced. If this lower bound on $\gdm$ contradicts the limits from spin-independent DD (spin-dependent DD, prompt-collider searches, perturbative unitarity, stoponium searches), it is colored in green (magenta, blue, black, cyan). We show the experimental limits obtained considering perturbative annihilations, the SE and the SE+BSF. The gray dashed line divides the parameter space into two regions. Above, the observed relic density can be generated via thermal freeze-out. Below, thermal freeze-out under-produces DM and accounting for the complete DM relic density requires production via conversion-driven freeze-out or freeze-in.}
    \label{fig:Summary_uR}
\end{figure}
In the following, we demonstrate how current and future experimental measurements constrain the parameter space of the $u_R$ $t$-channel simplified model and how the inclusion of Sommerfeld effect (SE) and bound-state formation (BSF) affect these exclusion limits.
We show the resulting exclusion limits in Fig.\,\ref{fig:Summary_uR}, obtained by considering only perturbative annihilations (a), with the inclusion of the SE (b), and by considering the SE as well as BSF (c), respectively. 
In order to arrive at these limits, we first determine the smallest coupling $g_{\text{DM},\Omega}$ that does not overproduce DM for a given $(\mdm,\Delta m$) point (cf. Fig.\@ \ref{fig:gDMRelic_uR}), and then compare it to the upper limits on $\gdm$ from direct detection (DD) and collider experiments $g_\text{DM,exp}$. 
If $g_{\text{DM},\Omega} > g_\text{DM,exp}$, we regard the data point as excluded by the experiments considered. 

While the precise value of $g_{\text{DM},\Omega}$ is determined by a scan with a modified version of micrOMEGAs, as described in section~\ref{sec:relicdensity}, the scaling behavior of $g_{\text{DM},\Omega}$ can be understood by means of a simple estimate for the relic density, $\Omega_\text{DM} \sim \left\langle \sigma v \right\rangle^{-1}$. For this estimate, it is useful to distinguish between three different regimes. 
Firstly, the region of parameter space where DM annihilations are mainly set by colored annihilations, which arises for DM masses below and around a TeV while $\Delta m \ll \mdm$. 
Thus, $g_{\text{DM},\Omega}$ can be chosen almost arbitrarily small, as long as chemical equilibrium in the dark sector is assured, such that the lower bound on $g_{\text{DM},\Omega}$ approaches zero.
Secondly, the region of parameter space where colored annihilations efficiently contribute to $\left\langle \sigma v \right\rangle$ but annihilations mediated by $\gdm$ are necessary to sufficiently deplete the relic density and provide the dominant contribution to $\left\langle \sigma v \right\rangle$. 
This happens for larger DM masses $\mdm \gtrsim \mathcal{O} \left( \mathrm{TeV} \right)$ and  $\Delta m \ll \mdm$. Lastly, the region where colored annihilations are suppressed by a large mass splitting $\Delta m \gg \mdm$ and the annihilation cross-section is purely set by direct DM annihilations. 
This estimate results in 
\begin{align}
\gdm^4 \propto 
\begin{cases}
  \rightarrow \,0 & \text{, if } \Delta m \ll \mdm \wedge \mdm \lesssim \mathcal{O} \left( \mathrm{TeV} \right), \\
  \mdm^2 & \text{, if } \Delta m \ll \mdm \wedge \mdm \gtrsim \mathcal{O} \left( \mathrm{TeV} \right),  \\
  \left( \Delta m \right)^4 \mdm^{-2} & \text{, if } \Delta m \gg \mdm.
\end{cases} \label{eq:gdmEstimate}
\end{align}
 
In the following, we comment in detail on the implications of the different constraints.
\subsection{Interplay of current experimental constraints}
\label{sec:constraints_uR}
\label{sec:Results_expconstraints}
\textbf{Spin-Independent Direct Detection (SI-DD). }
Following the discussion in Section~\ref{sec:DD}, we show in Fig.~\ref{fig:Summary_uR}, a green area corresponding to the resulting excluded regions from SI-DD searches in the $(\mdm, \Delta m)$-plane.
They lead to strong exclusion limits for small mass splittings ($\Delta m~\lesssim~100\,~\mathrm{GeV}$) and large DM masses ($\mdm~\gtrsim~1\,~\mathrm{TeV}$). 
It is noteworthy, that SI-DD is unable to constrain mass splittings $\Delta m \lesssim 50\,\mathrm{GeV}$ for smaller DM masses, $\mdm~\lesssim~1\,~\mathrm{TeV}$. 
The size of this region strongly depends on the inclusion of non-perturbative effects, as we discuss in the following section (cf. Section \ref{sec:Results_Combined}).
Additionally, we observe an excluded region around mass splittings equal to the top quark mass, which is a result of a top-quark resonance in the DM-nucleon scattering cross-section calculated at one-loop order \cite{Mohan_2019}. 
The scaling behavior of the SI-DD constraints can be understood directly from the form of the DM-nucleon cross-section \cite{Mohan_2019}
\begin{align}
    \sigma_\text{SI} \propto \begin{cases}
     \dfrac{\gdm^4}{\mdm^2 } \dfrac{m_N^4}{\left( \Delta m \right)^4} & \text{, if } m_q \ll \Delta m \ll \mdm,\\[5mm]
     \dfrac{\gdm^4}{\mdm^2} \dfrac{m_N^4}{\left( \Delta m \right)^4} \dfrac{\mdm^6}{\left( \Delta m \right)^6} & \text{, if } \Delta m \gg \mdm,
    \end{cases} \label{eq:SIDDEstimate}
\end{align}
where $m_q$ is the mass of the quark propagating in the loop mediating the interaction of DM with gluons and $m_N$ is the mass of the nucleon.
Additionally, the lower bound on $\gdm$ is fixed by the requirement of not overproducing the relic density. 
Using $\Omega_\text{DM} \sim \left\langle \sigma v \right\rangle^{-1}$ we arrive at Eq.~\eqref{eq:gdmEstimate}.
Inserting Eq.~\eqref{eq:gdmEstimate} into Eq.~\eqref{eq:SIDDEstimate}, we obtain
\begin{align}
    \sigma_\text{SI} \propto 
    \begin{cases}
    \rightarrow 0  & \text{, if } \Delta m \ll \mdm \wedge \mdm \lesssim \mathcal{O} \left( \mathrm{TeV} \right), \\
    m_N^4\left( \Delta m \right)^{-4} & \text{, if } \Delta m \ll \mdm \wedge \mdm \gtrsim \mathcal{O} \left( \mathrm{TeV} \right), \\
    m_N^4 \mdm^2 \left( \Delta m \right)^{-6} & \text{, if } \Delta m \gg \mdm.
    \end{cases} \, \label{eq:ScalingSIDD}
\end{align}
This result supports the observations made previously. 
Spin-independent DD is able to constrain areas of the parameter space involving small mass splittings that are not dominated by annihilations mediated by the strong gauge coupling, whereas regions of large mass splitting remain unconstrained due to the suppression by six powers of the relative mass splitting.
\\ \\
\textbf{Spin-Dependent Direct Detection (SD-DD).} Following the discussion in Section \ref{sec:DD}, we depict the resulting SD-DD excluded regions with magenta tones in the $(\mdm, \Delta m)$-plane in Fig.~\ref{fig:Summary_uR}. 
Spin-dependent DD is more constraining than spin-independent DD for larger mass splittings $\Delta m \gtrsim 200 \, \mathrm{GeV}$, while the exclusion limits are less stringent for mass splittings of $\Delta m \lesssim\,100 \mathrm{GeV}$ and, as for SI-DD, we find an unconstrained region of parameter space in the case of $\mdm~\lesssim~1\,~\mathrm{TeV}$ and $\Delta m \lesssim 50\,\mathrm{GeV}$.
Again, this fact can be understood from the scaling of the SD-DD cross-section:
\begin{align}
    \sigma_\text{SD} \propto
    \begin{cases}
      \dfrac{\gdm^4 m_N^2}{\left( \Delta m \right)^4} \rightarrow 0 & \text{, if } \Delta m \ll \mdm \lesssim \mathcal{O} \left( \mathrm{TeV} \right), \\[5mm]
      \dfrac{\gdm^4 m_N^2}{\left( \Delta m \right)^{2} \mdm^2} \propto \dfrac{m_N^2}{\left( \Delta m \right)^2} & \text{, if } \Delta m \ll \mdm \gtrsim \mathcal{O} \left( \mathrm{TeV} \right), \\[5mm]
     \dfrac{\gdm^4 m_N^2}{\left( \Delta m \right)^4} \propto \dfrac{m_N^2}{\mdm^2} & \text{, if } \Delta m \gg \mdm ,
    \end{cases} \, \label{eq:ScalingSDDD}
\end{align}
where in the second step we inserted the estimates for the lower bound on $\gdm$ given in Eq.~\eqref{eq:gdmEstimate}. 
Comparing the results for the SD and SI DM-nucleon cross-section reveals that for $\Delta m \ll \mdm$ the SD contribution is only enhanced by two powers of the relative mass splitting and thus less constraining than the SI contribution, which is enhanced by four powers of $\Delta m/\mdm$. 
On the other hand, for large mass splittings, the SD contribution does not depend on $\Delta m$ and thus is able to constrain these areas of the parameter space below a certain DM mass threshold of $\mdm \lesssim 200\, \mathrm{GeV}$.
\\ \\
\textbf{Partial-wave unitarity bounds. }\label{sec:Results_unitarity}
In order to ensure the perturbativity and unitarity of the model, we require a conservative limit of $\gdm < \sqrt{4\pi}$. We show with a black shaded area in Figs.\@ \ref{fig:Summary_uR}-\ref{fig:Summary_uR_proj} the range of the parameter space when this condition is violated. The limit on $\gdm$ from partial wave unitarity for this simplified $t$-channel DM model was given in \cite{Cahill-Rowley:2015aea}, relying on methods developed in \cite{Schuessler:2007av}.
\\ \\
\textbf{Prompt Collider Constraints. }
Following the discussion in Section \ref{sec:colliders}, we show with blue colours the resulting excluded regions in the $(\mdm, \Delta m)-$plane in Fig.~\ref{fig:Summary_uR}. 
They provide the most stringent constraints for $\mdm \lesssim 1.2 \, \mathrm{TeV} \land \Delta m \gtrsim 100 \, \mathrm{GeV}$. 
We do not consider exclusion limits from prompt collider searches if $\gdm > \sqrt{4 \pi}$, since it violates the limit obtained from perturbative unitarity and additionally the assumption of a narrow width for the colored scalar $X$ breaks down. 
For low mass gaps, the mono-jet constraints provide the strongest constraints as it relies on an emitted hard ISR jet recoiling against the missing energy. 
For large mass gaps, the jet multiplicity and the jet transverse momentum $p_{T}$ is significantly larger, and therefore the primary constraints originate from the multi-jet + $\cancel{E}_{T}$. 
Overall, the difference between the two searches is not significant as the multi-jet + $\cancel{E}_{T}$, this being an inclusive search, is generally competitive in most of the parameter space. The primary reason for this being the hard radiated jets from ISR that passes the somewhat strict pre-selection criteria for multi-jet searches. In summary, both multi-jet and mono-jet + $\cancel{E}_{T}$ searches yield similar limits in the low mass gap region, while the multi-jet + $\cancel{E}_{T}$ search yields stronger limits in the large mass gap region. Generally, the prompt searches have two effects that impact its shape in the $(m_{X},\Delta m)-$plane. The first, in regions with small mass gap, the cross-section is enough such that there is significant sensitivity to existing searches. 
The second, regions of large mass gap, there is a combined effect of rising sensitivity and falling cross-section resulting in a cliff. 

In addition, we provide limits from bound state searches at the LHC as described in section~\ref{sec:BSLHC}. 
We find that ATLAS excludes masses of the scalar mediators of $m_X = \mdm + \Delta \lesssim 290 \, \mathrm{GeV}$. 
The corresponding excluded data points are colored in cyan. 
In order to exclude a data point by this search, we further demand that the binding energy of the bound states exceeds the decay rate of the constituent, otherwise, bound state production at the colliders is suppressed \cite{Kats_2010}. 
We comment specifically on the potential of this type of search in Section~\ref{sec:ImplicationBSFatLHC}.

\subsection{Impact of Sommerfeld effect and bound state formation on the combined exclusion limits}
\label{sec:Results_Combined}
Considering the experimental constraints from direct detection and collider searches as well as the unitarity bound, the remaining unconstrained parameter space can be divided into two distinct areas, requiring very different values of the coupling $\gdm$ (cf. Fig.~\ref{fig:Summary_uR}). 
For large $\mdm$ and $\Delta m$, it is required to have $\gdm \gtrsim \mathcal{O} \left( 1 \right)$ in order to avoid the overproduction of the DM relic abundance;
on the contrary, for smaller masses, the couplings lie in the range $0 \leq \gdm \lesssim 0.1$.
Due to the considerable coupling strength $\gdm$ required, the first scenario is only mildly affected by SE and BSF. 

\begin{figure}[p]
    \begin{subfigure}{0.48\textwidth}
    \includegraphics[width=\textwidth]{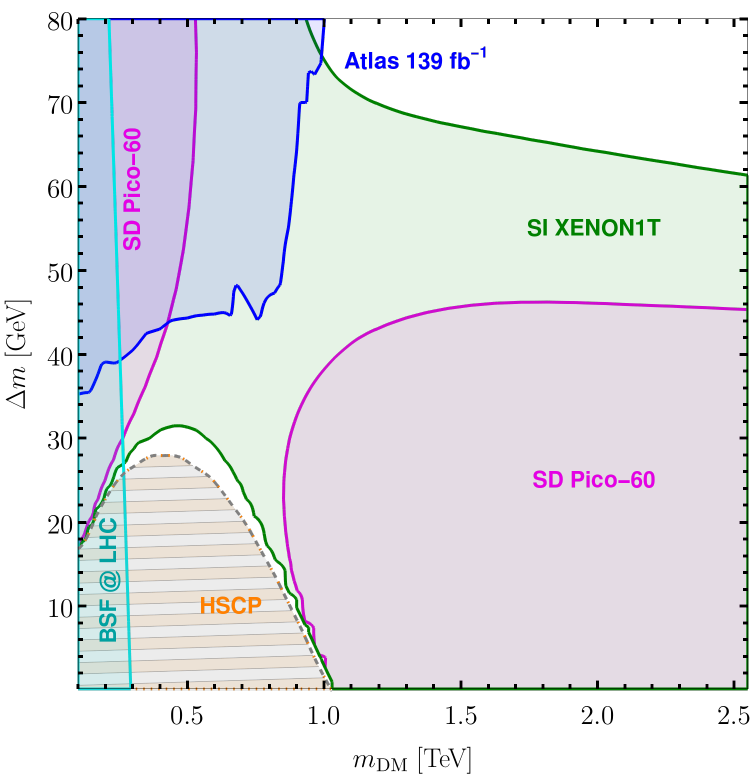}
    \caption{Perturbative Annihilations}
    \end{subfigure}
    \begin{subfigure}{0.48\textwidth}
    \includegraphics[width=\textwidth]{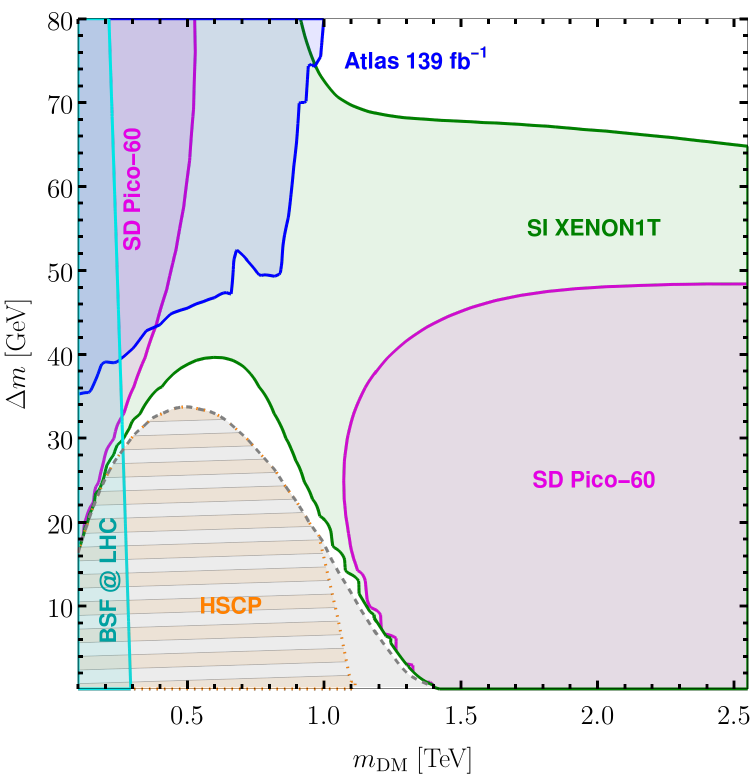}
    \caption{Sommerfeld}
    \end{subfigure}    \\ \vspace{2mm}
    \begin{subfigure}{0.48\textwidth}
    \includegraphics[width=\textwidth]{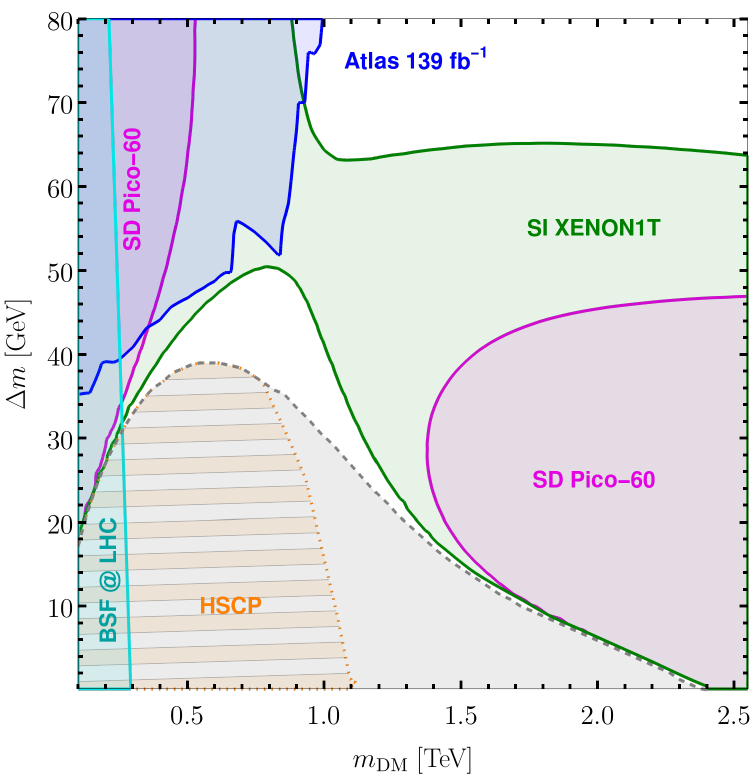}
    \caption{Sommerfeld+BSF}
    \end{subfigure}
    \caption{\small Exclusion limits from various experiments in the co-annihilating area. As before, the limits from spin-independent DD (spin-dependent DD, colliders, unitarity, stoponium searches) are colored in green (magenta, blue, black, cyan). In addition, we provide an estimate for the potential reach of LLP searches constraining the area of the parameter space where the correct relic density can be reproduced only via non-thermal mechanisms, such as conversion-driven freeze-out or freeze-in. 
    Those regions are marked in orange.}
    \label{fig:Summary_uR_zoom}
\end{figure}

As stressed in Section \ref{sec:relicdensity}, BSF is negligible for larger DM masses and mass splittings. 
In this region, however, in the case with $\gdm > g_s$, the particle-particle colored annihilation cross-sections are Sommerfeld-affected by the presence of a repulsive effective potential\footnote{The effect of Sommerfeld repulsion is only relevant when colored annihilations are most efficient, namely when $\Delta m\lesssim 0.2 \mdm$. As shown in Tab.~\ref{tab:processes}, the only velocity non-suppressed process that depends on $\gdm^4$ is $X_i X_i\rightarrow q_i q_i$ with same flavor $i$, whose color contribution results only in the sextet configuration, which is repulsive. As soon as $\gdm>g_s$ this process dominates.}. 
Hence, a larger $\gdm$ is necessary in order not to overproduce dark matter such that the limits arising from direct detection, collider experiments and unitarity are stronger.
The size of the effect is at the few $\%$ level and displays as a slight shift in, for instance, in the direct exclusion limits for large DM mass of $20$ TeV in Fig.~\ref{fig:Summary_uR}.

Remarkably, the parameter region featuring $\mdm \lesssim 2.5 \mathrm{TeV}$ and $\Delta m \lesssim 80 \, \mathrm{GeV}$ is strongly affected by SE and BSF. 
To highlight this effect, we refer to the exclusion limits in Fig.~\ref{fig:Summary_uR_zoom}.
In this region, both the SE and BSF significantly enhance the annihilation cross-section of DM, thereby, allowing for smaller values of $\gdm$.
Consequently, the experimental limits are relaxed.
For instance, we notice that the regions constrained by direct detection experiments are significantly reduced for almost all the mass splittings displayed in Fig. \@\ref{fig:Summary_uR_zoom}. 
Most notably, when including the SE and BSF, we find that parts of the parameter space for $500\,\mathrm{GeV} \lesssim \mdm \lesssim 1.5\,\mathrm{TeV}$, previously thought to be excluded, can indeed produce the observed relic density and not being ruled out by current experiments.

We observe that the largest viable DM mass in this region is shifted from $\mdm \lesssim 1 \, \mathrm{TeV}$ to $\mdm \lesssim 2.4 \, \mathrm{TeV}$, while the maximally viable mass splitting increases from $\Delta m \lesssim 30 \, \mathrm{GeV}$ to $\Delta m \lesssim 50 \, \mathrm{GeV}$. 
In conclusion, considering the SE and BSF leads to a $\mathcal{O} \left( 100 \% \right)$ correction on the exclusion limits in the strongly co-annihilating area. 
Finally, we would like to point out that only considering SE, although it provides a correction of $\mathcal{O} \left( 10 \% \right)$ to the exclusion limits, is insufficient in the strongly co-annihilating area, as can be clearly observed from Fig.~\ref{fig:Summary_uR_zoom}. 
Therefore, we want to emphasize that for a conclusive statement on the exclusion of such models, a full analysis including the Sommerfeld effect and bound state formation needs to be taken into account.

In the gray area shown in Figs.~\ref{fig:Summary_uR} and in the zoomed version Fig.~\ref{fig:Summary_uR_zoom}, thermal freeze-out always leads to underabundant DM, which is not experimentally excluded but requires an additional mechanism to explain the observed relic abundance.
For example, for couplings $\gdm$ small enough such that DM is not in chemical equilibrium with the colored scalars $X_i$, DM production can proceed via alternative mechanisms such as conversion-driven freeze-out or freeze-in.

Finally, thermal freeze-out is able to account for the observed DM abundance in the remaining white regions.
As clearly visible from Fig.~\ref{fig:Summary_uR_zoom}, SE and BSF lead to a significant increase of the allowed parameter space, which would have been underestimated when considering only a pure perturbative tree-level calculation and hence these effects need to be included for a comprehensive study.

As a final remark, since the couplings required to generate the observed relic density below the gray dashed line are tiny, $\gdm \lesssim 10^{-6}$, they can hardly be constrained by prompt collider signatures or direct detection experiments. 
However, for couplings of this size, the colored scalars $X$ are potentially collider-stable and therefore they could appear through long-lived-particle (LLP) signatures. 
In  what follows, we estimate the potential reach of LLP searches within these regions of the parameter space.

\subsection{Potential of long-lived particle searches}
\label{sec:Results_LLP}
In order to address the parameter space unable to account for 100\% of the DM relic density via thermal freeze-out (and possibly tested by LLP searches), we estimate the size $\tilde{g}_\text{DM}$ below which the interaction allows for an out-of-equilibrium dark sector.
To this end, we exploit the condition stated in Eq.~\eqref{eq:chemeqcond}, which ensures that the interaction rate of the (inverse) decays of the colored scalars is out-of-equilibrium at the typical timescales of thermal freeze-out.
In this sense, $\tilde{g}_\text{DM}$ provides a rough estimate of the couplings involved in scenarios relevant for the conversion-driven freeze-out, while it certainly acts as an upper bound on couplings involved in freeze-in scenarios.    

In contrast to the aforementioned constraints, the search for LLPs results in lower limits on the coupling $\gdm > \gdm^\text{LLP}$, since a feebler coupling implies more stable colored scalars $X_i$, which are therefore subject to tighter constraints from heavy stable charged particles (HSCP) searches, as described in section~\ref{sec:colliders_LLP}.
We mark the areas when $\tilde{g}_\text{DM} < \gdm^\text{LLP}$ in Fig.~\ref{fig:Summary_uR_zoom} with orange meshes: they indicate the potential of LLP searches to constrain DM production relying on an out-of-equilibrium dark sector\footnote{More precisely, we underestimate the exclusion regions for freeze-in, since $\gdm^\text{freeze-in} < \tilde{g}_\text{DM}$. 
On the other hand, the limits only provide a rough estimate for scenarios of conversion-driven freeze-out, which is relevant at $\gdm \approx \tilde{g}_\text{DM}$. 
To arrive at a conclusive statement, a dedicated study of this region is required. For the conversion-driven freeze-out scenario an analysis of the $d_R$-version of the model has been recently provided in \cite{garny2021bound}. 
The limits obtained therein are weaker since it is found that conversion-driven freeze-out can have sizable effects for couplings a factor $2$ to $5$ larger than obtained with our estimate in Eq.~\eqref{eq:chemeqcond}. 
Thus, in our work, we find decay lengths roughly a factor $10$ larger than in \cite{garny2021bound}, resulting in more stringent limits.}.

Overall, we find that HSCP searches could test DM masses up to roughly $1.2\,\mathrm{TeV}$ and decay lengths of $\mathcal{O} \left( 10 \, \mathrm{cm} \right)$. 
As a consequence, the complete area below the gray dashed-line is constrained when considering perturbative annihilations only.
Still, a large part of it would remain unconstrained when SE and BSF are taken into account, since it would extend to DM masses beyond the current reach of HSCP searches. 
In conclusion, we stress that the inclusion of the SE and BSF is essential for determining the parameter space of interest for LLP searches. 

\subsection{Projected exclusion limits of future experiments}
\label{sec:Results_projected}

\begin{figure}[p]
    \begin{subfigure}{0.48\textwidth}
    \centering
    \includegraphics[width=\textwidth]{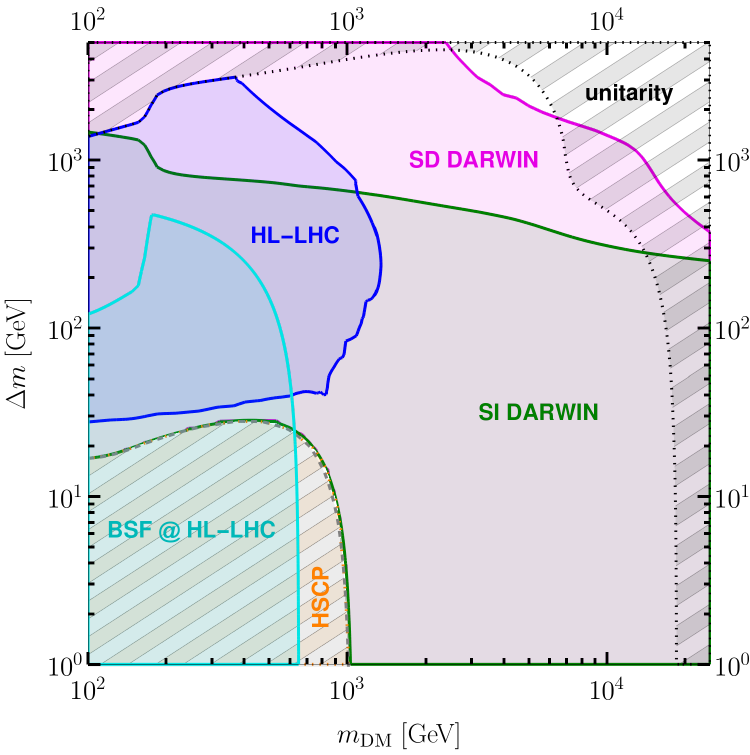}
    \caption{Perturbative Annihilations}
    \end{subfigure}
    \begin{subfigure}{0.48\textwidth}
    \centering
    \includegraphics[width=\textwidth]{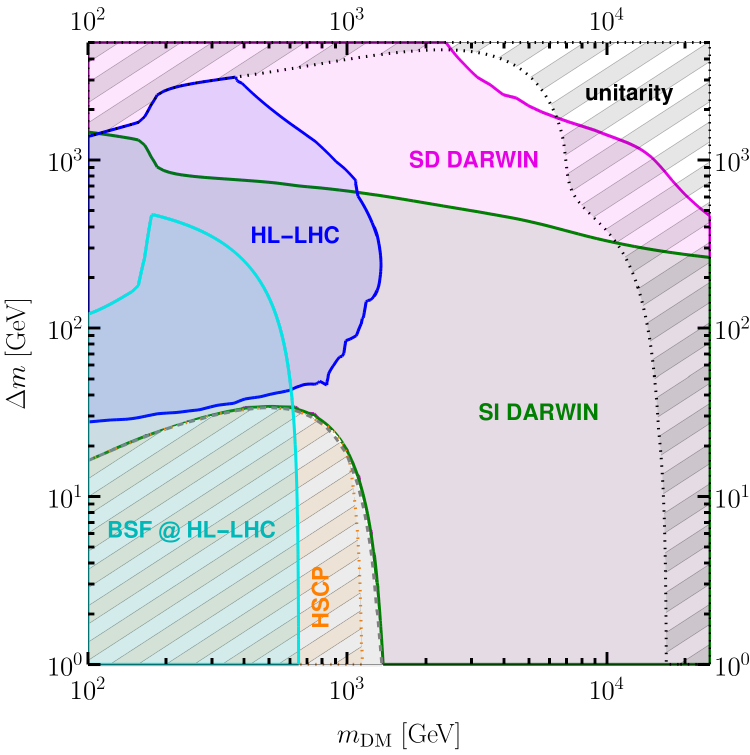}
    \caption{Sommerfeld}
    \end{subfigure}    \\ \vspace{2mm}
    \begin{subfigure}{0.48\textwidth}
    \includegraphics[width=\textwidth]{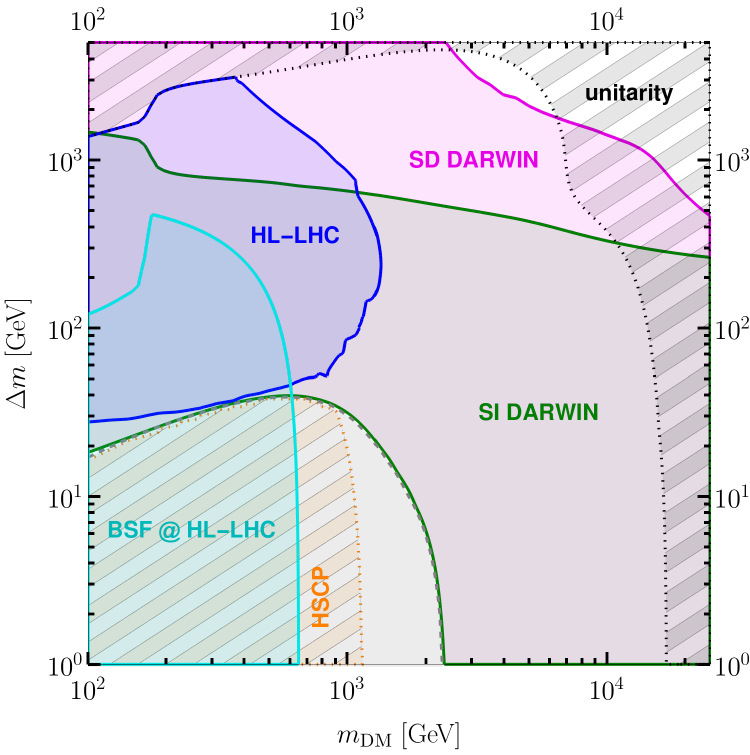}
    \caption{Sommerfeld+BSF}
    \end{subfigure}
    \caption{\small Projected exclusion limits from various future experiments on the $u_R$-version of the model in the $(\mdm,\Delta m)$-plane. For each point in the plane, we determine the smallest value of $\gdm$ such that DM is not overproduced. If this lower bound on $\gdm$ contradicts the projected limits from spin-independent DD (spin-dependent DD, prompt-collider searches, perturbative unitarity, stoponium searches, LLP searches), it is colored in green (magenta, blue, black, cyan, orange). We show the experimental limits obtained considering perturbative annihilations, the SE and the SE+BSF. The gray dashed line divides the parameter space into two regions. Above, the observed relic density can be generated via thermal freeze-out. Below, thermal freeze-out under-produces DM and accounting for the complete DM relic density requires production via conversion-driven freeze-out or freeze-in.}
    \label{fig:Summary_uR_proj}
\end{figure}
In Fig.~\ref{fig:Summary_uR_proj}, we illustrate the expected exclusion regions considering the projected sensitivities for the DARWIN experiment, as described in Section \ref{sec:DD}, and the HL-LHC, as discussed in Section \ref{sec:colliders}.
We find that, given the much larger amount of data expected for HL-LHC and considering the improvement of roughly two orders of magnitude on the excluded DM-nucleon cross-section with respect to XENON1T and PICO-60, the parameter space for freeze-out production of the observed DM relic density that was left unconstrained by the current experimental limits (cf. Fig.~\ref{fig:Summary_uR}) will be  tested almost completely (cf. Fig.~\ref{fig:Summary_uR_proj}).
Notice that, while HL-LHC will enlarge the probed parameter-space, the most dramatic improvement will in particular arise from DD experiments.

In particular, assuming that the Majorana particles account for 100 \% of the observed DM relic density,  both SI and SD constraints will be able to almost entirely test the currently-remaining parameter space of the $u_R$ model. 
Notice that, while Fig.~\ref{fig:Summary_uR_proj} seems to imply that all the parameter space above the gray dashed line is excluded by direct detection, in fact a tiny white space is remaining as the green and gray dashed line do not overlap exactly. 
The remaining region corresponds to small couplings, $10^{-6} \lesssim \gdm \lesssim 10^{-2}$, where direct detection is evaded but DM production still proceeds via thermal freeze-out\footnote{This region is narrow in $\Delta m$ (less or about a GeV), since the efficiency of the colored co-annihilations dominating the effective annihilation cross-section is regulated by $\exp \left( 2 \Delta m \cdot \mdm^{-1} \right)$.}. 
While the couplings involved are too small to allow for a signal in direct detection experiments or prompt searches, they are also too large to induce long-lived particle signatures. 
Remarkably, as visible in Figs.~\ref{fig:Summary_uR_zoom} and \ref{fig:Summary_uR_proj}, bound state searches at the LHC manage to close this gap below a certain mass threshold.
The latter is expected to increase significantly, from roughly $290$ GeV at the LHC to roughly $650$ GeV, at the HL-LHC. 

For larger mass splittings, the experimental limits improve drastically, with SD constraints being especially powerful in this region.
This is a result of the less pronounced suppression with the relative mass splitting for SD DM-nucleon scattering (cf. Eq.~\eqref{eq:ScalingSDDD}) compared to the suppression of the SI scattering (cf. Eq.~\eqref{eq:ScalingSIDD}).   

If future experiments do not observe a signal, three areas of the parameter space will still survive. 
Firstly, a small region of freeze-out DM involving $\gdm \sim \mathcal{O} \left( 1 \right)$ couplings with DM and mediator masses in the multi-TeV regime. 
This scenario could possibly be probed by future collider experiments with larger center-of-mass energies such as the FCC~\cite{FCC:2018vvp}. 
Secondly, a narrow belt involving couplings $10^{-6}\lesssim \gdm \lesssim 10^{-2}$ at the transition region to the non-thermal production area. 
Lastly, the region below the gray dashed line, where DM is either underproduced via freeze-out or created via conversion-driven freeze-out or freeze-in.
Importantly, we find that the latter two regions are significantly affected when Sommerfeld effects and bound state effects in the early universe are taken into account. 
If DM production in this region is non-thermal, it can be tested by LLP searches.
Again, we stress that LLP searches beyond the HL-LHC appear to be necessary to thoroughly probe this scenario.

To conclude, both in the absence or in the presence of a positive detection signal, it becomes evident that in order to accurately determine the remaining viable parameter space of simplified $t$-channel DM models, the inclusion of non-perturbative effects (SE and BSF) from long-range forces is mandatory when calculating the DM relic abundance.

\subsection{Potential of Bound State Searches at the LHC} \label{sec:ImplicationBSFatLHC}

\begin{figure}[p]
     \begin{subfigure}{0.48\textwidth}
    \centering
    \includegraphics[width=\textwidth]{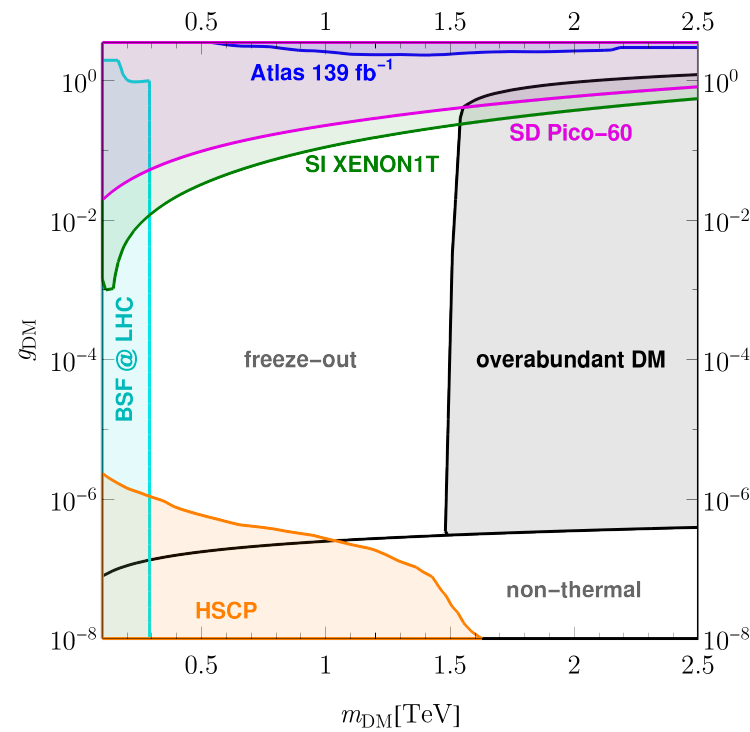}
    \caption{$\delta = 0.01$, current limits}
    \end{subfigure}
    \begin{subfigure}{0.48\textwidth}
    \centering
    \includegraphics[width=\textwidth]{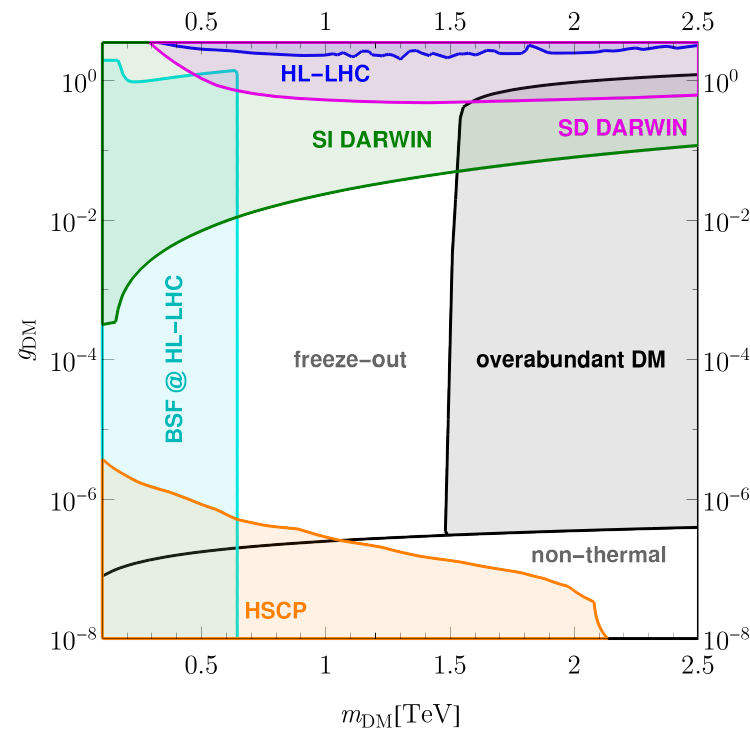}
    \caption{$\delta = 0.01$, projected limits}
    \end{subfigure}    \\ \vspace{2mm}
     \begin{subfigure}{0.48\textwidth}
    \centering
    \includegraphics[width=\textwidth]{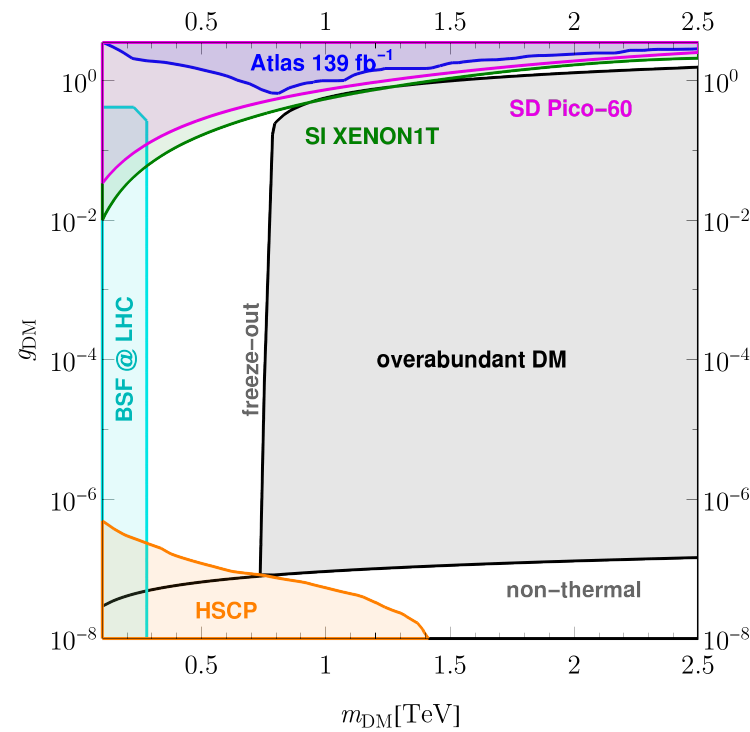}
    \caption{$\delta = 0.05$, current limits}
    \end{subfigure}
    \begin{subfigure}{0.48\textwidth}
    \centering
    \includegraphics[width=\textwidth]{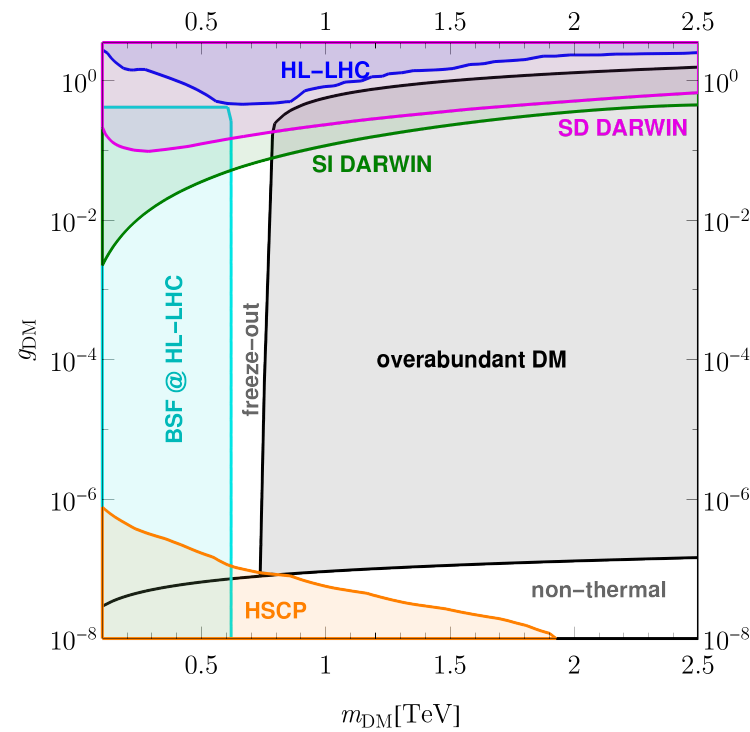}
    \caption{$\delta = 0.05$, projected limits}
    \end{subfigure}    
    \caption{Future and current limits are shown in the $(\mdm,\gdm)$-plane for a fixed relative mass splitting $\delta$. The limits from spin-independent direct detection (spin-dependent direct detection, prompt collider searches, long-lived particle searches, bound state searches at colliders) are shown in green (magenta, blue, orange, cyan). Additionally the plane is split into three regions. Firstly a region for small couplings where DM production proceeds non-thermally, secondly a freeze-out region where DM is either underproduced or matches the observed relic density, which coincides with the boundary of this region to the area of parameter space where DM is overabundant. }
    \label{fig:Summary_mvsg}
\end{figure}

In order to highlight the potential reach of searches for dark sector bound states at the LHC (cf. sections~ \ref{sec:BSLHC} and \ref{sec:Results_expconstraints}), we show in Fig.~\ref{fig:Summary_mvsg} the parameter space exclusion limits in the $(\mdm,\gdm)-$plane for fixed values of the relative mass splitting $\delta=\Delta m/\mdm=0.01$ (upper panels) and $\delta=0.05$ (lower panels).
The exclusion plots exploit the same color-scheme for the various experimental limits as in the $(\mdm,\Delta m)$-plane of Figs.~\ref{fig:Summary_uR}-\ref{fig:Summary_uR_proj}.
In contrast to the previous figures, it also features regions of parameter space where the DM abundance is overproduced, since the mass-splitting is fixed to the value $\Delta m = \delta \, \mdm$, while the coupling is varied\footnote{The parameter space presented in Fig.~\ref{fig:Summary_mvsg} corresponds to a straight line with $\Delta m = \delta \, \mdm$ in the Figs.~\ref{fig:Summary_uR}-\ref{fig:Summary_uR_proj}. In Fig.~\ref{fig:Summary_mvsg}, however, we vary the coupling $\gdm$, which in the Figs.~\ref{fig:Summary_uR}-\ref{fig:Summary_uR_proj} is fixed such that DM is not overproduced.}. 
The portion of parameter space excluded due to overabundant DM is bounded by solid black lines and colored in gray. 
The correct DM abundance in the freeze-out regime coincides with the black solid boundary between the overabundant and freeze-out region\footnote{The correct relic density can potentially be obtained in the non-thermal region. However, we did not analyze this situation in detail.}.   

Direct detection and prompt collider searches are able to constrain the DM-SM coupling $\gdm$ from above, since the production rate at the LHC and the DM-nucleon coupling increase with $\gdm$. 
Additionally, prompt collider searches require a minimal value for $\gdm$ to be efficient in order for the decay of the mediator to be prompt.
On the contrary, long-lived particle searches can constrain the coupling $\gdm$ from below but cease to be efficient when $\gdm$ becomes larger, as the decaying particle would disappear too quickly to be tagged as an LLP.

The production cross-section for a $X-X^\dagger$ bound state at a hadron collider is given by Eq.~\eqref{eq:BSFatLHCcs} and solely depends on the strong gauge coupling as long as the decay width of the constituent, regulated by $\gdm$, is smaller than the binding energy of the bound state \cite{Kats_2010}. 
For larger values of the decay width, however, the production of bound states at colliders is suppressed since it is more difficult for a bound state to form before one of the potential constituents decays and thus bound-state searches are only effective up to a certain coupling, roughly when $\gdm \lesssim g_s$. 
However, in contrast to prompt collider and long-lived particle searches, bound-state searches do not rely on a decay of the bound state mediated by $\gdm$. 
Since the bound state confines a particle $X$ and an antiparticle $X^\dagger$, it can decay via gauge-mediated annihilations of the constituents, a scenario not viable for the decay of a single particle. 
As a consequence, bound state searches can constrain a large range of couplings $\gdm$ as long as it is small enough to allow for the efficient production of bound states at colliders. 
This can be clearly seen in Fig.~\ref{fig:Summary_mvsg}, where the corresponding cyan region is the only search covering couplings in the range of $10^{-6} \lesssim \gdm \lesssim 10^{-2}$. 
Notably, the expected limit obtained via bound-state searches increases substantially at the HL-LHC, where it could potentially exclude DM production in the strongly-coannihilating region up to DM masses of $650$ GeV. 

We would like to stress that bound-state searches at colliders appear to be one of the most suitable search strategies to fully test the last remaining areas of the parameter space for colored co-annihilation scenarios featuring couplings of $10^{-6} \lesssim \gdm \lesssim 10^{-2}$.

\section{Conclusions} \label{sec:Conclusions}
We have studied the phenomenological implications of non-perturbative effects arising from long-range forces between color-charged dark sector particles, in particular the Sommerfeld effect (SE) and Bound State Formation (BSF) on the interpretation of experimental exclusion limits based on the example of a simplified $t$-channel dark matter (DM) model.
Hereby, we analyzed a class of simplified $t$-channel dark matter models featuring a Majorana fermion $\chi$, the DM candidate, coupled via a Yukawa interaction with SM quarks and a color triplet scalar $X$ with the same quantum numbers of the quark it interacts with.
In the strongly co-annihilating regime, the DM abundance is mainly set by annihilations of the colored scalars, which are affected by the SE that can enhance or diminish the effective DM annihilation cross section depending on the corresponding potential. Additionally, bound states can form in the dark sector and their decays into SM particles provide another channel for depleting dark sector particles and hence DM.
These effects are a natural consequence of the existence of color charge in such models and hence need to be taken into account.

We have shown that the size and the sign of the corrections to the DM annihilation cross section crucially depend on the coupling structure ($\gdm$ vs $g_s$) of the model (cf. Fig.~\ref{fig:BSFvspert_uR}). The SE affects any annihilation channel with two colored particles in the inital state, independently of the size of $\gdm$. However, it enhances the annihilation cross section for $\gdm \lesssim g_s$, while it tends to decrease the cross section otherwise, especially for $\gdm > g_s$. In contrast, BSF always increases the annihilation cross section since it acts as an additional annihilation channel. Being purely mediated by $g_s$, BSF becomes irrelevant as soon as $\gdm > g_s$. Consequently, non-perturbative effects increase the DM annihilation cross section for $\gdm \lesssim g_s$, while they can be irrelevant or even decrease the annihilation cross section otherwise. This clearly illustrates that a naive analysis of the non-perturbative corrections with simple flat factors is not an accurate solution, whereas a more detailed analysis of the processes and relevant model is crucial.

It is important to stress that when including SE and BSF, we predict larger dark matter masses than naively expected from tree-level calculations only. 
This is in particular interesting with respect to future indirect detection searches such as CTA. For instance, for a coupling of $\gdm = 10^{-2} \ll g_s$, we found that the consideration of SE and BSF leads to an increase of the expected DM mass $\mdm$ required in order not exceed the observed relic density. In the limit of a degenerate dark sector we predict as the maximal dark matter mass $2.4 \, \mathrm{TeV}$ instead of $1.0\, \mathrm{TeV}$ (correction of around $140\%$) for DM couplings to right-handed quarks and $2.0 \, \mathrm{TeV}$ instead of $1.0\, \mathrm{TeV}$ (correction of around $100\%$) for DM couplings to left-handed quarks, cf. Fig.~\ref{fig:3models_RD}.

Most importantly, we find that by considering SE and BSF, parameter space thought to be excluded by direct detection experiments and LHC searches remains still viable, cf. Fig.~\ref{fig:Summary_uR} and Fig.~\ref{fig:Summary_uR_zoom}. This concerns especially the strongly coannihilating region, where non-perturbative effects have the largest impact. Based on tree-level calculations only, the viable parameter space was thought to span an area around $(\mdm,\Delta m) \lesssim (1 \, \mathrm{TeV},30 \, \mathrm{GeV})$, however, when considering SE and BSF it opens up again to $(\mdm,\Delta m) \lesssim (2.4 \, \mathrm{TeV}, 50 \, \mathrm{GeV})$. Additionally, we find that DM freeze-out can lead to the observed relic abundance while not yet being experimentally excluded for mass splittings ranging roughly from $50\,\text{GeV}\lesssim\Delta m\lesssim 3\,\text{TeV}$ and DM masses ranging roughly from $200\,\text{GeV}\lesssim\mdm\lesssim 17\,\text{TeV}$ for the $u_R$ model, apart for a region around the $\Delta m\simeq m_t$, where the one-loop-induced SI DM-nucleon scattering is resonant. 

SE and BSF extend the parameter space which leads to an underabundant relic via the freeze-out mechanism, featuring very small couplings $\gdm \approx 10^{-7} - 10^{-8}$. We demonstrated that this parameter space can be probed by long-lived particle (LLP) searches, in particular heavy-stable-charged particle (HSCP) searches are able to probe $\Delta m\lesssim 40$ GeV and $\mdm\lesssim1.1$ TeV, cf. Fig.~\ref{fig:Summary_uR_zoom}.

In the strongly coannihilating region, thermal production of DM can generate the observed DM relic abundance within a narrow slice in the $(\Delta m,\mdm)$-plane with couplings varying over four orders of magnitude ($10^{-6}\lesssim\gdm\lesssim10^{-2}$). This coupling range is currently inaccessible to prompt collider searches, direct detection, as well as long-lived particle searches (LLPs).
Remarkably, we found that searches for dark-sector bound-state resonances at colliders provide a complementary probe in this region of parameter space. This is due to the fact that prompt collider searches rely on large $\gdm$ necessary for a sizeable production cross section and the successive prompt decay of the colored mediator. This is in contrast to the production and decay of the bound state which proceeds via gauge couplings. 
Hence, bound state searches at the LHC are necessary to probe the complete parameter space relevant to colored coannihilation.

Finally, we investigated the reach of future DD and collider experiments, such as DARWIN and HL-LHC, and found that almost the entire remaining parameter space will be the tested in the near future (cf. Fig.~\ref{fig:Summary_uR_proj}).
Only two small regions would still allow for thermal freeze-out as a viable solution: a corner around $\Delta m\sim2-4$ TeV and $\mdm\sim3-6$ TeV
and an even narrower slice in the strongly co-annihilating regime, featuring tiny DM couplings.
While the first one could be finally probed by colliders with higher center-of-mass energies (e.g., the FCC), the latter is potentially (and possibly uniquely) testable by dedicated bound state searches at colliders. This will be complemented by future HSCP searches at the HL-LHC, improving the limits in the underabundant/non-thermal region, c.f. Figs.~\ref{fig:Summary_uR_proj}-\ref{fig:Summary_mvsg}). 

In conclusion, in order to arrive at a final statement on the feasibility of the class of simplified $t$-channel models here analyzed to describe the observed DM relic density in the universe, non-perturbative effects arising from long-range forces must be taken into account. These are naturally present in dark sectors featuring color-charged particles and lead to a substantial impact on the theoretically predicted DM abundance. Therefore, similar consequences are expected for comparable scenarios in relevant theoretical or experimental analyses, which we would like to raise awareness for.

\section*{Acknowledgments}
We thank Alexander Pukhov for his support during the implementation in \texttt{micrOMEGAs}. M.~B., E.~C. and J.~H. acknowledge support from the DFG Emmy Noether Grant No. HA 8555/1-1. E.~C. acknowledges also support from the DFG Collaborative Research Centre “Neutrinos and Dark Matter in Astro- and Particle Physics” (SFB 1258). DS is supported by the National Science Foundation under Grant No. PHY- 1915147 and in part by the ARC Discovery Project DP180102209, the ARC Centre of Excellence for Dark Matter Particle Physics CE200100008 and the Centre for the Subatomic Structure of Matter (CSSM).
\appendix
\section{Amplitude color-decomposition for $\mathbf{3}\otimes\mathbf{3}$ initial state} \label{sec:AppendixA}
\renewcommand{\theequation}{\thesection.\arabic{equation}}
\setcounter{equation}{0}
The bound-state formation cross section can be associated with a matrix element $\mathcal{M}_\text{trans}$ as elaborated on in more detail in reference \cite{Harz:2018csl}. This matrix element can be decomposed according to its color structure~\cite{Beneke:2009rj}
\begin{align}
    \left[ \mathcal{M}_\text{trans} \right]^a_{ii'jj'}&= -i f^{abc} \left( T_1^b \right)_{i'i}  \left( T_2^c \right)_{j'j} \mathcal{M}_1 \nonumber \\ &+ \left( \eta_2  \left( T_1^a \right)_{i'i} \delta_{j'j} - \eta_1  \left( T_2^a \right)_{j'j} \delta_{i'i}  \right) \mathcal{M}_2 \nonumber \\ &= \left[ C_1 \right]^a_{ii'jj'} \mathcal{M}_1 + \left[ C_2 \right]^a_{ii'jj'} \mathcal{M}_2 \, .
\end{align}
Here, $a$ is an adjoint index associated to the emitted gluon while $i,j$ ($i',j'$) refer to the incoming (outgoing) colored particles in the fundamental representation and $\eta_{i} = \frac{m_{i}}{m_1+m_2}$.   
The strength of the potential acting on the initial/final state particles depends on the color configuration $(i,j)$/$(i',j')$ of the product representation according to Eq.~\eqref{eq:ColorPotential}. Thus, for an unambiguous initial and final state potential, the amplitude has to be decomposed into its irreducible representations. For the two relevant processes discussed in this article, we find
\begin{align}
\mathcal{M} &= P_{\left[ \mathbf{1} \right]} \mathcal{M} + P_{\left[ \mathbf{8} \right]} \mathcal{M} \, , \text{ for } 3 \times \bar{3} \, , \\ 
\mathcal{M} &= P_{\left[ \mathbf{\bar{3}} \right]} \mathcal{M} + P_{\left[ \mathbf{6} \right]} \mathcal{M} \, , \text{ for } 3 \times 3 \, ,
\end{align}
with
\begin{align}
    \left[ P_{\left[ \mathbf{1} \right]} \right]_{ijmn} &= \frac{1}{N_C} \delta_{ij} \delta_{mn} \, , \\
    \left[ P_{\left[ \mathbf{8} \right]} \right]_{ijmn} &= 2 T^\alpha_{ij} T^\alpha_{nm} \, , \\
    \left[ P_{\left[ \mathbf{\bar{3}} \right]} \right]_{ijmn} &= \frac{1}{2} \left( \delta_{im} \delta_{jn} - \delta_{in} \delta_{jm}  \right) \, , \\
    \left[ P_{\left[ \mathbf{6} \right]} \right]_{ijmn} &= \frac{1}{2} \left( \delta_{im} \delta_{jn} + \delta_{in} \delta_{jm}  \right) \, .
\end{align}
Using the fact that $\mathcal{M}_1 \approx \frac{\alpha_s^\text{NA}}{\alpha_g^\text{B}} \mathcal{M}_2 \approx \mathcal{M}_2 \equiv \mathcal{M}_0$ \cite{Harz:2018csl}, we find for the squared matrix element of the process $(X+X)_{[\bar{\mathbf{3}}]}~\rightarrow~\big\{\mathcal{B}(XX)_{[\bar{\mathbf{3}}]}+g\big\}_{[\bar{\mathbf{3}}]}$
\begin{align}
    \bigg|\left[ P_{\left[ \mathbf{\bar{3}} \right]} \right]_{ijmn} \left[ P_{\left[ \mathbf{\bar{3}} \right]} \right]_{i'j'm'n'} \left[\mathcal{M}_\text{trans} \right]^a_{i,i',j,j'} \bigg|^2 
    = \frac{1}{8} \left( N^3 - 2 N^2 - N + 2 \right) \left( \eta_1 - \eta_2 \right)^2 |\mathcal{M}_0|^2 \, .
    \label{eq:appA_3bar3bar}
\end{align}
Since we only consider bound state formed from particles equal in mass, we have $\eta_1 = \eta_2$, which in turn implies that the squared amplitude for the BSF $(X+X)_{[\bar{\mathbf{3}}]}\rightarrow\big\{\mathcal{B}(XX)_{[\bar{\mathbf{3}}]}+g\big\}_{[\bar{\mathbf{3}}]}$ vanishes. Note that this result is independent from the approximation $\mathcal{M}_1=\mathcal{M}_2$, since any contribution involving $\mathcal{M}_1$ vanishes individually. 
The remaining BSF process $(X+X)_{[\mathbf{6}]}\rightarrow\big\{\mathcal{B}(XX)_{[\bar{\mathbf{3}}]}+g\big\}_{[\mathbf{6}]}$ yields 
\begin{align}
    \bigg|\left[ P_{\left[ \mathbf{6} \right]} \right]_{ijmn} \left[ P_{\left[ \mathbf{\bar{3}} \right]} \right]_{i'j'm'n'} \left[\mathcal{M}_\text{trans} \right]^a_{i,i',j,j'} \bigg|^2 
    = \frac{1}{8} N \left(N^2 - 1 \right) \left( \eta_1 + \eta_2 - 1 \right)^2 |\mathcal{M}_0|^2 \, .
    \label{eq:appA_6anti3_Meq}
\end{align}
Since $\eta_1+\eta_2 = 1$, also this contribution to the BSF process vanishes. However, in this case, we find a non-zero result if we do not assume $\mathcal{M}_1=\mathcal{M}_2$. In fact, if we employ $\mathcal{M}_1 \approx \frac{\alpha_s^\text{NA}}{\alpha_g^\text{B}} \mathcal{M}_2$, we instead find 
\begin{align}
    \bigg|\left[ P_{\left[ \mathbf{6} \right]} \right]_{ijmn} \left[ P_{\left[ \mathbf{\bar{3}} \right]} \right]_{i'j'm'n'} \left[\mathcal{M}_\text{trans} \right]^a_{i,i',j,j'} \bigg|^2 
    = \frac{1}{8} N \left(N^2 - 1 \right) \left( \frac{\alpha_s^\text{NA}}{\alpha_g^\text{B}} - 1 \right)^2 |\mathcal{M}_0|^2 \, .
     \label{eq:appA_6anti3_Mnoteq}
\end{align}
Nevertheless, as discussed in \cite{Harz:2018csl}, $\alpha_s^\text{NA}$ and $\alpha_g^B$ typically differ only up to $\mathcal{O} \left( 10 \% \right)$, resulting in a suppression of this contribution to the BSF cross section by a factor of $\mathcal{O}\left(100\right)$. As a consequence, we do not include particle-particle bound states in our analysis.

The same conclusion can be derived by considering $\bar{\mathbf{3}}\otimes\bar{\mathbf{3}}$ initial states.

\section{Boltzmann Equations including Bound State Decays via Constituent Decays}\label{app:B}
\renewcommand{\theequation}{\thesection.\arabic{equation}}
\setcounter{equation}{0}
For a single generation of colored mediators, we need three Boltzmann equations to describe the system: one for the DM particle $\chi$, one for the mediator $X$ and one for the bound states $\mathcal{B}$. The discussion follows the arguments firstly pointed out in \cite{Ellis_2015}:
\begin{align}
    \frac{dY_\chi}{dx} &= \left( \frac{dY_\chi}{dx}\right)_\text{w/o BSF} + 2 \frac{c \sqrt{g_{*,\text{eff}}}}{s x^2} \left \langle \Gamma_{\mathcal{B} \rightarrow X \chi q} \right \rangle \left( Y_\mathcal{B} - \frac{Y_\chi}{Y_\chi^\text{eq}} \frac{Y_X}{Y_X^\text{eq}} Y_\mathcal{B}^\text{eq}  \right), \\[2pt]
    \frac{d\left(Y_X+Y_{X^\dagger} \right)}{dx} &= 2 \left( \frac{dY_X}{dx}\right)_\text{w/o BSF} + 2\frac{c \sqrt{g_{*,\text{eff}}}}{s x^2} \left \langle \Gamma_{\mathcal{B} \rightarrow X \chi q} \right \rangle \left( Y_\mathcal{B} - \frac{Y_\chi}{Y_\chi^\text{eq}} \frac{Y_X}{Y_X^\text{eq}} Y_\mathcal{B}^\text{eq}  \right) \nonumber \\[1pt] &\phantom{=} - 2 \frac{c \sqrt{g_{*,\text{eff}}}}{x^2} \left \langle \sigma_\text{BSF} v \right \rangle \left( Y_X^2 - \left( Y_X^\text{eq} \right)^2 \frac{Y_\mathcal{B}}{Y_\mathcal{B}^\text{eq}} \right), \\[2pt]
    \frac{dY_\mathcal{B}}{dx} &= - \frac{c \sqrt{g_{*,\text{eff}}}}{x^2 s} \left \langle \Gamma_{\mathcal{B}\rightarrow SM} \right \rangle \left( Y_\mathcal{B} - Y_\mathcal{B}^\text{eq} \right) - 2 \frac{c \sqrt{g_{*,\text{eff}}}}{x^2 s} \left \langle \Gamma_{\mathcal{B}\rightarrow X \chi q} \right \rangle \left( Y_\mathcal{B} - Y_\mathcal{B}^\text{eq} \frac{Y_\chi}{Y_\chi^\text{eq}} \frac{Y_X}{Y_X^\text{eq}} \right) \nonumber \\[1pt] &\phantom{=} + \frac{c \sqrt{g_{*,\text{eff}}}}{x^2} \left \langle \sigma_\text{BSF} v \right \rangle \left( Y_X^2 - \frac{Y_\mathcal{B}}{Y_\mathcal{B}^\text{eq}} \left( Y_X^\text{eq} \right)^2 \right) \, , 
\end{align}
where $c = \sqrt{\frac{\pi}{45}} m_\text{Pl} m_\chi$ and we assume the absence of any particle anti-particle asymmetries in the dark sector.
In the equations above, we include the effect of BSF. Firstly, the process of radiative bound state formation and its inverse process bound state dissociation, $X X^\dagger \leftrightarrow \mathcal{B} g $. Secondly, (inverse) decays of the bound state induced by the annihilation of the decay products into SM particles, $\mathcal{B} \rightarrow SM$. Lastly, we consider the (inverse) decay of the bound state caused by the instability of one of its constituents, $X \leftrightarrow \chi q$, leading to $\mathcal{B} \leftrightarrow X \chi q$. The last type of process has not been considered previously and we study its effects on the resulting effective Boltzmann equation in the following.
If the total interaction rate of the bound state $ \Gamma_{\mathcal{B},\text{tot}} $ satisfies $\left \langle \Gamma_{\mathcal{B},\text{tot}} \right \rangle \gtrsim 2 x H$ \cite{Binder:2021vfo}, we can assume $\frac{dY_\mathcal{B}}{dx} \approx 0$. This assumption results in
\begin{align}
    \frac{Y_\mathcal{B}}{Y_\mathcal{B}^\text{eq}} = \frac{\left \langle \Gamma_{\mathcal{B}\rightarrow SM} \right \rangle + 2 \left \langle \Gamma_{\mathcal{B} \rightarrow X \chi q} \right \rangle \frac{Y_\chi}{Y_\chi^\text{eq}} \frac{Y_X}{Y_X^\text{eq}} + \left \langle \Gamma_\text{dis} \right \rangle \left( \frac{Y_X}{Y_X^\text{eq}} \right)^2}{\left \langle \Gamma_{\mathcal{B},\text{tot}} \right \rangle} \, ,
\end{align}
where $\left \langle \Gamma_\text{dis} \right \rangle = \left \langle \sigma_\text{BSF} v \right \rangle s \frac{\left( Y_X^\text{eq}\right)^2}{Y_\mathcal{B}^\text{eq}}$ and $\left \langle \Gamma_{\mathcal{B},\text{tot}} \right \rangle = \left \langle \Gamma_{\mathcal{B}\rightarrow SM} \right \rangle + 2 \left \langle \Gamma_{\mathcal{B} \rightarrow X \chi q} \right \rangle + \left \langle \Gamma_\text{dis} \right \rangle$.
With this expression and the assumption of chemical equilibrium in the dark sector, $Y_\chi/Y_\chi^\text{eq}=Y_X/Y_X^\text{eq}$, we can reduce the system of Boltzmann equations to a single evolution equation for the dark sector number density
\begin{align}
    &\frac{d \left( Y_\chi + 2 Y_X \right)}{dx} = \left( \frac{d \left( Y_\chi + 2 Y_X \right)}{dx} \right)_\text{w/o BSF} \nonumber \\ &\phantom{=} - \frac{2 c \sqrt{g_{*,\text{eff}}}}{x^2 s} \frac{\left \langle \Gamma_{\mathcal{B}\rightarrow SM} \right \rangle}{\left \langle \Gamma_{\mathcal{B},\text{tot}} \right \rangle} \left[ \frac{Y_\mathcal{B}^\text{eq}}{\left( Y_X^\text{eq} \right)^2} \left \langle \Gamma_\text{dis} \right \rangle \left( Y_X^2 - \left( Y_X^\text{eq} \right)^2 \right) + 2 \frac{Y_\mathcal{B}^\text{eq}}{ Y_X^\text{eq}Y_\chi^\text{eq}} \left \langle \Gamma_{\mathcal{B} \rightarrow X \chi q} \right \rangle \left( Y_\chi Y_X - Y_\chi^\text{eq} Y_X^\text{eq} \right) \right] \\ &= \left( \frac{d \left( Y_\chi + 2 Y_X \right)}{dx} \right)_\text{w/o BSF} -2 \frac{c \sqrt{g_{*,\text{eff}}}}{x^2} \frac{\left \langle \Gamma_{\mathcal{B}\rightarrow SM} \right \rangle}{\left \langle \Gamma_{\mathcal{B},\text{tot}} \right \rangle} \left[ \left \langle \sigma_\text{BSF} v \right \rangle + \frac{2 \left \langle \Gamma_{\mathcal{B} \rightarrow X \chi q} \right \rangle}{s} \frac{Y_\mathcal{B}^\text{eq}}{\left( Y_X^\text{eq} \right)^2}  \right] \left( Y_X^2 - \left( Y_X^\text{eq} \right)^2 \right) \label{eq:B6} \\
    &= \left( \frac{d \left( Y_\chi + 2 Y_X \right)}{dx} \right)_\text{w/o BSF} - \frac{c \sqrt{g_{*,\text{eff}}}}{x^2} 2 \left \langle \sigma_\text{BSF} v \right \rangle_\text{eff}  \left( Y_X^2 - \left( Y_X^\text{eq} \right)^2 \right)  \, ,
    \label{eq:appB_BoltzmannComplete}
\end{align}
where we have absorbed the interaction rates in Eq.~\ref{eq:B6} into $\left \langle \sigma_\text{BSF} v \right \rangle_\text{eff}$ in Eq.~\ref{eq:appB_BoltzmannComplete}. This expression can be rewritten in the spirit of the Eqs. \ref{eq:BoltzmannEq} and \ref{eq:effective_sigmav_coannih}, yielding
\begin{align}
    \langle\sigma_{\text{eff}} v_{\text{rel}}\rangle=\sum_{ij}\langle\sigma_{ij}v_{ij}\rangle_\text{eff} \dfrac{Y_i^{\text{eq}}}{\tilde{Y}^{\text{eq}}}\dfrac{Y_j^{\text{eq}}}{\tilde{Y}^{\text{eq}}},
\end{align}
where $\left \lbrace i,j\right \rbrace \in \left \lbrace \chi, X, X^\dagger \right \rbrace$ and 
\begin{align}
    \langle\sigma_{X X^\dagger}v\rangle_\text{eff} = \langle\sigma_{X^\dagger X}v\rangle_\text{eff} =  \langle\sigma_{X X^\dagger}v_{X X^\dagger}\rangle + \left \langle \sigma_\text{BSF} v \right \rangle_\text{eff} \, ,
\end{align}
while all other channels are unaffected, i.e. $\langle\sigma_{ij}v_{ij}\rangle_\text{eff} = \langle\sigma_{ij}v_{ij}\rangle$, for $(i,j)\neq \left \lbrace (X X^\dagger),(X^\dagger X) \right \rbrace$.

We are now in the position to investigate the impact of a non-zero DM-SM coupling $\gdm$, which allows for $\Gamma_{\mathcal{B} \rightarrow X \chi q} \neq 0$. On the one hand, $\Gamma_{\mathcal{B} \rightarrow X \chi q} \neq 0$ gives an additional positive contribution to the effective annihilation cross section of DM in Eq.~\eqref{eq:appB_BoltzmannComplete}. On the other hand, $\Gamma_{\mathcal{B} \rightarrow X \chi q} \neq 0$ increases the total bound state interaction rate $\left \langle \Gamma_{\mathcal{B},\text{tot}} \right \rangle$, thereby decreasing the effective annihilation cross section of DM. In order to analyze those two competing effects, we inspect the variation of the bound state contribution to the effective annihilation cross section with $\left \langle \Gamma_{\mathcal{B} \rightarrow X \chi q} \right \rangle$
\begin{align}
\frac{\dd \left \langle \sigma_\text{BSF} v \right \rangle_\text{eff}}{\dd \, \left \langle \Gamma_{\mathcal{B} \rightarrow X \chi q} \right \rangle} &= \frac{\dd}{\dd \, \left \langle \Gamma_{\mathcal{B} \rightarrow X \chi q} \right \rangle} \left( \frac{\left \langle \Gamma_{\mathcal{B}\rightarrow SM} \right \rangle}{\left \langle \Gamma_{\mathcal{B},\text{tot}} \right \rangle} \left[ \left \langle \sigma_\text{BSF} v \right \rangle + \frac{2 \left \langle \Gamma_{\mathcal{B} \rightarrow X \chi q} \right \rangle}{s} \frac{Y_\mathcal{B}^\text{eq}}{\left( Y_X^\text{eq} \right)^2}  \right]  \right) \\  &= \frac{\left \langle \Gamma_{\mathcal{B}\rightarrow SM} \right \rangle}{\left \langle \Gamma_{\mathcal{B},\text{tot}} \right \rangle^2} \left(2 \left \langle  \Gamma_{\mathcal{B}\rightarrow SM} \right \rangle \frac{n_\mathcal{B}^\text{eq}}{\left( n_X^\text{eq} \right)^2} + \underbrace{2 \left \langle \Gamma_\text{dis} \right \rangle \frac{n_\mathcal{B}^\text{eq}}{\left( n_X^\text{eq} \right)^2} - 2 \left \langle \sigma_\text{BSF} v \right \rangle}_{=0} \right) > 0
\label{eq:appB_dsigmadgDM}
\end{align}
Thus, a larger $\left \langle \Gamma_{\mathcal{B} \rightarrow X \chi q} \right \rangle$ increases the effective annihilation cross section for all $T$, $M$ and $\Delta$. As a consequence, $\left \langle \sigma_\text{BSF} v \right \rangle_\text{eff} \left( \gdm = 0 \right) = 0$ provides the smallest possible bound state contribution to the effective annihilation cross section of DM.

\begin{figure}[!t]
    \centering
    \includegraphics[width=0.8\textwidth]{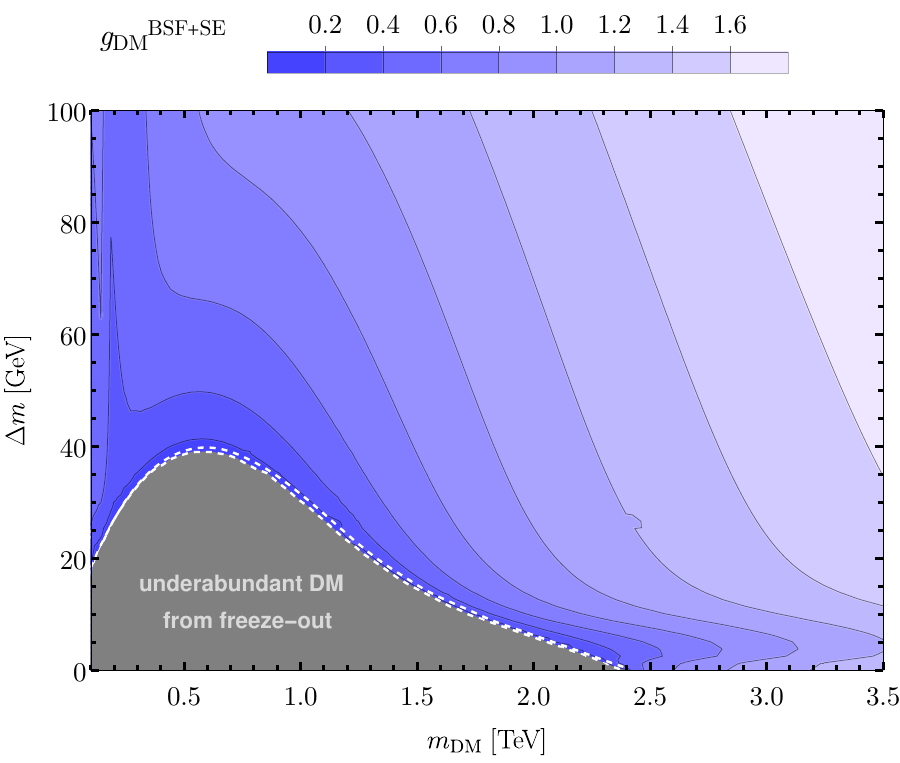}
    \caption{\small Values of the Yukawa coupling that ensures the correct $\Omega_\text{DM}$ for different mass-splitting $\Delta m = m_X-M_\text{DM}$ and DM mass combinations. 
    Here, we take into account BSF+Sommerfeld processes and, in comparison with Fig.~ \ref{fig:gDMRelic_uR}, also include effects of $\Gamma_{\mathcal{B} \rightarrow X \chi q} \neq 0$, more precisely we use $\Gamma_{\mathcal{B} \rightarrow X \chi q} \simeq 2 \Gamma_{X \rightarrow \chi q}$.
    The dashed white lines are in correspondence of the region where $10^{-7}\lesssim g_\text{DM}\lesssim 10^{-2}$. Below this interval, it is not possible to ensure chemical equilibrium between the unbound particles in the dark sector and the co-annihilation freeze-out assumption breaks down.}
    \label{fig:gDMRelic_uR_app}
\end{figure}

In this sense, the results presented in this paper, assuming a negligible three-body decay width of the bound state, provide the most conservative estimate of the effects of bound states in the context of the DM relic density. 
Furthermore, we verified that even a sizable three-body decay width does not alter our results presented in section~\ref{sec:Results}. 
In order to do so, we approximate the three-body decay width by the sum of the decay width of the constituent particles $\Gamma_{\mathcal{B} \rightarrow X \chi q} \simeq 2 \Gamma_{X \rightarrow \chi q}$. We present the results for the coupling $\gdm$ that leads to the observed relic density in Fig.~\ref{fig:gDMRelic_uR_app}.

In comparison with Fig.~\ref{fig:gDMRelic_uR}, which presents the same results assuming $\Gamma_{B \rightarrow X \chi q}=0$, we only find sizable effects for small mass splittings $\Delta m \lesssim 30\,\mathrm{GeV}$ and DM masses above a few TeV. These regions are already clearly excluded by direct detection experiments as can be seen from Figs.~\ref{fig:Summary_uR}, \ref{fig:Summary_uR_zoom} and \ref{fig:Summary_uR_proj}. 
\FloatBarrier
\bibliographystyle{JHEP}
\bibliography{biblio}

\providecommand{\href}[2]{#2}\begingroup\raggedright\begin{thebibliography}{10}

\bibitem{Planck2018}
{\scshape Planck} collaboration, \emph{{Planck 2018 results. VI. Cosmological
  parameters}},
  \href{https://doi.org/10.1051/0004-6361/201833910}{\emph{Astron. Astrophys.}
  {\bfseries 641} (2020) A6}
  [\href{https://arxiv.org/abs/1807.06209}{{\ttfamily 1807.06209}}].

\bibitem{Arcadi:2017kky}
G.~Arcadi, M.~Dutra, P.~Ghosh, M.~Lindner, Y.~Mambrini, M.~Pierre et~al.,
  \emph{{The waning of the WIMP? A review of models, searches, and
  constraints}},
  \href{https://doi.org/10.1140/epjc/s10052-018-5662-y}{\emph{Eur. Phys. J. C}
  {\bfseries 78} (2018) 203}
  [\href{https://arxiv.org/abs/1703.07364}{{\ttfamily 1703.07364}}].

\bibitem{Abulaiti:2799299}
{\scshape ATLAS Collaboration} collaboration, \emph{{Status of searches for
  dark matter at the LHC}},  tech. rep., CERN, Geneva, Jan, 2022.

\bibitem{XENON:2017vdw}
{\scshape XENON} collaboration, \emph{{First Dark Matter Search Results from
  the XENON1T Experiment}},
  \href{https://doi.org/10.1103/PhysRevLett.119.181301}{\emph{Phys. Rev. Lett.}
  {\bfseries 119} (2017) 181301}
  [\href{https://arxiv.org/abs/1705.06655}{{\ttfamily 1705.06655}}].

\bibitem{PICO:2017tgi}
{\scshape PICO} collaboration, \emph{{Dark Matter Search Results from the
  PICO-60 C$_3$F$_8$ Bubble Chamber}},
  \href{https://doi.org/10.1103/PhysRevLett.118.251301}{\emph{Phys. Rev. Lett.}
  {\bfseries 118} (2017) 251301}
  [\href{https://arxiv.org/abs/1702.07666}{{\ttfamily 1702.07666}}].

\bibitem{HESS:2013rld}
{\scshape H.E.S.S.} collaboration, \emph{{Search for Photon-Linelike Signatures
  from Dark Matter Annihilations with H.E.S.S.}},
  \href{https://doi.org/10.1103/PhysRevLett.110.041301}{\emph{Phys. Rev. Lett.}
  {\bfseries 110} (2013) 041301}
  [\href{https://arxiv.org/abs/1301.1173}{{\ttfamily 1301.1173}}].

\bibitem{Harz:2012fz}
J.~Harz, B.~Herrmann, M.~Klasen, K.~Kovarik and Q.~L. Boulc'h,
  \emph{{Neutralino-stop coannihilation into electroweak gauge and Higgs bosons
  at one loop}}, \href{https://doi.org/10.1103/PhysRevD.87.054031}{\emph{Phys.
  Rev. D} {\bfseries 87} (2013) 054031}
  [\href{https://arxiv.org/abs/1212.5241}{{\ttfamily 1212.5241}}].

\bibitem{Ellis:2014ipa}
J.~Ellis, K.~A. Olive and J.~Zheng, \emph{{The Extent of the Stop
  Coannihilation Strip}},
  \href{https://doi.org/10.1140/epjc/s10052-014-2947-7}{\emph{Eur. Phys. J. C}
  {\bfseries 74} (2014) 2947}
  [\href{https://arxiv.org/abs/1404.5571}{{\ttfamily 1404.5571}}].

\bibitem{Harz:2014gaa}
J.~Harz, B.~Herrmann, M.~Klasen, K.~Kova\v{r}\'\i{}k and M.~Meinecke,
  \emph{{SUSY-QCD corrections to stop annihilation into electroweak final
  states including Coulomb enhancement effects}},
  \href{https://doi.org/10.1103/PhysRevD.91.034012}{\emph{Phys. Rev. D}
  {\bfseries 91} (2015) 034012}
  [\href{https://arxiv.org/abs/1410.8063}{{\ttfamily 1410.8063}}].

\bibitem{Ibarra:2015nca}
A.~Ibarra, A.~Pierce, N.~R. Shah and S.~Vogl, \emph{{Anatomy of Coannihilation
  with a Scalar Top Partner}},
  \href{https://doi.org/10.1103/PhysRevD.91.095018}{\emph{Phys. Rev. D}
  {\bfseries 91} (2015) 095018}
  [\href{https://arxiv.org/abs/1501.03164}{{\ttfamily 1501.03164}}].

\bibitem{Baker:2015qna}
M.~J. Baker et~al., \emph{{The Coannihilation Codex}},
  \href{https://doi.org/10.1007/JHEP12(2015)120}{\emph{JHEP} {\bfseries 12}
  (2015) 120} [\href{https://arxiv.org/abs/1510.03434}{{\ttfamily
  1510.03434}}].

\bibitem{Harz:2016dql}
J.~Harz, B.~Herrmann, M.~Klasen, K.~Kovarik and P.~Steppeler,
  \emph{{Theoretical uncertainty of the supersymmetric dark matter relic
  density from scheme and scale variations}},
  \href{https://doi.org/10.1103/PhysRevD.93.114023}{\emph{Phys. Rev. D}
  {\bfseries 93} (2016) 114023}
  [\href{https://arxiv.org/abs/1602.08103}{{\ttfamily 1602.08103}}].

\bibitem{ElHedri:2018atj}
S.~El~Hedri and M.~de~Vries, \emph{{Cornering Colored Coannihilation}},
  \href{https://doi.org/10.1007/JHEP10(2018)102}{\emph{JHEP} {\bfseries 10}
  (2018) 102} [\href{https://arxiv.org/abs/1806.03325}{{\ttfamily
  1806.03325}}].

\bibitem{Schmiemann:2019czm}
S.~Schmiemann, J.~Harz, B.~Herrmann, M.~Klasen and K.~Kova\v{r}\'\i{}k,
  \emph{{Squark-pair annihilation into quarks at next-to-leading order}},
  \href{https://doi.org/10.1103/PhysRevD.99.095015}{\emph{Phys. Rev. D}
  {\bfseries 99} (2019) 095015}
  [\href{https://arxiv.org/abs/1903.10998}{{\ttfamily 1903.10998}}].

\bibitem{Branahl:2019yot}
J.~Branahl, J.~Harz, B.~Herrmann, M.~Klasen, K.~Kova\v{r}\'\i{}k and
  S.~Schmiemann, \emph{{SUSY-QCD corrected and Sommerfeld enhanced stau
  annihilation into heavy quarks with scheme and scale uncertainties}},
  \href{https://doi.org/10.1103/PhysRevD.100.115003}{\emph{Phys. Rev. D}
  {\bfseries 100} (2019) 115003}
  [\href{https://arxiv.org/abs/1909.09527}{{\ttfamily 1909.09527}}].

\bibitem{Arina:2020tuw}
C.~Arina, B.~Fuks, L.~Mantani, H.~Mies, L.~Panizzi and J.~Salko, \emph{{Closing
  in on $t$-channel simplified dark matter models}},
  \href{https://doi.org/10.1016/j.physletb.2020.136038}{\emph{Phys. Lett. B}
  {\bfseries 813} (2021) 136038}
  [\href{https://arxiv.org/abs/2010.07559}{{\ttfamily 2010.07559}}].

\bibitem{Arina:2020udz}
C.~Arina, B.~Fuks and L.~Mantani, \emph{{A universal framework for t-channel
  dark matter models}},
  \href{https://doi.org/10.1140/epjc/s10052-020-7933-7}{\emph{Eur. Phys. J. C}
  {\bfseries 80} (2020) 409}
  [\href{https://arxiv.org/abs/2001.05024}{{\ttfamily 2001.05024}}].

\bibitem{Abdallah:2015ter}
J.~Abdallah et~al., \emph{{Simplified Models for Dark Matter Searches at the
  LHC}}, \href{https://doi.org/10.1016/j.dark.2015.08.001}{\emph{Phys. Dark
  Univ.} {\bfseries 9-10} (2015) 8}
  [\href{https://arxiv.org/abs/1506.03116}{{\ttfamily 1506.03116}}].

\bibitem{Sommerfeld:1931qaf}
A.~Sommerfeld, \emph{{\"Uber die Beugung und Bremsung der Elektronen}},
  \href{https://doi.org/10.1002/andp.19314030302}{\emph{Annalen Phys.}
  {\bfseries 403} (1931) 257}.

\bibitem{Sakharov:1948plh}
A.~D. Sakharov, \emph{{Interaction of an Electron and Positron in Pair
  Production}},
  \href{https://doi.org/10.1070/PU1991v034n05ABEH002492}{\emph{Zh. Eksp. Teor.
  Fiz.} {\bfseries 18} (1948) 631}.

\bibitem{Hisano:2002fk}
J.~Hisano, S.~Matsumoto and M.~M. Nojiri, \emph{{Unitarity and higher order
  corrections in neutralino dark matter annihilation into two photons}},
  \href{https://doi.org/10.1103/PhysRevD.67.075014}{\emph{Phys. Rev. D}
  {\bfseries 67} (2003) 075014}
  [\href{https://arxiv.org/abs/hep-ph/0212022}{{\ttfamily hep-ph/0212022}}].

\bibitem{Drees:2009gt}
M.~Drees, J.~M. Kim and K.~I. Nagao, \emph{{Potentially Large One-loop
  Corrections to WIMP Annihilation}},
  \href{https://doi.org/10.1103/PhysRevD.81.105004}{\emph{Phys. Rev. D}
  {\bfseries 81} (2010) 105004}
  [\href{https://arxiv.org/abs/0911.3795}{{\ttfamily 0911.3795}}].

\bibitem{Beneke:2014hja}
M.~Beneke, C.~Hellmann and P.~Ruiz-Femenia, \emph{{Heavy neutralino relic
  abundance with Sommerfeld enhancements - a study of pMSSM scenarios}},
  \href{https://doi.org/10.1007/JHEP03(2015)162}{\emph{JHEP} {\bfseries 03}
  (2015) 162} [\href{https://arxiv.org/abs/1411.6930}{{\ttfamily 1411.6930}}].

\bibitem{Beneke:2019qaa}
M.~Beneke, R.~Szafron and K.~Urban, \emph{{Wino potential and Sommerfeld effect
  at NLO}}, \href{https://doi.org/10.1016/j.physletb.2019.135112}{\emph{Phys.
  Lett. B} {\bfseries 800} (2020) 135112}
  [\href{https://arxiv.org/abs/1909.04584}{{\ttfamily 1909.04584}}].

\bibitem{Hisano:2006nn}
J.~Hisano, S.~Matsumoto, M.~Nagai, O.~Saito and M.~Senami,
  \emph{{Non-perturbative effect on thermal relic abundance of dark matter}},
  \href{https://doi.org/10.1016/j.physletb.2007.01.012}{\emph{Phys. Lett. B}
  {\bfseries 646} (2007) 34}
  [\href{https://arxiv.org/abs/hep-ph/0610249}{{\ttfamily hep-ph/0610249}}].

\bibitem{vonHarling:2014kha}
B.~von Harling and K.~Petraki, \emph{{Bound-state formation for thermal relic
  dark matter and unitarity}},
  \href{https://doi.org/10.1088/1475-7516/2014/12/033}{\emph{JCAP} {\bfseries
  12} (2014) 033} [\href{https://arxiv.org/abs/1407.7874}{{\ttfamily
  1407.7874}}].

\bibitem{Ellis_2015}
J.~Ellis, F.~Luo and K.~A. Olive, \emph{Gluino coannihilation revisited},
  \href{https://doi.org/10.1007/jhep09(2015)127}{\emph{JHEP} {\bfseries 2015}
  (2015) }.

\bibitem{An:2016gad}
H.~An, M.~B. Wise and Y.~Zhang, \emph{{Effects of Bound States on Dark Matter
  Annihilation}}, \href{https://doi.org/10.1103/PhysRevD.93.115020}{\emph{Phys.
  Rev. D} {\bfseries 93} (2016) 115020}
  [\href{https://arxiv.org/abs/1604.01776}{{\ttfamily 1604.01776}}].

\bibitem{Asadi:2016ybp}
P.~Asadi, M.~Baumgart, P.~J. Fitzpatrick, E.~Krupczak and T.~R. Slatyer,
  \emph{{Capture and Decay of Electroweak WIMPonium}},
  \href{https://doi.org/10.1088/1475-7516/2017/02/005}{\emph{JCAP} {\bfseries
  02} (2017) 005} [\href{https://arxiv.org/abs/1610.07617}{{\ttfamily
  1610.07617}}].

\bibitem{Petraki:2016cnz}
K.~Petraki, M.~Postma and J.~de~Vries, \emph{{Radiative bound-state-formation
  cross-sections for dark matter interacting via a Yukawa potential}},
  \href{https://doi.org/10.1007/JHEP04(2017)077}{\emph{JHEP} {\bfseries 04}
  (2017) 077} [\href{https://arxiv.org/abs/1611.01394}{{\ttfamily
  1611.01394}}].

\bibitem{Liew:2016hqo}
S.~P. Liew and F.~Luo, \emph{{Effects of QCD bound states on dark matter relic
  abundance}}, \href{https://doi.org/10.1007/JHEP02(2017)091}{\emph{JHEP}
  {\bfseries 02} (2017) 091}
  [\href{https://arxiv.org/abs/1611.08133}{{\ttfamily 1611.08133}}].

\bibitem{Binder:2020efn}
T.~Binder, B.~Blobel, J.~Harz and K.~Mukaida, \emph{{Dark matter bound-state
  formation at higher order: a non-equilibrium quantum field theory approach}},
  \href{https://doi.org/10.1007/JHEP09(2020)086}{\emph{JHEP} {\bfseries 09}
  (2020) 086} [\href{https://arxiv.org/abs/2002.07145}{{\ttfamily
  2002.07145}}].

\bibitem{Petraki_2015}
K.~Petraki, M.~Postma and M.~Wiechers, \emph{Dark-matter bound states from
  feynman diagrams}, \href{https://doi.org/10.1007/jhep06(2015)128}{\emph{JHEP}
  {\bfseries 2015} (2015) }.

\bibitem{Harz:2018csl}
J.~Harz and K.~Petraki, \emph{{Radiative bound-state formation in unbroken
  perturbative non-Abelian theories and implications for dark matter}},
  \href{https://doi.org/10.1007/JHEP07(2018)096}{\emph{JHEP} {\bfseries 07}
  (2018) 096} [\href{https://arxiv.org/abs/1805.01200}{{\ttfamily
  1805.01200}}].

\bibitem{toolbox}
S.~El~Hedri, A.~Kaminska and M.~de~Vries, \emph{A sommerfeld toolbox for
  colored dark sectors},
  \href{https://doi.org/10.1140/epjc/s10052-017-5168-z}{\emph{Eur. Phys. J. C}
  {\bfseries 77} (2017) }.

\bibitem{ElHedri:2017nny}
S.~El~Hedri, A.~Kaminska, M.~de~Vries and J.~Zurita, \emph{{Simplified
  Phenomenology for Colored Dark Sectors}},
  \href{https://doi.org/10.1007/JHEP04(2017)118}{\emph{JHEP} {\bfseries 04}
  (2017) 118} [\href{https://arxiv.org/abs/1703.00452}{{\ttfamily
  1703.00452}}].

\bibitem{Mohan_2019}
K.~A. Mohan, D.~Sengupta, T.~M.~P. Tait, B.~Yan and C.-P. Yuan, \emph{Direct
  detection and lhc constraints on a t-channel simplified model of majorana
  dark matter at one loop},
  \href{https://doi.org/10.1007/jhep05(2019)115}{\emph{JHEP} {\bfseries 2019}
  (2019) }.

\bibitem{garny2021bound}
M.~Garny and J.~Heisig, \emph{{Bound state effects on dark matter
  coannihilation: pushing the boundaries of conversion-driven freeze-out}},
  \href{https://arxiv.org/abs/2112.01499}{{\ttfamily 2112.01499}}.

\bibitem{Bollig:2021psb}
J.~Bollig and S.~Vogl, \emph{{Impact of bound states on non-thermal dark matter
  production}},  \href{https://arxiv.org/abs/2112.01491}{{\ttfamily
  2112.01491}}.

\bibitem{DiFranzo_2013}
A.~DiFranzo, K.~I. Nagao, A.~Rajaraman and T.~M. Tait, \emph{Simplified models
  for dark matter interacting with quarks},
  \href{https://doi.org/10.1007/jhep11(2013)014}{\emph{JHEP} {\bfseries 2013}
  (2013) }.

\bibitem{Liu:2021crr}
Y.~Liu, B.~Yan and R.~Zhang, \emph{{Loop induced top quark FCNC through top
  quark and dark matter interactions}},
  \href{https://doi.org/10.1016/j.physletb.2022.136964}{\emph{Phys. Lett. B}
  {\bfseries 827} (2022) 136964}
  [\href{https://arxiv.org/abs/2103.07859}{{\ttfamily 2103.07859}}].

\bibitem{Harz_2018_HiggsEnh}
J.~Harz and K.~Petraki, \emph{Higgs enhancement for the dark matter relic
  density}, \href{https://doi.org/10.1103/physrevd.97.075041}{\emph{Phys. Rev.
  D} {\bfseries 97} (2018) }.

\bibitem{Harz_2019}
J.~Harz and K.~Petraki, \emph{Higgs-mediated bound states in dark-matter
  models}, \href{https://doi.org/10.1007/jhep04(2019)130}{\emph{JHEP}
  {\bfseries 2019} (2019) }.

\bibitem{Biondini:2018xor}
S.~Biondini, \emph{{Bound-state effects for dark matter with Higgs-like
  mediators}}, \href{https://doi.org/10.1007/JHEP06(2018)104}{\emph{JHEP}
  {\bfseries 06} (2018) 104}
  [\href{https://arxiv.org/abs/1805.00353}{{\ttfamily 1805.00353}}].

\bibitem{Becker_?}
M.~Becker, E.~Copello and J.~Harz : In preparation.

\bibitem{Edsj__1997}
J.~Edsjö and P.~Gondolo, \emph{Neutralino relic density including
  coannihilations}, \href{https://doi.org/10.1103/physrevd.56.1879}{\emph{Phys.
  Rev. D} {\bfseries 56} (1997) 1879–1894}.

\bibitem{Garny_2017}
M.~Garny, J.~Heisig, B.~Lülf and S.~Vogl, \emph{Coannihilation without
  chemical equilibrium},
  \href{https://doi.org/10.1103/physrevd.96.103521}{\emph{Phys. Rev. D}
  {\bfseries 96} (2017) }.

\bibitem{Garny_2018}
M.~Garny, J.~Heisig, M.~Hufnagel and B.~Lülf, \emph{Top-philic dark matter
  within and beyond the wimp paradigm},
  \href{https://doi.org/10.1103/physrevd.97.075002}{\emph{Phys. Rev. D}
  {\bfseries 97} (2018) }.

\bibitem{Dagnolo_2017}
R.~T. D’Agnolo, D.~Pappadopulo and J.~T. Ruderman, \emph{Fourth exception in
  the calculation of relic abundances},
  \href{https://doi.org/10.1103/physrevlett.119.061102}{\emph{Phys. Rev. Lett.}
  {\bfseries 119} (2017) }.

\bibitem{Just2008}
O.~Just, \emph{The partial wave formalism and its application to neutralino
  dark matter},  Master's thesis, 2008.

\bibitem{Zyla:2020zbs}
{\scshape Particle Data Group} collaboration, \emph{{Review of Particle
  Physics}}, \href{https://doi.org/10.1093/ptep/ptaa104}{\emph{PTEP} {\bfseries
  2020} (2020) 083C01}.

\bibitem{Belanger:2006is}
G.~Belanger, F.~Boudjema, A.~Pukhov and A.~Semenov, \emph{{MicrOMEGAs 2.0: A
  Program to calculate the relic density of dark matter in a generic model}},
  \href{https://doi.org/10.1016/j.cpc.2006.11.008}{\emph{Comput. Phys. Commun.}
  {\bfseries 176} (2007) 367}
  [\href{https://arxiv.org/abs/hep-ph/0607059}{{\ttfamily hep-ph/0607059}}].

\bibitem{Beneke:2009rj}
M.~Beneke, P.~Falgari and C.~Schwinn, \emph{{Soft radiation in heavy-particle
  pair production: All-order colour structure and two-loop anomalous
  dimension}},
  \href{https://doi.org/10.1016/j.nuclphysb.2009.11.004}{\emph{Nucl. Phys. B}
  {\bfseries 828} (2010) 69} [\href{https://arxiv.org/abs/0907.1443}{{\ttfamily
  0907.1443}}].

\bibitem{Oncala:2021tkz}
R.~Oncala and K.~Petraki, \emph{{Bound states of WIMP dark matter in
  Higgs-portal models. Part I. Cross-sections and transition rates}},
  \href{https://doi.org/10.1007/JHEP06(2021)124}{\emph{JHEP} {\bfseries 06}
  (2021) 124} [\href{https://arxiv.org/abs/2101.08666}{{\ttfamily
  2101.08666}}].

\bibitem{Cassel_2010}
S.~Cassel, \emph{Sommerfeld factor for arbitrary partial wave processes},
  \href{https://doi.org/10.1088/0954-3899/37/10/105009}{\emph{J. Phys. G}
  {\bfseries 37} (2010) 105009}.

\bibitem{Iengo2009}
R.~Iengo, \emph{Sommerfeld enhancement: general results from field theory
  diagrams}, \href{https://doi.org/10.1088/1126-6708/2009/05/024}{\emph{JHEP}
  {\bfseries 2009} (2009) 024–024}.

\bibitem{Biondini:2018pwp}
S.~Biondini and M.~Laine, \emph{{Thermal dark matter co-annihilating with a
  strongly interacting scalar}},
  \href{https://doi.org/10.1007/JHEP04(2018)072}{\emph{JHEP} {\bfseries 04}
  (2018) 072} [\href{https://arxiv.org/abs/1801.05821}{{\ttfamily
  1801.05821}}].

\bibitem{Biondini:2018ovz}
S.~Biondini and S.~Vogl, \emph{{Coloured coannihilations: Dark matter
  phenomenology meets non-relativistic EFTs}},
  \href{https://doi.org/10.1007/JHEP02(2019)016}{\emph{JHEP} {\bfseries 02}
  (2019) 016} [\href{https://arxiv.org/abs/1811.02581}{{\ttfamily
  1811.02581}}].

\bibitem{Biondini:2019int}
S.~Biondini and S.~Vogl, \emph{{Scalar dark matter coannihilating with a
  coloured fermion}},
  \href{https://doi.org/10.1007/JHEP11(2019)147}{\emph{JHEP} {\bfseries 11}
  (2019) 147} [\href{https://arxiv.org/abs/1907.05766}{{\ttfamily
  1907.05766}}].

\bibitem{Binder:2018znk}
T.~Binder, L.~Covi and K.~Mukaida, \emph{{Dark Matter Sommerfeld-enhanced
  annihilation and Bound-state decay at finite temperature}},
  \href{https://doi.org/10.1103/PhysRevD.98.115023}{\emph{Phys. Rev. D}
  {\bfseries 98} (2018) 115023}
  [\href{https://arxiv.org/abs/1808.06472}{{\ttfamily 1808.06472}}].

\bibitem{Binder:2019erp}
T.~Binder, K.~Mukaida and K.~Petraki, \emph{{Rapid bound-state formation of
  Dark Matter in the Early Universe}},
  \href{https://doi.org/10.1103/PhysRevLett.124.161102}{\emph{Phys. Rev. Lett.}
  {\bfseries 124} (2020) 161102}
  [\href{https://arxiv.org/abs/1910.11288}{{\ttfamily 1910.11288}}].

\bibitem{Binder:2021otw}
T.~Binder, K.~Mukaida, B.~Scheihing-Hitschfeld and X.~Yao, \emph{{Non-Abelian
  electric field correlator at NLO for dark matter relic abundance and
  quarkonium transport}},
  \href{https://doi.org/10.1007/JHEP01(2022)137}{\emph{JHEP} {\bfseries 01}
  (2022) 137} [\href{https://arxiv.org/abs/2107.03945}{{\ttfamily
  2107.03945}}].

\bibitem{Binder:2021vfo}
T.~Binder, A.~Filimonova, K.~Petraki and G.~White, \emph{{Saha equilibrium for
  metastable bound states and dark matter freeze-out}},
  \href{https://arxiv.org/abs/2112.00042}{{\ttfamily 2112.00042}}.

\bibitem{Belyaev_2013}
A.~Belyaev, N.~D. Christensen and A.~Pukhov, \emph{Calchep 3.4 for collider
  physics within and beyond the standard model},
  \href{https://doi.org/10.1016/j.cpc.2013.01.014}{\emph{Comput. Phys. Commun.}
  {\bfseries 184} (2013) 1729–1769}.

\bibitem{Garny:2018ali}
M.~Garny and J.~Heisig, \emph{{Interplay of super-WIMP and freeze-in production
  of dark matter}},
  \href{https://doi.org/10.1103/PhysRevD.98.095031}{\emph{Phys. Rev. D}
  {\bfseries 98} (2018) 095031}
  [\href{https://arxiv.org/abs/1809.10135}{{\ttfamily 1809.10135}}].

\bibitem{Belanger:2021smw}
G.~B\'elanger et~al., \emph{{Leptoquark Manoeuvres in the Dark: a simultaneous
  solution of the dark matter problem and the $R_D$ anomalies}},
  \href{https://arxiv.org/abs/2111.08027}{{\ttfamily 2111.08027}}.

\bibitem{DARWIN:2016hyl}
{\scshape DARWIN} collaboration, \emph{{DARWIN: towards the ultimate dark
  matter detector}},
  \href{https://doi.org/10.1088/1475-7516/2016/11/017}{\emph{JCAP} {\bfseries
  11} (2016) 017} [\href{https://arxiv.org/abs/1606.07001}{{\ttfamily
  1606.07001}}].

\bibitem{Alwall:2014hca}
J.~Alwall, R.~Frederix, S.~Frixione, V.~Hirschi, F.~Maltoni, O.~Mattelaer
  et~al., \emph{{The automated computation of tree-level and next-to-leading
  order differential cross sections, and their matching to parton shower
  simulations}}, \href{https://doi.org/10.1007/JHEP07(2014)079}{\emph{JHEP}
  {\bfseries 07} (2014) 079} [\href{https://arxiv.org/abs/1405.0301}{{\ttfamily
  1405.0301}}].

\bibitem{Dulat:2015mca}
S.~Dulat, T.-J. Hou, J.~Gao, M.~Guzzi, J.~Huston, P.~Nadolsky et~al.,
  \emph{{New parton distribution functions from a global analysis of quantum
  chromodynamics}},
  \href{https://doi.org/10.1103/PhysRevD.93.033006}{\emph{Phys. Rev. D}
  {\bfseries 93} (2016) 033006}
  [\href{https://arxiv.org/abs/1506.07443}{{\ttfamily 1506.07443}}].

\bibitem{ATLAS:2020syg}
{\scshape ATLAS} collaboration, \emph{{Search for squarks and gluinos in final
  states with jets and missing transverse momentum using 139 fb$^{-1}$ of
  $\sqrt{s}$ =13 TeV $pp$ collision data with the ATLAS detector}},
  \href{https://doi.org/10.1007/JHEP02(2021)143}{\emph{JHEP} {\bfseries 02}
  (2021) 143} [\href{https://arxiv.org/abs/2010.14293}{{\ttfamily
  2010.14293}}].

\bibitem{Dumont:2014tja}
B.~Dumont, B.~Fuks, S.~Kraml, S.~Bein, G.~Chalons, E.~Conte et~al.,
  \emph{{Toward a public analysis database for LHC new physics searches using
  MADANALYSIS 5}},
  \href{https://doi.org/10.1140/epjc/s10052-014-3242-3}{\emph{Eur. Phys. J. C}
  {\bfseries 75} (2015) 56} [\href{https://arxiv.org/abs/1407.3278}{{\ttfamily
  1407.3278}}].

\bibitem{ATLAS:2021kxv}
{\scshape ATLAS} collaboration, \emph{{Search for new phenomena in events with
  an energetic jet and missing transverse momentum in $pp$ collisions at $\sqrt
  {s}$ =13 TeV with the ATLAS detector}},
  \href{https://doi.org/10.1103/PhysRevD.103.112006}{\emph{Phys. Rev. D}
  {\bfseries 103} (2021) 112006}
  [\href{https://arxiv.org/abs/2102.10874}{{\ttfamily 2102.10874}}].

\bibitem{ma5url}
\href{https://madanalysis.irmp.ucl.ac.be/wiki/PublicAnalysisDatabase}{https://madanalysis.irmp.ucl.ac.be/wiki/PublicAnalysisDatabase}.

\bibitem{Cacciari:2011ma}
M.~Cacciari, G.~P. Salam and G.~Soyez, \emph{{FastJet User Manual}},
  \href{https://doi.org/10.1140/epjc/s10052-012-1896-2}{\emph{Eur. Phys. J. C}
  {\bfseries 72} (2012) 1896}
  [\href{https://arxiv.org/abs/1111.6097}{{\ttfamily 1111.6097}}].

\bibitem{Dokshitzer:1997in}
Y.~L. Dokshitzer, G.~D. Leder, S.~Moretti and B.~R. Webber, \emph{{Better jet
  clustering algorithms}},
  \href{https://doi.org/10.1088/1126-6708/1997/08/001}{\emph{JHEP} {\bfseries
  08} (1997) 001} [\href{https://arxiv.org/abs/hep-ph/9707323}{{\ttfamily
  hep-ph/9707323}}].

\bibitem{deFavereau:2013fsa}
{\scshape DELPHES 3} collaboration, \emph{{DELPHES 3, A modular framework for
  fast simulation of a generic collider experiment}},
  \href{https://doi.org/10.1007/JHEP02(2014)057}{\emph{JHEP} {\bfseries 02}
  (2014) 057} [\href{https://arxiv.org/abs/1307.6346}{{\ttfamily 1307.6346}}].

\bibitem{Sjostrand:2014zea}
T.~Sj\"ostrand, S.~Ask, J.~R. Christiansen, R.~Corke, N.~Desai, P.~Ilten
  et~al., \emph{{An introduction to PYTHIA 8.2}},
  \href{https://doi.org/10.1016/j.cpc.2015.01.024}{\emph{Comput. Phys. Commun.}
  {\bfseries 191} (2015) 159}
  [\href{https://arxiv.org/abs/1410.3012}{{\ttfamily 1410.3012}}].

\bibitem{Hoeche:2005vzu}
S.~Hoeche, F.~Krauss, N.~Lavesson, L.~Lonnblad, M.~Mangano, A.~Schalicke
  et~al., \emph{{Matching parton showers and matrix elements}}, .

\bibitem{Read_2002}
A.~L. Read, \emph{Presentation of search results: theclstechnique},
  \href{https://doi.org/10.1088/0954-3899/28/10/313}{\emph{J. Phys. G}
  {\bfseries 28} (2002) 2693}.

\bibitem{ZurbanoFernandez:2020cco}
I.~Zurbano~Fernandez et~al., \emph{{High-Luminosity Large Hadron Collider
  (HL-LHC): Technical design report}}, .

\bibitem{ATLAS:2017ayi}
{\scshape ATLAS} collaboration, \emph{{Search for new phenomena in high-mass
  diphoton final states using 37 fb$^{-1}$ of proton--proton collisions
  collected at $\sqrt{s}=13$ TeV with the ATLAS detector}},
  \href{https://doi.org/10.1016/j.physletb.2017.10.039}{\emph{Phys. Lett. B}
  {\bfseries 775} (2017) 105}
  [\href{https://arxiv.org/abs/1707.04147}{{\ttfamily 1707.04147}}].

\bibitem{ATLAS:2016gzy}
{\scshape ATLAS} collaboration, \emph{{Search for resonances in diphoton events
  at $\sqrt{s}$=13 TeV with the ATLAS detector}},
  \href{https://doi.org/10.1007/JHEP09(2016)001}{\emph{JHEP} {\bfseries 09}
  (2016) 001} [\href{https://arxiv.org/abs/1606.03833}{{\ttfamily
  1606.03833}}].

\bibitem{Martin:2008sv}
S.~P. Martin, \emph{{Diphoton decays of stoponium at the Large Hadron
  Collider}}, \href{https://doi.org/10.1103/PhysRevD.77.075002}{\emph{Phys.
  Rev. D} {\bfseries 77} (2008) 075002}
  [\href{https://arxiv.org/abs/0801.0237}{{\ttfamily 0801.0237}}].

\bibitem{Batell:2015zla}
B.~Batell and S.~Jung, \emph{{Probing Light Stops with Stoponium}},
  \href{https://doi.org/10.1007/JHEP07(2015)061}{\emph{JHEP} {\bfseries 07}
  (2015) 061} [\href{https://arxiv.org/abs/1504.01740}{{\ttfamily
  1504.01740}}].

\bibitem{CMS:2013czn}
{\scshape CMS} collaboration, \emph{{Searches for Long-Lived Charged Particles
  in $pp$ Collisions at $\sqrt{s}$=7 and 8 TeV}},
  \href{https://doi.org/10.1007/JHEP07(2013)122}{\emph{JHEP} {\bfseries 07}
  (2013) 122} [\href{https://arxiv.org/abs/1305.0491}{{\ttfamily 1305.0491}}].

\bibitem{Belanger:2018sti}
G.~B\'elanger et~al., \emph{{LHC-friendly minimal freeze-in models}},
  \href{https://doi.org/10.1007/JHEP02(2019)186}{\emph{JHEP} {\bfseries 02}
  (2019) 186} [\href{https://arxiv.org/abs/1811.05478}{{\ttfamily
  1811.05478}}].

\bibitem{CMS-PAS-EXO-16-036}
{\scshape CMS Collaboration} collaboration, \emph{{Search for heavy stable
  charged particles with $12.9~\mathrm{fb}^{-1}$ of 2016 data}},  tech. rep.,
  CERN, Geneva, 2016.

\bibitem{CMS:2015lsu}
{\scshape CMS} collaboration, \emph{{Constraints on the pMSSM, AMSB model and
  on other models from the search for long-lived charged particles in
  proton-proton collisions at sqrt(s) = 8 TeV}},
  \href{https://doi.org/10.1140/epjc/s10052-015-3533-3}{\emph{Eur. Phys. J. C}
  {\bfseries 75} (2015) 325}
  [\href{https://arxiv.org/abs/1502.02522}{{\ttfamily 1502.02522}}].

\bibitem{Cahill-Rowley:2015aea}
M.~Cahill-Rowley, S.~El~Hedri, W.~Shepherd and D.~G.~E. Walker,
  \emph{{Perturbative Unitarity Constraints on Charged/Colored Portals}},
  \href{https://doi.org/10.1016/j.dark.2018.04.003}{\emph{Phys. Dark Univ.}
  {\bfseries 22} (2018) 48} [\href{https://arxiv.org/abs/1501.03153}{{\ttfamily
  1501.03153}}].

\bibitem{Schuessler:2007av}
A.~Schuessler and D.~Zeppenfeld, \emph{{Unitarity constraints on MSSM trilinear
  couplings}},  in \emph{{15th International Conference on Supersymmetry and
  the Unification of Fundamental Interactions}}, pp.~236--239, 10, 2007,
  \href{https://arxiv.org/abs/0710.5175}{{\ttfamily 0710.5175}}.

\bibitem{Kats_2010}
Y.~Kats and M.~D. Schwartz, \emph{Annihilation decays of bound states at the
  lhc}, \href{https://doi.org/10.1007/jhep04(2010)016}{\emph{JHEP} {\bfseries
  2010} (2010) }.

\bibitem{FCC:2018vvp}
{\scshape FCC} collaboration, \emph{{FCC-hh: The Hadron Collider}: {Future
  Circular Collider Conceptual Design Report Volume 3}},
  \href{https://doi.org/10.1140/epjst/e2019-900087-0}{\emph{Eur. Phys. J. ST}
  {\bfseries 228} (2019) 755}.

\end{thebibliography}\endgroup
\end{document}